\documentclass[12pt]{scrartcl}

 \usepackage{bezier}
 \usepackage{latexsym}
 \usepackage{amsmath}
 \usepackage{amssymb}
 \usepackage{graphicx}

\newfont{\lae}{cmssdc10 scaled 1200}
\newfont{\N}{cmssdc10 scaled 1200}
\newfont{\MA}{cmmi12 scaled 1000}
\newfont{\MAK}{cmff10 scaled 1200}
\newcommand{\T}{\textstyle} 
\newcommand{\C}{\cal} 
\newcommand{\AO}{\mbox{\lae a}} 
\newcommand{\VO}{\mbox{\lae a}^\dagger} 
\newcommand{\VB}{\mbox{\lae b}} 
\newcommand{\EB}{\mbox{\lae b}^\dagger} 
\newcommand{\modi}{\varphi_i (\vec{r})}
\newcommand{\K}{\bf k_\xi}    
\newcommand{\ii}{\mbox i}       
\newcommand{\OA}{\mbox{\N A}}
\newcommand{\OQ}{\mbox{\N q}}
\newcommand{\OP}{\mbox{\N p}}
\newcommand{\OH}{\mbox{\N H}}
\newcommand{\ON}{\mbox{\N N}}
\newcommand{\oE}{\mbox{\N E}}
\newcommand{\OD}{\mbox{\N D}}
\newcommand{\OO}{\mbox{\N O}}
\newcommand{\OX}{\mbox{\N X}}
\newcommand{\OY}{\mbox{\N Y}}
\newcommand{\OR}{\mbox{\N R}}

\newcommand{\1}{\mbox{\N 1}}
\newcommand{\Or}{\mbox{\N r}}
\newcommand{\Ov}{\mbox{\N v}}

\newcommand{\MOX}{\mbox{\MAK X}}
\newcommand{\MOY}{\mbox{\MAK Y}}
\newcommand{\Omu}{\mbox{\boldmath$\mu$\unboldmath}}
\newcommand{\Orho}{\mbox{\boldmath$\rho$\unboldmath}}
\newcommand{\OrhoA}{\mbox{\boldmath$\rho^{(A)}$\unboldmath}}
\newcommand{\Vr}{\vec{r}}
\newcommand{\Vmu}{\vec{\mu}}
\newcommand{\dd}{{\rm d}} 
\newcommand{\ee}{{\rm e}} 

\newcommand{\matrize}[1]{\boldsymbol{#1}} 
\newcommand{\HSO}[1]{\mbox{\N #1}} 
\newcommand{\bra}[1]{{\bigl\langle{#1}\bigr\rvert}} 
\newcommand{\ket}[1]{{\bigl\lvert{#1}\bigr\rangle}} 

\newcommand{\abs}[1]{\lvert#1\rvert} 

 \usepackage{hyperref}

\begin{document}
\title{Nanoporous compound materials for optical applications --\\
Microlasers and microresonators}
\author{
\href{http://www.physik.tu-darmstadt.de/lto/lp/}
{F. Laeri}\\  Darmstadt University of Technology\\ D-64289 Darmstadt
  (Germany)
  \and
\href{http://darkwing.uoregon.edu/~noeckel}{J. U. N\"ockel} \\ 
Department of Physics,
    University of Oregon\\
1371 E 13th Avenue, Eugene, OR 97403
}

\date{\footnotesize Published in Vol. 6 of \textit{Handbook of Advanced Electronic and
Photonic Materials}, edited by H.~S.~Nalwa, Academic Press, San
Diego, 2001}

\maketitle
\newpage
\tableofcontents


\section{Introduction}

Since many decades nanoporous materials, for example zeolites, play
an eminently important role in the catalysis of oil refining and
petrochemistry. On the other hand, molecular sieve materials began
also to attract some attention as optical material in last years.
It was realized that their nanometer size pores allow to host guest
molecules giving so substance to a new class of optical material
with properties which neither the host, nor the guest alone could
ever possess. In this way new pigments and luminophores were
realized as well as novel optically nonlinear and switching
materials.  The various actually realized materials are reviewed in
this book in chapter \textit{Nanoporous compound materials for
optical applications -- Material design and properties}.

A closer look at the approach of arranging molecules in an ordered
framework of pores reveals a series of aspects of fundamental
interest: For example, the stereometric restrictions which the pore
framework imposes on the motional degrees of freedom of the
enclosed molecules reduces their diffusion to \textit{one
dimensional diffusion} in channel pores \cite{KRA98}. Or, as with
the concentration also the average distance between two guest
molecules is controlled, their dipolar \textit{near field
interaction} can so be adjusted. In optimal circumstances optical
excitation energy can then be transferred nonradiatively over
distances of several micrometers \cite{GFE98}. That are just two
examples of new phenomena in molecular sieves, which at this moment
are still studied to achieve full understanding, and which will
soon find their way into applications in science and technology.

\begin{figure}[!ht]
\begin{center}
 \resizebox{0.5\textwidth}{!}{%
 \includegraphics{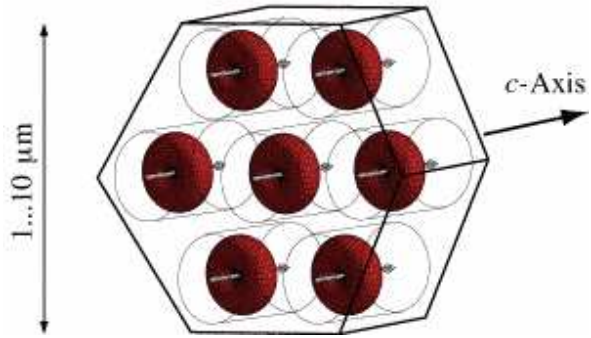}
 }
\end{center}
\caption{Schematic picture of the arrangement of the molecular dye
dipoles in the channel pores of the hexagonal molecular sieve host
crystal AlPO$_4$-5. The spatial cosine-square emission efficiency
of a dipole is represented as doughnut shaped surface around the
dipole axes.}
 \label{Microlasersketch}
\end{figure}

\medskip
In this chapter we will be concerned with another new optical
application of nano\-po\-rous compound materials, namely
microlasers in which light is generated by organic dyes embedded in
wavelength size resonators of molecular sieve material. In these
laser devices the dye molecules are enclosed in nanometer size
channel pores of the molecular sieve, whereas the crystalline sieve
material itself which forms the resonator has exterior dimensions
on the order of a few micrometers. Figure \ref{Microlasersketch}
illustrates the arrangement of the molecular dye dipoles which are
aligned in the pores of a hexagonally shaped AlPO$_4$-5 molecular
sieve host.
%
In this example of molecular sieve material the channel pores point
to the direction of the crystal $c$-axis. If the enclosed species
of dye molecules have an elongated shape and exhibit a transition
dipole moment along their elongation axis, then the dipoles end up
oriented parallel to the host $c$-axis as well.

\begin{figure}[!ht]
\begin{center}
\begin{minipage}{0.35\textwidth}
 \resizebox{1.0\textwidth}{!}{%
 \includegraphics{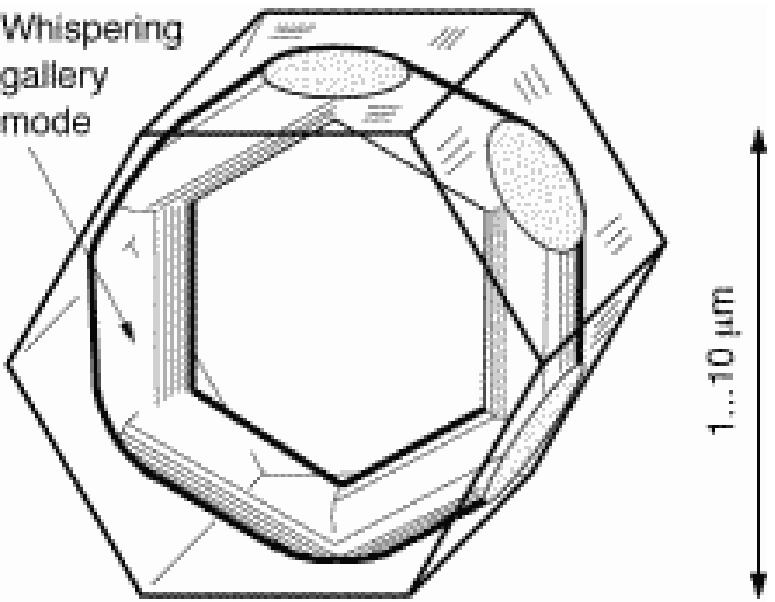}
 }
\end{minipage}
\hfill
\begin{minipage}{0.55\textwidth}
\resizebox{1.0\textwidth}{!}{%
 \includegraphics{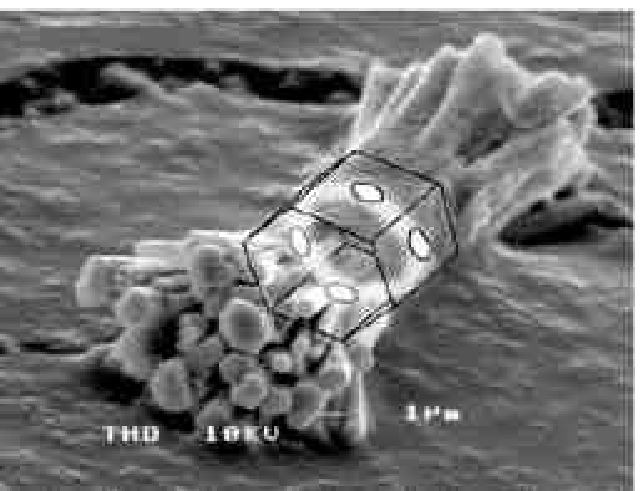}
 }
\end{minipage}
\end{center}
 \caption{
 \label{whisp}
 Light which is confined by total internal reflection at the
 hexagonal boundaries can circulate inside the molecular sieve crystal as
 \textit{whispering gallery mode}.
  }
 \label{TIR}
\end{figure}

A dipole can not emit in direction of its axis. Therefore the
emission of the dipoles shown in Fig.~\ref{Microlasersketch} occurs
in a plane normal to the host crystal $c$-axis. Once the light
emitted by the dipoles arrives at the boundary of the molecular
sieve crystal, part of it is reflected back into the material. In
fact, given the hexagonal geometry there is even a bundle of
directions for which \textit{total internal reflection} occurs.
Figure \ref{TIR} illustrates how optical rays of this bundle loop
around the crystal, and, reflected by the hexagonal sides, form a
\textit{whispering gallery mode}.
%
%
In that way regularly shaped crystals of molecular sieve hosts form
microresonator environments for the light emission of inclosed dye
molecule guests, and if the molecules provide sufficient optical
gain, laser oscillations will build up \cite{VIE98}.

\bigskip
Light emission in a microcavity environment, however, is in many
respect different from the familiar emission into free space, such
as fluorescence. In fact, this difference can be striking. For
example, to achieve laser action in a conventional millimeter size
(or larger) laser resonator the pump must overcome a certain
threshold. In a microlaser in which the resonator is of the size of
a few wavelengths, however, lasing can occur without threshold.
That means that every absorbed pump quantum is transformed into a
laser photon. Therefore, in order to appreciate the significance of
light generation in molecular sieve based microlasers, we need to
understand the differences between the emission processes of a
molecule in a free space environment as opposed to emission in a
microcavity environment.

In the following we will review spontaneous as well as stimulated
light emission of molecules, and we will show that the respective
emission rate is not an inherent molecular property, but is a
function of its environment, or more precisely, of the mode density
of the electromagnetic field. This becomes apparent when in the
distance of a half to a few wavelengths of the molecule mirrors
exist. But this is indeed exactly the situation of here discussed
dye molecules which are enclosed in a molecular sieve
microresonator.

On the other hand, we know that emission and absorption of
radiation is accompanied by a transition of the molecule from a
state with energy $E_1$ to a state with energy $E_2$. The frequency
of the emitted or absorbed light is then
$\omega=\abs{E_1-E_2}/\hbar$, where $\hbar$ is Planck's constant.
The presence of $\hbar$ clearly shows that emission and absorption
of light is intrinsically a quantum mechanical process. Therefore
our discussion will have to involve the quantum aspects of the
interaction of the dye molecules with the light field, as well as
the quantum nature of the light field itself.

The situation is even more peculiar because the resonators of the
molecular sieve microlasers have a hexagonal outline. For a
hexagonal resonator one can not find an orthogonal coordinate
system in which the wave equation can be solved by the usual method
of separation of variables. Thus we will have to discuss the
properties of microresonators in view of this impediment.

\bigskip
In consideration of these facts we have organized the discussion in
the following way: In the first section we introduce the concept of
\textit{modes} of the electromagnetic field as its countable
degrees of freedom, and based on this, we introduce the quantized
optical field. In the next section fluorescence, i.e.\ spontaneous
emission in a free space environment is discussed, and the frame is
set for the  treatment of cavity effects in the next section, in
which spontaneous emission in a resonator environment is examined.
Then the effect of a resonator on stimulated emission and laser
action are surveyed. After this we characterize the peculiarities
of microresonators, and finally we present the most recent
achievements and realizations of molecular sieve microlasers.

\section{The concept of ``modes'' of the electromagnetic
field and its quantization}\label{sect-mode}

In optics in general, and particularly when lasers are involved,
the notion of \textit{modes} is ubiquitous. As this term is used in
many different and disparate circumstances it is necessary to
define the term for further usage. In this section we give a short
tutorial introduction of the concept of \textit{modes} of the
electromagnetic field, and we show the important role the mode
concept plays in the procedure of the canonical quantization of the
field. An account of the mode structure of resonators and
particularly microresonators is then presented in section
\ref{microresonators}.

\subsection{The dynamics of the classical field}

A convenient way to introduce the mode concept is to consider a
simple realization of the electromagnetic field, for example a
\textit{source-free field}. We can think of it as the field that
subsists after a source located far away has stopped to emit. After
the emission stopped, the classical field evolution is governed by
Maxwell's equation in the following form (cgs-units):
\begin{eqnarray}
\vec{\nabla} \cdot \vec{E} = 0 & \qquad & \vec{\nabla} \times
\vec{E}  = - \frac{1}{c} \frac{\partial \vec{B}}{\partial t}
\nonumber \\ \vec{\nabla} \cdot \vec{B} = 0 & \qquad & \vec{\nabla}
\times \vec{B}  = \; \; \frac{1}{c} \frac{\partial
\vec{E}}{\partial t} \label{Max}
\end{eqnarray}

\subsection{Discretizing fields -- Random fields} \label{rand}

Most practical light sources, such as incandescent lamps or light
emitting diodes LEDs (though \textbf{not} lasers), emit a
nondeterministic, chaotic field, i.e.\ a field whose
spatio-temporal evolution can only be described in statistical
terms. In communication engineering terms such a field is referred
to as a \textit{noise field}. We like to point out that we can
understand important features of those fields in classical terms.
It is thus not necessary to enter the quantum world in order to
encounter nondeterministic fields. In order to keep the mathematics
simple, it is convenient to deal with a field occupying a discrete,
instead of a continuous number of degrees of freedom.

The trick usually used to achieve discretization consists in
confining the field into a finite volume $V$. At the end, the limit
$V\rightarrow\infty$ can be carried out, if necessary. Confinement
to a finite volume $V$ allows us to represent the \textit{spatial}
component of the field as a series (superposition) of a discrete
number of functions. As \eqref{Max} is a system of linear equations
of the fields $\vec{E}$ and $\vec{B}$ we can always represent a
solution as a linear superposition of field functions.  Of course
it is convenient to choose a complete and orthonormal set of
functions $\{\modi\}$, that also fulfill eventually given boundary
conditions. These functions are called (spatial) \textit{modes}.
Thus a mode of the classical electromagnetic field is characterized
by the following properties:
 \begin{itemize}
\item $\{\modi\}$ orthonormal:\\[-0ex]
\begin{equation}
\int_V \varphi_m^\ast(\vec{r}) \varphi_n(\vec{r}) {\rm d}V =
\delta_{mn}
\end{equation}
\item $\{\modi\}$ complete: \\[-0ex]
\nolinebreak
\begin{eqnarray}
E_x(\vec{r},t) &=& \sum_i C_i \modi {\rm e}^{-{\rm i} \omega_i t} +
\sum_i C_i^\ast \varphi_i^\ast (\vec{r}) {\rm e}^{{\rm i} \omega_i
t} \\ E_x(\vec{r},t) &=&\quad \;\qquad E_x^{(+)}\qquad + \qquad
E_x^{(-)}
\end{eqnarray}
\item $\{\modi\}$ satisfies the spatial boundary conditions given by
the shape of volume $V$.
 \end{itemize}
Obviously $E_x^{(-)} = \bigl(E_x^{(+)}\bigr)^\ast$ (with $^\ast$ we
denote complex conjugation). The sets $\{C_i\}_{x,y,z}$ are
discrete, and they now represent the complete information about the
field. In a deterministic field the individual $\{C_i\}_{x,y,z}$
are fixed complex numbers. For a nondeterministic field, however,
the $\{C_i\}_{x,y,z}$ represent random variables. Thus for a
stationary field they are defined in terms of probability
functions:
\begin{equation}
p\,(\,\{C_i\}\,) = p\,(C_1,C_2,C_3, \; \ldots\;)
\end{equation}
If we consider a function $F$ that depends on the random field, say
$E$, or $E^{(+)}$, then we can only express $F$ in statistical
terms, that means we can only assign expectation values:
\begin{equation}
\langle \, F(E^{(+)}) \, \rangle = \int_V  p\,(\,\{C_k\}\,) \; F \,
[E^{(+)}(\{C_k\})]\, \prod_k \, {\rm d}^2C_k
\end{equation}
In contrast to a thermodynamic situation in which the expectation
values are sharply peaked, an experimental realization of $F$ in
optics can significantly differ from the expectation value $\langle
\, F \, \rangle$.

\subsection{The classical Hamiltonian of the source-free
field}\label{class-HAM-sourcefree}

For a source-free field the classical Hamiltonian $H$ can be
interpreted as the total energy of the field. As we have
constrained the field to the Volume $V$, the total energy is given
by \cite{FEY64}:
\begin{equation}
 \label{KlHam}
H = \frac{1}{2} \, \int_V [ \varepsilon_0 \vec{E}^2(\vec r,t) +
\frac{1}{\mu_0} \vec{B}^2(\vec r,t)] \; {\rm d}V \quad .
\end{equation}
In the following we will show that the Hamiltonian (\ref{KlHam})
can be represented as a sum of terms that are analogous to a
harmonic oscillator (The thoughtful reader anticipates the
reason...). For that purpose we express the fields $\vec E$ and
$\vec B$ in terms of their potential.

\subsubsection{The potential of the free field}

We recall that Maxwell's equations of a source free field are gauge
invariant. In the case we discuss here we choose the Coulomb gauge,
and as a result we obtain a purely transverse field potential $\vec
A$:
\begin{equation}
\vec \nabla \; \cdot \; \vec A \; = \; 0 \quad .
 \label{eich}
\end{equation}
The electric and magnetic fields are related to the potential
according to:
\begin{equation}
\vec E (\vec r,t) = - \frac{\partial}{\partial t} \vec A(\vec r,t)
\quad , \quad \vec B (\vec r,t) =  \vec \nabla \, \times \, \vec A
(\vec r,t) \quad .
 \label{EBA}
\end{equation}
Inserting this into (\ref{Max}) we obtain the wave equation
\begin{equation}
\vec \nabla^2 \vec A\,-\, \frac{1}{c^2}\,
\frac{\partial^2}{\partial t^2} \vec A \, = \, \vec 0 \quad .
 \label{well}
\end{equation}

\subsubsection{Discretization procedure for the potential A}

We follow the thread outlined in section \ref{rand}, and to keep
things simple, we consider a cube shaped volume $V$ with an edge
length of $L$. With this conditions the discretization of $\vec A$
acquires the form of a Fourier series (plane wave expansion):
\begin{equation}
\vec A(\vec r,t) = \frac{1}{ \sqrt{\varepsilon_0}\; L^{3/2}}
\sum_{\vec k} \vec{\C A}_{\vec k}(t) \exp({\rm i} \vec k \cdot \vec
r) \quad ,
 \label{PLANE-WAVE-EXPAN}
\end{equation}
where the wave vector $\vec k$ has the components
\begin{eqnarray}
k_1 = 2 \pi n_1 / L \quad &,& \quad n_1=0,\;\pm 1,\; \pm 2, \;
\ldots \nonumber \\ k_2 = 2 \pi n_2 / L \quad &,& \quad n_2=0,\;\pm
1,\; \pm 2, \; \ldots \label{zeiger}\\ k_3 = 2 \pi n_3 / L \quad
&,& \quad n_3=0,\;\pm 1,\; \pm 2, \; \ldots \quad ,\nonumber
\end{eqnarray}
and $\sum_{\vec k}$ extends over the modes indexed by $n_1, n_2,
n_3$. The chosen normalizing factor will soon prove to be useful.
Evaluation of the gauge relation (\ref{eich}) results in
\begin{equation}
\frac{\rm i}{\sqrt{\varepsilon_0}\; L^{3/2}} \sum_{\vec k} \vec k\,
\cdot \,\vec{\C A}_{\vec k}(t) \exp({\rm i} \vec k \cdot \vec r)
\;=\;0
\end{equation}
for all $\vec r$. This is only possible when
\begin{equation}
\vec k \; \cdot \; \vec{\C A}_{\vec k}(t) \; = \; 0 \;,
 \label{trans}
\end{equation}
thus $\vec k \bot \vec{\C A}_{\vec k}$ (i.e.\ transversal field).
As the potential $\vec A(\vec r,t)$ assumes real values, the
coefficients $\vec{\C A}_{\vec k}$ observe
\begin{equation}
\vec{\C A}_{-\vec{k}}(t) = \vec{\C A}_{\vec{k}}^\ast(t) \;.
\end{equation}
In addition, wave equation (\ref{well}) must be satisfied,
resulting in
\begin{equation}
\left( \frac{\partial^2}{\partial t^2} + \omega_{\vec k}^2
\right)\; \vec{\C A}_{\vec k}(t) = 0  \;,
\end{equation}
with $\omega_{\vec k} = ck$. The general solution of this ordinary
differential equation is represented as
\begin{equation}
\vec{\C A}_{\vec k}(t) = \: \vec c_{\vec k} \; {\rm e}^{-{\rm
i}\omega_{\vec k} t} \;+\; \vec c_{-\vec k}^{\: \ast}\; {\rm
e}^{{\rm i}\omega_{\vec k} t} \;.
\end{equation}
As (\ref{trans}) fixes the transversal character of each plane wave
mode, only two components (the polarization components) of the
vector $\vec{\C A}_{\vec k}(t)$ are at free disposition. In order
to simplify the notation let us agree to let index $_{\vec k}$
point to $n_1,n_2,n_3$ (see (\ref{zeiger})), as well as to the
polarization components $\vec \xi_1,\vec \xi_2$. To remember this,
we will from now on refer to the corresponding index as $_{\K}$.
With this notation we can write for the potential
\begin{equation}
\vec A(\vec r,t)=\frac{1}{\sqrt{\varepsilon_0} L^{3/2} } \sum_{\K}
[c_{\K}  {\rm e}^{{-\rm i}\omega_{\K} t} +
c_{-\K}^{\ast} {\rm e}^{{\rm i}\omega_{\K} t}]  {\rm e}^{{\rm i}%
 \vec k \cdot \vec r} \quad ,
\end{equation}
and with
\begin{equation}
u_{\K}(t) = c_{\K}\, {\rm e}^{-{\rm i}\omega_{\K}t} \label{u}
\end{equation}
we get
\begin{equation}
\vec A(\vec r,t)=\frac{1}{\sqrt{\varepsilon_0} L^{3/2}} \sum_{\K}
[u_{\K}(t) {\rm e}^{{\rm i}\vec k \cdot \vec r } + u_{\K}^{\ast}(t)
{\rm e}^{{-\rm i}\vec k \cdot \vec r }]. \label{Au}
\end{equation}

\medskip
By inserting (\ref{Au}) in (\ref{EBA}), we can express the electric
and magnetic field as a sum of mode functions
\begin{eqnarray}
\vec E(\vec r,t)&=&\frac{\rm i}{\sqrt{\varepsilon_0} L^{3/2}}
\sum_{\K} \omega_{\K}[u_{\K}(t) {\rm e}^{{\rm i}\vec k \cdot \vec r
} - {\rm c.c.}] \label{Epot} \\ \vec B(\vec r,t)&=&\frac{\rm
i}{\sqrt{\varepsilon_0} L^{3/2}} \sum_{\K} [u_{\K}(t) (\vec k
\times \vec \xi) {\rm e}^{{\rm i}\vec k \cdot \vec r } - {\rm
c.c.}]
 \label{Bpot}
\end{eqnarray}
where with $\vec \xi$ we represent the polarization unit-vector. We
note that the total information on the (classical) field is now
contained in the functions $u_{\K}(t)$ (\ref{u}).
\\[2ex]
{\bf Note:\label{notemag}} To avoid confusion, we have to point to
a minor inconsistency in the notation: As is typical for
expressions of the magnetic field, the vector product in
(\ref{Bpot}) reshuffles spatial and polarization indices so that
the correct expression in fact consists of two sums containing the
polarization vector. The resulting expression looks bulky, and
requires some consideration. For this tutorial we prefer to
emphasize the basic mathematical structure. So we choose this
visually intuitive representation although the indices are not
correctly rendered. To obtain the correct result one may work out
the procedure in the component notation of (\ref{Au}) and
(\ref{EBA}).

\subsubsection{The mode density}

Let us return for a moment to the volume $L^3$ we considered for
deriving the mode expansion \eqref{PLANE-WAVE-EXPAN}, which was a
cube with edges oriented along the coordinate axes $(x_1,x_2,x_3)$.
According to \eqref{PLANE-WAVE-EXPAN} we can expand the field in
this cube in a 3-dimensional set of running modes
$\{k_1\},\{k_2\},\{k_3\}$ [cf.\ \eqref{zeiger}]. Along the
$x_1$-axis we have the modes labeled by $k_1=2\pi n_1/L; \;
(n_1=\pm 1,\pm 2, \dots)$; because we consider only running modes
(travelling waves), we omit $n_1=0$. In the interval between $k_1$
and $k_1+\dd k_1$ we find $\dd n_1$ modes, where $\dd n_1 = \dd k_1
\; L / 2\pi$. The analogous applies for the other directions.
\begin{figure}[!ht]
 \setlength{\unitlength}{0.75em}
\begin{center}
\begin{picture}(25.5,20)

\multiput(11.1,8)(-0.78,-0.52){10}{
\begin{picture}(0,0)\multiput(0,0)(1,0){10}{
\begin{picture}(0,0)\multiput(0,0)(0,1){10}{\circle*{0.1}}
\end{picture}}\end{picture}}

\thinlines
 \put(12,8){\vector(0,1){12}}
 \put(12,8){\vector(1,0){13}}
 \put(12,8){\vector(-3,-2){9}}
 \linethickness{0.5pt}
 \bezier{25}(12,16)(8.7,16)(6.34,13,66)
 \bezier{25}(6.34,13,66)(4,11.3)(4,8)
 \bezier{25}(4,8)(4,4.7)(6.34,2.34)
 \bezier{25}(6.34,2.34)(8.7,0)(12,0)
 \bezier{25}(12,0)(15.3,0)(17.66,2.34)
 \bezier{25}(17.66,2.34)(20,4.7)(20,8)
 \linethickness{1pt}
 \bezier{300}(20,8)(20,11.3)(17.66,13.66)
 \bezier{300}(17.66,13.66)(15.3,16)(12,16)
 \bezier{300}(12,16)(6,13)(6,4)
 \bezier{300}(6,4)(15,4)(20,8)
 \bezier{300}(19.7,8)(19.7,11)(17.46,13.46)
 \bezier{300}(17.46,13.46)(15.1,15.7)(12,15.7)
 \bezier{300}(12,15.7)(6.3,12.8)(6.3,4.2)
 \bezier{300}(6.3,4.2)(14.8,4.2)(19.7,8)
 \put(1,1){$n_1=\frac{k_1 L}{2\pi}$}
 \put(20.5,6.2){$n_2=\frac{k_2 L}{2\pi}$}
 \put(12.5,18){$n_3=\frac{k_3 L}{2\pi}$}
\end{picture}
\end{center}
\caption{
 \label{ModeDensity}
 Mode distribution in $\vec{k}$-space scaled by multiplication
 with $\frac{L}{2\pi}$. Each dot corresponds to a wavevector $\vec{k}$ which
 represents two independent, orthogonally polarized waves.
  }
\end{figure}
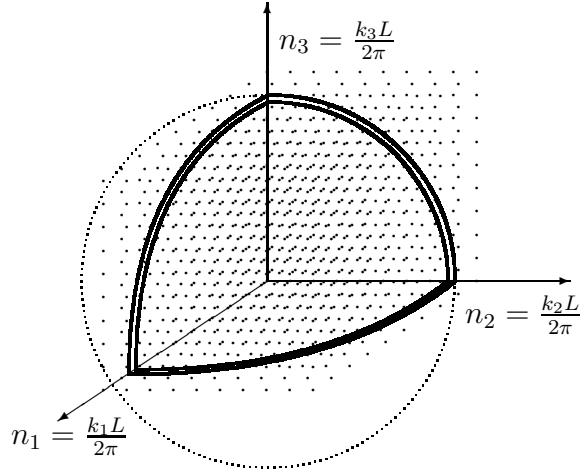
%
%
According to Fig.~\ref{ModeDensity} we can specify the number of
modes in the interval between $\abs{\vec{k}}$ and
$\abs{\vec{k}+\vec{\dd k}}$ simply by counting the dots in the
volume of the corresponding spherical shell, which amounts to $4\pi
k^2\;\dd k \cdot 2$ (factor 2 because there are two independent
polarization states associated with each $\vec{k}$). Considering
the scaling of the axes in Fig.~\ref{ModeDensity} we can use this
to express the number of modes in Volume $V=L^3$ as
\begin{equation}
\text{Number of modes in $V$} = 2 \cdot 4\pi k^2 \dd k
\;\left(\frac{L}{2\pi}\right)^3 \; .
\end{equation}
The \textit{mode density} $\rho(k)\dd k$, that is the number of
modes in interval $\abs{\vec{\dd k}}$ per Volume $V=L^3$, is then
given by
\begin{equation}
 \rho(k)\dd k = \frac{\text{\# of modes in $V$}}{V}=
 \frac{k^2 \; \dd k}{\pi^2} \; .
\end{equation}
With $k=\frac{\omega}{c}$ we obtain
\begin{equation}
\rho(\omega)\dd\omega = \frac{\omega^2}{c^3\pi^2}\;\dd\omega\; .
\label{FREE-SPACE-MODE-DENSITY}
\end{equation}
This is known as the \textit{free space mode density}. Note that
$\rho(\omega)$ represents the volume normalized density of modes,
i.e.\ the units of $\rho(\omega)$ are number of modes / (frequency
$\times$ volume).

Let us now consider a linear function $\zeta$ of the field, which
thus can be represented as a sum over the field modes $\vec{k}$ as
\begin{equation}
\zeta = \sum_{k_1,k_2,k_3} \zeta(\vec{k}) \;.
\end{equation}
For a large cube, $L\to\infty$, the sum over the discrete set of
modes has to be transformed into an integral, which has to sum over
a measure that is a density
\begin{equation}
\frac{1}{L^3} \sum_{k_1,k_2,k_3} \zeta(\vec{k}) \longrightarrow
\frac{1}{(2\pi)^3}\,\int \dd^3k\; \zeta({\vec{k}}) \; .
\end{equation}
Switching to spherical coordinates $\dd^3 k= k^2\dd
k\;\sin\theta\,\dd\theta\;\dd\phi$, where $\phi$ denotes the
azimuthal and $\theta$ the polar angle of $\vec{k}$, and
transforming this equation to frequency space $\omega$ we obtain
observing $k=\omega/c$
\begin{equation}
\frac{1}{L^3} \sum_{k_1,k_2,k_3} \zeta(\vec{k}) \longrightarrow
\frac{1}{(2\pi)^3}\,\int
\dd\omega\;\frac{\omega^2}{c^3}\;\int_0^\pi\dd\theta\,\sin\theta\,\int_0^{2\pi}
\dd\phi\; \zeta({\vec{k}})
 \label{SUM-TO-OMEGAINT}
\end{equation}

In this representation we take into account that $\zeta$ can depend
on the polarization, which can be expressed as $\zeta
=\zeta(\phi,\theta,k)$. Also, $\zeta$ allows us to introduce
$\vec{k}$-space structure functions into the mode counting sum, so
that the mode density can be calculated for arbitrary boundary
conditions. Note that the procedure relies on certain properties of
the field, one of which is \eqref{KlHam}. This equation expresses
energy conservation, thus the above calculated mode density refers
to undamped modes. The situation with microlasers, where sources of
the field are present and the modes are damped, is more intricate
and is discussed in section \ref{microresonators}.

\subsubsection{The classical Hamiltonian of the source free field}

Let us insert the electric field (\ref{Epot}) and the magnetic
field (\ref{Bpot}) into the energy (\ref{KlHam}) (note that
$\int_{L^3} {\rm e}^{{\rm i}(\vec k - \vec{k^\prime})\cdot \vec
r}{\rm d}^3r = L^3\, \delta_{\vec k \vec{k^\prime}}$), then we
obtain the Hamiltonian
\begin{equation}
H = 2 \sum_{\K}\omega^2 \, | u_{\K}(t) |^2 \;.
 \label{H}
\end{equation}
The energy appears here as the sum of the energy of each mode. This
is intuitive, and at this point we could stop satisfied with the
result. However, there is one thing that experience revealed: As
elegant it might be, expression (\ref{H}) does not lend itself to
an idea that opens a viable way for the quantization of the
electromagnetic field.

Although at this time it looks utterly artificial, but the
following substitution proved to be a wonderfully prolific device:
\begin{eqnarray}
q_{\K}(t) &=& \qquad \;\;[u_{\K}(t) +  u_{\K}^\ast(t)]
 \label{q}\\
p_{\K}(t) &=& -{\rm i}\omega_{\K} [u_{\K}(t) -  u_{\K}^\ast(t)]
 \label{p}
\end{eqnarray}
At this time this expressions can be regarded as the definition for
the functions $q$ and $p$ that, as can be shown, satisfy the
relations $\dot{q}_{\K}=p_{\K}$ and $\dot{p}_{\K}=-\omega^2q_{\K}$.
Thus the functions $q$ and $p$ evolve in analogy to the canonical
position and momentum variables of the harmonic oscillator: In
fact, inserting (\ref{q}) and (\ref{p}) in (\ref{H}), we obtain the
expression
\begin{equation}
H =\frac{1}{2} \sum_{\K} [p_{\K}^2(t) + \omega_{\K}^2
q_{\K}^2(t)]\quad , \label{HHO}
\end{equation}
which has the same form as the Hamiltonian of a series of
independent (not coupled) harmonic oscillators. \\[2ex]
What do we have achieved so far?
\begin{enumerate}
  \item By restricting the fields to a finite Volume $V$, the
  potential of the electromagnetic field was decomposed in a
  discrete, although infinite, series of spatial modes.
  \item Each spatial mode was associated with a harmonic oscillator.
  \item That this is a sound result can be seen by verifying the
  canonical equations of motion, which indeed show that
  $\partial H / \partial p_{\K}
  = \dot{q}_{\K}$ and $\partial H / \partial q_{\K} =
  -\dot{p}_{\K}$.
\end{enumerate}

To be complete we write the fields expressed in terms of $q$ and
$p$ :
\begin{eqnarray}
\vec A(\vec r,t) &=&  {\T \frac{1}{2 \sqrt{\epsilon_0}
L^{\frac{3}{2}}}  } \sum_{\K}  \left\{ \!  \left[ q_{\K}(t) + \!\!
{\T \frac{\rm i}{\omega_{\K}}   } p_{\K}(t)\right] {\rm e}^{{\rm i}
\vec k \cdot \vec r} \!\! + {\rm c.c.}\! \right\}
 \label{AHO} \\
\vec E(\vec r,t)  &=&  {\T \frac{\rm i}{2 \sqrt{\epsilon_0}
L^{\frac{3}{2}}}  }  \sum_{\K} \left\{ \! \left[\omega_{\K}
q_{\K}(t) + \!\! {\rm i} p_{\K}(t)\right] {\rm e}^{{\rm i} \vec k
\cdot \vec r} \!\! - {\rm c.c.}\! \right\}
 \label{EHO} \\
\vec B(\vec r,t)  &=&  {\T \frac{1}{2 \sqrt{\epsilon_0}
L^{\frac{3}{2}}}  } \sum_{\K} \left\{ \!  \left[ q_{\K}(t) + \!\!
{\T \frac{\rm i}{\omega_{\K}}   } p_{\K}(t)\right]  (\vec k \times
\vec \xi {\rm e}^{{\rm i} \vec k \cdot \vec r})  - {\rm c.c.}
\right\}
 \label{BHO}
\end{eqnarray}
{\bf Note:} As in (\ref{Bpot}) the same inconsistency arises here
with (\ref{BHO}). See the note on p.~\pageref{notemag}.

\bigskip
Thinking about quantization of the electromagnetic field? Well,
quantizing a harmonic oscillator - we know how to do this, don't
we?

\subsection{Canonical quantization of the electromagnetic
field}\label{CanonicalQieldQuantization}

We assume that the reader is familiar with the postulates of
quantum mechanics in general, and with the quantum mechanics of the
harmonic oscillator in particular (see for example \cite{MER70}).
In anticipation of the following procedures, we have already put
together the main constituents we need to apply the
\textit{correspondence principle} \cite[p.\ 337ff]{MER70} for
quantizing the field. As first step we associate with each of the
classical dynamic variables a Hilbert-space operator. At this time
we choose operators in the Heisenberg picture. In this way the
correspondence to the classical formulas is more obvious. As always
in quantum mechanics we also have to carefully consider commutation
of the operators.

According to the correspondence principle intent
\cite{HEI54,LOU73}, we associate the following operators with the
classical canonical variables (\ref{q}) and (\ref{p}):
\begin{eqnarray}
q_{\K}(t) & \longrightarrow & \OQ_{\, \K}(t) \quad\mbox{and} \\
 \label{oopq}
 p_{\K}(t) &
\longrightarrow & \OP_{\K}(t) \quad .
 \label{oopp}
\end{eqnarray}
The operators $\OQ$ and $\OP$ of the same mode represent
noncompatible operators. We use that the commutator of a
canonically conjugated pair of operators amounts to $\ii \hbar$. On
the other hand, equation (\ref{HHO}) indicates that the  modes
(oscillators) are uncoupled. Therefore the associated Hilbert-space
operators of different modes will commute. Thus we obtain the
following commutation relations:
\begin{eqnarray}
[ \OQ_{\,\K}(t), \OP_{\K^\prime}(t) ] &=&
  \ii \hbar \, \delta^3_{\K \K^\prime}
  \label{KommPQ} \\
 { [ } \OQ_{\,\K}(t), \OQ_{\,\K^\prime}(t){ ] } &=& 0
 \label{KommQQ} \\
 { [ } \OP_{\K}(t), \OP_{\K^\prime}(t){ ] } &=& 0 \qquad ,
 \label{KommPP}
\end{eqnarray}
and accordingly, the following uncertainty relation:
\begin{eqnarray}
\Delta q_{\K} \Delta p_{\K^\prime} &\ge& {\T
\frac{1}{2}}\;|\langle[\OQ_{\,\K},\OP_{\K^\prime}]\rangle | \\
&\ge& {\T \frac{\hbar}{2}}\; \delta_{\K\K^\prime}^3  \qquad .
 \label{DqDp}
\end{eqnarray}

\bigskip
The state of a quantum mechanical system -- here the
electromagnetic field -- is characterized by a state vector, say
the ket $|\psi\rangle$. The measurement of the observable $O$ will
produce a value coinciding with an eigenvalue of the Hilbert space
operator {\N O} associated with the observable $O$. However, only
when the state of the field $|\psi\rangle$ happens to be an
eigenstate of the operator {\N O}, we can precisely predict the
outcome of the measurement of $O$. If the state of the system does
not coincide with an eigenstate of {\N O}, then it is only possible
to predict the probability to obtain a certain eigenvalue. Thus the
outcome of the measurement of $O$ is predicted by the probability
distribution given through the scalar product $\langle \psi |
\mbox{\N O} | \psi \rangle$.

At this point the profit we gained by reinterpreting the classical
Hamiltonian (\ref{H}) in the form of (\ref{HHO}) becomes apparent.
By identifying the classical observables in (\ref{HHO}) with the
operators (\ref{oopq}) and (\ref{oopp}), we obtain a Hamiltonian
operator that exhibits required quantum properties. Thus
\begin{equation}
\OH = \frac{1}{2}\sum_{\K} [\OP^2_{\K}(t) + \omega^2_{\K}
\OQ^2_{\K}(t) ] \;. \label{QH}
\end{equation}
Note that with the transposition of the classical problem into
Hilbert space, the variable space was transposed as well: The new
dynamic variables now are the operators $\OQ_{\,\K}(t)$,
$\OP_{\K}(t)$, as well as the field operators $\vec{\mbox{\N
A}}(\vec r,t)$, $\vec{\mbox{\N E}}(\vec r,t)$, $\vec{\mbox{\N
B}}(\vec r,t)$, etc.. The classical variables $\vec r$ and $t$ are
now relegated to play the role of mere parameters.

\bigskip
In the following we try to find out why the analogous transposition
of the classical Hamiltonian (\ref{H}) fails to produce a
Hamiltonian operator that is consistent with the quantum mechanical
postulates. We will start to seek the quantum mechanical equivalent
of mode functions $u_{\K}(t)$ appearing in (\ref{q}) and (\ref{p}).
Obviously these must be operators.

\subsection{Creation and annihilation operators}

The operators that play the analogous role to the functions
$u_{\K}(t)$ in (\ref{q}) and (\ref{p}) are called annihilation
operator $\VO_{\K}(t)$ and creation operator
$\AO_{\K}(t)$\footnote{Although they play an analogous role, they
do not \textit{correspond} to the classical mode functions in the
sense of the correspondence principle, as we will soon discover.}.
They can be expressed in terms of the Hermitian operators $\OQ_{\,
\K}(t)$ and $\OP_{\K}(t)$ as the following pair (the normalization
will be justified later):
\begin{eqnarray}
\AO_{\K}(t) &=& {\T \frac{1}{\sqrt{2 \hbar \omega_{\K}}}}
[\omega_{\K} \OQ_{\, \K}(t) + \ii \OP_{\K}(t) ]
 \label{defAO} \\
\VO_{\K}(t) &=& {\T \frac{1}{\sqrt{2 \hbar \omega_{\K}}}}
[\omega_{\K} \OQ_{\, \K}(t) - \ii \OP_{\K}(t) ] \quad .
 \label{defVO}
\end{eqnarray}
The comparison with (\ref{q}) and (\ref{p}) also shows that these
operators exhibit the the time dependence
\begin{eqnarray}
\AO_{\K}(t) &=& \AO_{\K}(0) {\rm e}^{- \ii \omega_{\K}t}
 \label{At}\\
\VO_{\K}(t) &=& \VO_{\K}(0) {\rm e}^{\ii \omega_{\K}t} \quad .
 \label{Vt}
\end{eqnarray}

Similarly as the classically equivalent functions in (\ref{q}) and
(\ref{p}) are complex conjugates, the annihilation and creation
operators are Hermitian adjoints, which is expressed by the symbol
$^\dagger$. Since $\VO_{\K}(t) \not= \AO_{\K}(t)$, they are not
themselves Hermitian but Hermitian adjoints, and thus \textit{can
not qualify as physical observables}. Like $\OQ_{\, \K}(t)$ and
$\OP_{\K}(t)$ they are not compatible (do not commute). From
(\ref{KommPQ}) and (\ref{KommPP}) we obtain the commutation
relations
\begin{eqnarray}
{[} \AO_{\K}(t), \VO_{\K^\prime}(t) {]} &=& \delta^3_{\K \K^\prime}
\quad ,
 \label{KommVE} \\
{[} \AO_{\K}(t) , \AO_{\K^\prime}(t) {]} &=& 0 \quad ,
 \label{KommVV} \\
{[} \VO_{\K}(t) , \VO_{\K^\prime}(t) {]} &=& 0    \quad .
\label{KommEE}
\end{eqnarray}
However the dynamic variables $\OQ_{\, \K}(t)$ and $\OP_{\K}(t)$
are Hermitian, and can consequently be associated with observables
\begin{eqnarray}
\OQ_{\, \K}(t) &=& {\T \sqrt{\frac{\hbar}{2 \omega_{\K}}}}
[\AO_{\K}(t) + \VO_{\K}(t) ] \quad ,
 \label{qq} \\
\OP_{\K}(t) &=& \ii \; {\T \sqrt{\frac{\hbar \omega_{\K}}{2}}}
[\VO_{\K}(t) - \AO_{\K}(t) ] \quad .
 \label{qp}
\end{eqnarray}

\bigskip
After comparison of these two equations with (\ref{q}) and
(\ref{p}), we are confident that $\AO_{\K}(t)$ and $\VO_{\K}(t)$
play an analogous role as the classical mode functions $u_{\K}(t)$
and $u_{\K}^\ast(t)$. Therefore, it is the set of operators
$\{\AO_{\K}(t)\}$ that in the quantum analogy carries the total
information about the field. As the information of each quantized
field mode is now expressed in terms of an operator, instead of
simple $c$-numbers, we must expect a correspondingly richer
structure of the quantized field -- richness in the sense that in
the quantum description, \textit{non classical field}
configurations arise that do not have a classical equivalent.

\bigskip
We can now express the Hamiltonian operator in terms of
annihilation and creation operators. We insert (\ref{qq}) and
(\ref{qp}) into (\ref{QH}) and obtain
\begin{equation}
\OH = \frac{1}{2} \sum_{\K} \hbar \omega_{\K} [\AO_{\K}(t)
\VO_{\K}(t) + \VO_{\K}(t) \AO_{\K}(t)] \label{SYMHO}
\end{equation}
In this equation the annihilation and the creation operators appear
in a symmetric way. From its visual appearance (\ref{SYMHO}) can be
considered to closely resemble the classical Hamiltonian (\ref{H}).
An the other side we have to consider that annihilation and
creation operators do not commute, and the apparent symmetry may
only be a typographical one. There is no doubt that the Hamiltonian
operator per se is Hermitian and thus represents an observable, the
energy of the field. The energy of an optical field is usually
measured by absorbing the light in a photo detector, say for
example a photo diode. As will become obvious below [c.f.\
(\ref{EZ})-(\ref{TZ})] for the description of absorption processes
it is practical to have the Hamiltonian operator in a form in which
the annihilation operators stand to the right of the creation
operators. This \textit{normal ordering form} is achieved by
successive application of the commutation relations
(\ref{KommVE})-(\ref{KommEE}). As a result we obtain
\begin{equation}
\OH = \sum_{\K} \hbar \omega_{\K} [ \VO_{\K}(t) \AO_{\K}(t) + {\T
\frac{1}{2}}]  \quad.
 \label{NGHO}
\end{equation}
Obviously the operator product  $\VO_{\K}(t) \AO_{\K}(t)$ is
Hermitian, and represents the \textit{number operator} $\ON_{\K}$
\begin{equation}
\ON_{\K} = \VO_{\K}(t) \AO_{\K}(t)  \quad .
 \label{ONumb}
\end{equation}
As its name indicates, $\ON_{\K}$ counts the number of photons in
each mode ${\K}$. Note that because of  (\ref{At}) and (\ref{Vt})
the number operator $\ON_{\K}$ is constant in time -- a  mandatory
property for a conserved quantity of the field, like  $\OH$
(\ref{NGHO}). The eigenvalues of the Hamiltonian operator
(\ref{NGHO}) are $\hbar \omega_{\K}\times ( n_{\K}+ \frac{1}{2}$),
with $n_{\K} = 0,1,2,3,\ldots,\infty$. It is evident that $n_{\K}$
represents the photon occupation number of mode ${\K}$. In
(\ref{NGHO}) the contribution of $\frac{1}{2} \hbar \omega_{\K}$ to
each mode represents the \textit{vacuum fluctuations} of the field.
Clearly, this term is not present in the classical Hamiltonian
(\ref{H}).

At this point we should have discussed the eigenfunctions of the
annihilation and creation operators. We postpone the discussion to
the following section. in which we introduce the coherent states,
and were we are able to better illustrate the special meaning of
those eigenstates.

\bigskip
Summarizing we can now reconstruct the reasons why the direct
application of the correspondence principle to the classical
Hamiltonian expression (\ref{H}) fails, but works on (\ref{HHO}):
\begin{itemize}
  \item The correspondence principle refers to physical
  observables. On the other side, from the perspective of classical
  physics, the concept of observables is not well defined, and therefore it
  is not possible to anticipate that the operators $\AO$ and $\VO$
  associated with the classical mode functions $u_{\K}(t)$ and
  $u_{\K}^\ast(t)$ are not Hermitian and thus do not represent
  observables.
  \item In the representations of the classical Hamiltonian the
  mode functions $u_{\K}(t)$ and $u_{\K}^\ast(t)$ enter in a
  symmetric way. The normal ordering procedure of the operators
  $\AO$ and $\VO$ in the Hamiltonian operator shows, however, that
  the equivalent operators do not contribute equally to the
  Hamiltonian. In (\ref{NGHO}) the factor $\frac{1}{2} \hbar
  \omega_{\K}$ representing the vacuum fluctuations is a result of
  this asymmetry. Since vacuum fluctuations do not exist in the classical
  picture, the term could not be anticipated.
\end{itemize}

\subsection{States of the optical field}

In classical physics the information of the electrodynamic field is
contained in the electric and magnetic fields, or their associated
potential functions. In the quantum description, however, the
information is contained in a state vector (wave function). In the
following we discuss two important states of the field, the
\textit{Fock states} (or number states,  states of fixed photon
number), and the \textit{coherent states} (harmonically oscillating
fields).

A practical way to represent a state vector is in terms of a series
of eigenstates of a suitable operator, and this is what we are
going to do as next.

\subsubsection{The Fock states (number states)}

The Fock states are defined as states with a fixed photon number.
Thus they can be represented as eigenstates of the number operator
(\ref{ONumb}). Considering the mode ${\K}$, we have the following
eigenvalue equation
\begin{equation}
\ON_{\K} |n_{\K}\rangle = \VO_{\K}(t) \AO_{\K}(t) |n_{\K}\rangle =
n_{\K} |n_{\K}\rangle \;, \label{EWGN}
\end{equation}
in which $n_{\K}$ represents the number of photons in mode ${\K}$.
The Fock states are stationary because of (\ref{At}) and
(\ref{Vt}). In fact, we can see that a ground state $|0\rangle$
(vacuum state) is associated to each mode by
\begin{equation}
\AO_{\K} |0\rangle = 0 \quad .
\end{equation}
The following relations motivate the names of the annihilation
operator $\AO$ and creation operator $\VO$ :
\begin{eqnarray}
\AO_{\K} |n_{\K}\rangle &=& \sqrt{n_{\K}} \;\; |n_{\K} - 1 \rangle
 \label{EZ}\\
\VO_{\K} |n_{\K}\rangle &=& \sqrt{n_{\K}+1} \;\;
|n_{\K} + 1 \rangle
 \label{VN}\\
\VO_{\K} \AO_{\K} |n_{\K}\rangle &=& n_{\K} \;\; |n_{\K} \rangle
 \label{TZ}
\end{eqnarray}
We can construct an arbitrary Fock state $|n_{\K}\rangle$ by an
$n_{\K}$-fold application of the creation operator to the ground
state,
\begin{equation}
|n_{\K}\rangle = {\T \frac{1}{\sqrt{n_{\K}!}}} (\VO_{\K})^{n_{\K}}
|0\rangle \quad .
 \label{Rek}
\end{equation}
The Fock states are orthogonal, and with the factor introduced in
(\ref{defAO}) and (\ref{defVO}) they are normalized,
\begin{equation}
\langle n_{\K}|m_{\K}\rangle = \delta_{nm} \quad ,
\end{equation}
and complete
\begin{equation}
\sum_{n_{\K}}^\infty  |n_{\K}\rangle \langle n_{\K}|  = \mbox{\N 1}
\quad . \label{VRFZ}
\end{equation}
They therefore form a complete system of base vectors. The Fock
state base is often used to describe fields with few photons of
high energy, for example $\gamma$-radiation. Optical fields of
visible radiation, like fields emitted by a laser, are more
suitably described by \textit{coherent states}.

\subsubsection{Coherent states}

The coherent state represents the quantum mechanical analogue of a
classical, harmonically oscillating field. To reduce clutter in the
following discussion we will pick out one spatial mode of the
field: $|\alpha\rangle_{\K} \rightarrow |\alpha\rangle$. In the
following we adhere to Glauber's ``classical'' presentation
\cite{GLA63}. The coherent state $|\alpha\rangle$ of a mode is
defined as eigenstate of the annihilation operator $\AO$:
\begin{equation}
\AO |\alpha\rangle = \alpha |\alpha\rangle
 \label{DefA}
\end{equation}
with the \textit{complex} eigenvalue $\alpha$ (remember, $\AO$ is
not Hermitian)
\begin{equation}
\alpha = |\alpha| \; {\rm e}^{\ii \phi}
\end{equation}
$\AO$ being not Hermitian, it is not surprising that the associated
eigenstates, the coherent states $|\alpha\rangle$, are not
orthogonal. Thus they can not provide a universally suitable base
system. Although coherent states can therefore not be used to
expand an arbitrary electromagnetic field in the conventional way
(see below), they represent an important class of fields, namely
harmonically oscillating fields, like radio frequency fields, or
fields emitted by well stabilized lasers, and for this reason are
interesting to characterize.

\bigskip
Fock states as eigenstates of the number operator have an intuitive
meaning, and therefore we will seek a representation of the
coherent states as a superposition of Fock states. As the coherent
state is the quantum analogue of a harmonically oscillating field,
we shall determine the corresponding electric field operator $\oE$
as well.

Formally we find the expansion of the coherent state in the Fock
base with the usual trick of inserting the unit operator
(\ref{VRFZ}):
\begin{equation}
|\alpha\rangle = \sum_{|n\rangle}  |n\rangle \langle
n|\alpha\rangle
 \label{trick}
\end{equation}
The scalar product $\langle n|\alpha\rangle$ therefore represents
the expansion coefficients. Their evaluation is performed with
(\ref{DefA}), and after we multiply from left with $\langle n|$ we
obtain
\begin{equation}
\langle n|\AO |\alpha\rangle = \alpha \langle n|\alpha\rangle
\end{equation}
Inserting the Hermite adjoint of (\ref{VN}) gives
\begin{equation} \sqrt{n+1} \langle n+1|\alpha\rangle = \alpha
\langle n|\alpha\rangle \quad .
\end{equation}
The expansion coefficients are then obtained by the analogous
application of the recursion (\ref{Rek}) as
\begin{equation}
\langle n|\alpha \rangle = \frac{\alpha^n}{\sqrt{n!}}\;\langle
0|\alpha \rangle \quad .
\end{equation}
After inserting this into (\ref{trick}), we obtain the
representation
\begin{equation}
| \alpha \rangle = \langle 0 |\alpha \rangle \sum_n
\frac{\alpha^n}{\sqrt{n!}}  \; | n \rangle \quad ,
\end{equation}
which now is to be normalized. Normalization delivers the value of
$\langle 0 |\alpha \rangle$:
\begin{equation}
1 = \langle \alpha | \alpha \rangle = |\langle 0 | \alpha
\rangle|^2 \sum_n \frac{|\alpha|^{2n}}{n!} = |\langle 0 | \alpha
\rangle|^2  \; {\rm e}^{|\alpha|^2} \quad \mbox{, thus}
\end{equation}
\begin{equation}
\langle 0 | \alpha \rangle = {\rm e}^{-\frac{1}{2}|\alpha|^2} \quad
.
\end{equation}
As a result we obtain the coherent states in the Fock base
expansion as
\begin{eqnarray}
|\alpha \rangle &=& {\rm e}^{-\frac{1}{2}|\alpha|^2} \sum_n
\frac{\alpha^n}{\sqrt{n!}}  \; | n \rangle
 \label{la} \\
\langle \alpha | &=& {\rm e}^{-\frac{1}{2}|\alpha|^2} \sum_n
\frac{(\alpha^\ast)^n}{\sqrt{n!}}  \; \langle n |
 \label{ra}
\end{eqnarray}
At this point we are able to calculate the expectation value for
finding the considered mode in a state  $|\alpha \rangle$ with $n$
photons:
\begin{equation}
p(\alpha,n)=|\langle n| \alpha \rangle |^2 =
\frac{|\alpha|^{2n}}{n!} {\rm e}^{-|\alpha|^2} \quad ,
 \label{Poisson}
\end{equation}
where $|\alpha|^2$ denotes the average photon number $\bar{n}$ of
the mode,
\begin{equation}
\bar{n} = |\alpha|^2 = \langle \alpha |\VO \AO | \alpha \rangle
\quad .
\end{equation}
The expression (\ref{Poisson}) represents a Poisson distribution
with mean value $\bar{n}= |\alpha|^2$ and variance $(\Delta n)^2 =
|\alpha|^2=\bar{n}$.

\bigskip
Equations (\ref{la})--(\ref{Poisson}) show that the the coherent
state $|\alpha \rangle$ associated with the eigenvalue $\alpha = 0$
coincides with Fock state $| 0 \rangle$, i.e.\ a state with
vanishing photon number expectation, thus called \textit{vacuum
state}.

We derived the coherent states as eigenvectors of a nonhermitian
operator. According to the postulates of quantum mechanics we can
therefore not expect that they span a base of the Hilbert space. In
fact, their scalar product does not define an orthogonal relation
\begin{equation}
\langle \alpha | \beta \rangle = \sum_n \frac{(\alpha^\ast)^n
\beta^n}{n!} {\rm e}^{-\frac{1}{2}|\alpha|^2} {\rm
e}^{-\frac{1}{2}|\beta|^2} = {\rm e}^{[\alpha^\ast
\beta-\frac{1}{2}(|\alpha|^2 +|\beta|^2)]} \quad .
\end{equation}
(Note: Although the coherent states are not orthogonal this does
not automatically imply that it is impossible to expand an
arbitrary state in a series of coherent states \cite{GLA63}.) The
absolute value of the scalar product is
\begin{equation}
|\langle \alpha | \beta \rangle|^2 = {\rm e}^{-|\alpha-\beta|^2}
\quad .
\end{equation}
Loosly speaking, the coherent states $|\alpha \rangle$ and $|\beta
\rangle$ become increasingly more orthogonal the more they are
apart. This is due to the overlap of the state vectors, which is
given by the zero point fluctuations. In fact, the coherent states
can also be represented as a displaced vacuum state. Intuitively
this corresponds to a classical field amplitude with added zero
point fluctuations. In operator notation:
\begin{equation} |\alpha \rangle = \OD(\alpha)
|0 \rangle \quad ,
 \label{Displ}
\end{equation}
where the displacement operator $\OD(\alpha)$ is represented as
\cite{GLA63}:
\begin{equation}
\OD(\alpha) = \exp{(\alpha \VO - \alpha^\ast \AO)} \quad .
\end{equation}

\subsubsection{The operator of the electric field E}

According to the correspondence principle we construct the electric
field operator $\oE$ corresponding to (\ref{EHO}) by replacing the
classical canonical variables $q$ and $p$ with the operators
(\ref{qq}) and (\ref{qp}). For the mode we consider this leads to
\begin{equation}
\oE(t) = {\T \frac{1}{\sqrt{\epsilon_0 L^3}}} \; [\OP(0) \cos(\vec
k \cdot \vec r - \omega t) -
 \omega \OQ(0) \sin(\vec k \cdot \vec r - \omega t)] \quad ,
 \label{OpE}
\end{equation}
and when we insert (\ref{defAO}) and (\ref{defVO})
\begin{equation}
\oE(t) = {\T \ii \sqrt{\frac{\hbar \omega}{2 \epsilon_0 L^3}}} \;
[\AO(t){\rm e}^{{\rm i}\vec k \cdot \vec r}
 \;-\;  \VO(t){\rm e}^{-{\rm i}\vec k \cdot \vec r}] \quad ,
 \label{E-FIELDOP}
\end{equation}
where in the literature the factor
\begin{equation}
E_p = \sqrt{\frac{\hbar \omega}{2 \epsilon_0 L^3}}
 \label{E-FIELD-PER-PHOTON}
\end{equation}
is often called \textit{electric field per photon}. Expression
(\ref{OpE}) particularly illustrates the following points:
\begin{itemize}
\item The field consists of two components oscillating \textit{in
quadrature}.
\item The operators $\OQ(t=0)$ and
$\OP(t=0)$ associated with the quadratures are Hermitian and
constant in time.
\item However, $\OQ(t=0)$ and $\OP(t=0)$ are not compatible. That
means the quadratures are not simultaneously measurable quantities,
a result that from a classical point of view can not be
anticipated.
\item The quantum properties of the field, including its fluctuations,
are determined at one time, for example $t=0$, and do not change at
later times (provided the field is not further manipulated).
\item As the two field quadrature operators are not compatible, we
should not be surprised that it is not possible to construct a
Hermitian operator for the phase of the field.
\end{itemize}

\subsubsection{The uncertainties of the coherent state}

Remember that the coherent states are defined as eigenstates of the
annihilation operator and not of the quadrature operators  $\OQ$
and $\OP$. Therefore, the expectation value of latter observables
can not be exactly defined, but will exhibit an amount of
uncertainty. In addition, we have seen that the quadrature
operators are incompatible, thus, they obey the uncertainty
relation (\ref{DqDp}). Let us now determine the uncertainty product
$\Delta q \Delta p$ for a coherent state. (Note that $(\Delta
A)^2=\langle(\mbox{\N A}-\langle \mbox{\N A} \rangle \mbox{\N
1})^2\rangle = \langle \mbox{\N A}^2 \rangle - \langle \mbox{\N
A}\rangle^2$.)

We calculate the expectation value of the observables $q$ and $p$
for a coherent state $|\alpha\rangle$ by inserting (\ref{qq}) and
(\ref{qp}). Referring to (\ref{DefA}), we obtain
\begin{eqnarray}
\langle \OQ \rangle_{\rm coh.st.} = \langle \alpha |\OQ(0) |\alpha
\rangle &=& \sqrt{\hbar / 2 \omega} \; [\alpha + \alpha^\ast]
 \label{EWq}\\
\langle \OP \rangle_{\rm coh.st.} = \langle \alpha | \OP(0) |\alpha
\rangle &=& \ii \sqrt{\hbar \omega/2} \; [-\alpha + \alpha^\ast]
\quad ,
 \label{EWp}
\end{eqnarray}

After application of the commutation relations
(\ref{KommPQ})-(\ref{KommPP}) and (\ref{qq}) we find for the
operator $\OQ^2$
\begin{eqnarray}
\OQ^2 &=& {\T \frac{\hbar}{2 \omega}}\; [\AO^2 + {\VO}^{2} + \AO
\VO + \VO \AO] \\ &=& {\T \frac{\hbar}{2 \omega}}\; [\AO^2 +
{\VO}^{2} + 2 \VO \AO + 1] \: \mbox{, \quad and}
\end{eqnarray}
\begin{equation}
\langle \OQ^2 \rangle_{\rm coh.st.} = \langle \alpha | \OQ^2|\alpha
\rangle = {\T \frac{\hbar}{2 \omega}}\; [\alpha^2 + {\alpha^\ast}^2
+ 2 \alpha^\ast \alpha +1 ] \quad .
\end{equation}
The width of $\Delta q$ for a coherent state $|\alpha\rangle$ is
then given by
\begin{equation}
(\Delta q)^2_{\rm coh.st.} = \langle \OQ^2 \rangle -  \langle \OQ
\rangle^2 = \frac{\hbar}{2 \omega} \quad .
\end{equation}
Analogously we obtain for the conjugated observable $\OP$
\begin{equation}
(\Delta p)^2_{\rm coh.st.} = \langle \OP^2 \rangle -  \langle \OP
\rangle^2 = \frac{\hbar \omega}{2} \quad ,
\end{equation}
and for their product
\begin{equation}
(\Delta q)_{\rm coh.st.}(\Delta p)_{\rm coh.st.} = \frac{\hbar}{2}
\quad .
\end{equation}
This results shows that this uncertainty product assumes the
minimum value allowed by Heisenberg's famous relation; cf.\
(\ref{DqDp}). Thus, the coherent state is a \textit{minimum
uncertainty state}, and can be considered to be as close to the
equivalent classical state as quantum mechanics permits.

\bigskip
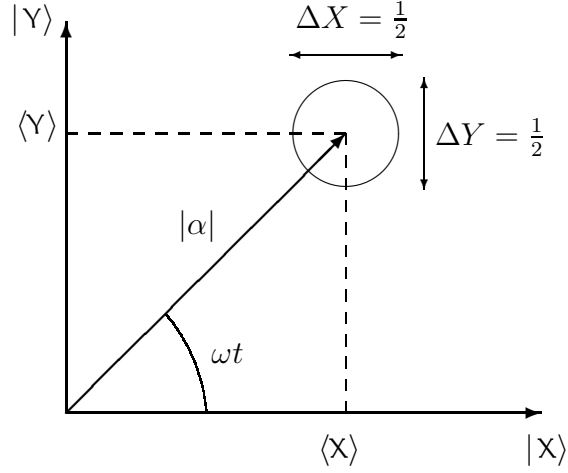
\begin{figure}[!ht]
 \setlength{\unitlength}{0.18em}
\begin{center}


\begin{picture}(100,80)
\thicklines
\put(0,0){\vector(1,0){85}}         
\put(0,0){\vector(0,1){70}}         
\put(0,0){\vector(1,1){50}}       
\thinlines \put(64,40){\vector(0,1){20}}
\put(64,60){\vector(0,-1){20}} \put(40,64){\vector(1,0){20}}
\put(60,64){\vector(-1,0){20}} \put(50,50){\circle{20}}
\put(0,0){\dashbox{2}(50,50){}}
\bezier{200}(17.7,17.7)(24.1,10)(25,0)  
\put(26,9){$\MA \omega t$} \put(45,-8){$\MA \langle {\MOX} \rangle
$} \put(-9,50){$\MA \langle {\MOY} \rangle $} \put(82,-8){$\MA |\,
{\MOX} \rangle $} \put(-10,69){$\MA |\, {\MOY} \rangle $}
\put(20,32){$\MA |\alpha|$} \put(41,69){$\MA \Delta X=
\frac{1}{2}$} \put(66,48){$\MA \Delta Y = \frac{1}{2}$}

\end{picture}
\end{center}
\caption{ \label{Fig-Coh-State} Quadrature representation of a
coherent state $|\alpha\rangle$ according to equations
(\protect\ref{QX})--(\protect\ref{UY}). Note that the quadrature
operators are not compatible, and therefore, can not be
simultaneously determined. Thus, the uncertainty circle in this
picture should not be interpreted as the contour of a classical
joint probability distribution of observables $X$ and $Y$, but as
Wigner distribution function.
 }
\end{figure}
To conclude this section we use (\ref{EWq}) and (\ref{EWp}) to
visualize the nature of the coherent state. For that purpose we
write the expectation values of the quadratures in their time
dependent form. In (\ref{OpE}) we can see that already the
operators exhibit a purely scalar time dependence ($c$-number).
Thus
\begin{eqnarray}
\langle \OQ(t) \rangle_{\rm coh.st.} = \langle \alpha |\OQ(t)
|\alpha \rangle &=& \sqrt{2 \hbar / \omega} \; |\alpha| \cos{\omega
t}
 \label{Bq}\\
\langle \OP(t) \rangle_{\rm coh.st.} = \langle \alpha | \OP(t)
|\alpha \rangle &=& -\sqrt{2 \hbar \omega} \; |\alpha| \sin{\omega
t} \;.
\label{Bp}
\end{eqnarray}
The expectation values for the quadrature observables $q$ and $p$
of the coherent state thus evolve in the way of a classical,
harmonic field. With operators normalized to the photon energy we
get
\begin{equation}
\OX = \OQ \sqrt{\frac{\omega}{2 \hbar}}\;, \qquad \OY = - \OP
\sqrt{\frac{1}{2 \hbar \omega}} \quad ,
 \label{Quadraturen}
\end{equation}
and for the expectation values respectively
\begin{eqnarray}
\langle \OX \rangle_{\rm coh.st.} &=&  |\alpha| \cos{\omega t}
\label{QX} \\ \langle \OY \rangle_{\rm coh.st.} &=&  |\alpha|
\sin{\omega t}
 \label{QY} \\
(\Delta X)_{\rm coh.st.} &=& {\T \frac{1}{2}}
 \label{UX}\\
(\Delta Y)_{\rm coh.st.} &=& {\T \frac{1}{2}}
 \label{UY}
\end{eqnarray}

We picture these equations in Fig.\ \ref{Fig-Coh-State}  (the
quadratures are represented in units of the photon energy $\hbar
\omega$). The uncertainty is equal to the vacuum fluctuation energy
$\frac{1}{2}\hbar \omega$ of the considered field mode, and the
amplitude $|\alpha|$ is characterized by the average photon number.
In summary, Fig.\ \ref{Fig-Coh-State} illustrates that the coherent
state can be interpreted as a classical, harmonic field (amplitude
$|\alpha|$) with added vacuum fluctuations (uncertainty circle).

\bigskip
We said that the coherent state represents the quantum mechanical
analogue of a classical, harmonically oscillating field. Most
people connect this with images of same sort of antenna and radio
frequency fields. In fact, we are used to modern electronic devices
which process radio frequency cycle times of up to some tens of ps
($10^{-10}$\,s). For optical frequencies, however, electrons would
have to respond on the $10^{-15}$\,s-scale. Circuits with such fast
responses are not available yet. The standard way to receive
optical signals is therefore not with an antenna, but with a photo
detector, such as a photodiode or a photomultiplier. In a photo
detector an incident photon liberates an electron. The generated
\textit{photoelectrons} may then be  processed with available
electronic devices. For the ideal detector (photon to electron
conversion efficiency, or quantum efficiency, $\eta = 1$) we can
thus regard the electron current at the detector output as an exact
image of the photon current at the input.

In practice we are thus faced with the following question: how does
the field state translates into practically observable features of
the photo current? This question can only be answered in
statistical terms. If we assume that the optical field consists of
one mode in a coherent state, then (\ref{Poisson}) contains the
answer. The probability to find the detector output releasing $n$
electrons is therefore given by the same Poisson distribution
\begin{equation}
p(n) = \frac{\bar{n}^{2n}}{n!} {\rm e}^{-\bar{n}^2} \quad ,
 \label{El-Poisson}
\end{equation}
The Poisson distribution characterizes a point process (here the
process of outputting photons, or electrons respectively), that
means that the quanta are emitted independently and therefore only
exhibit delta function like correlations. Loosely speaking, the
Poisson distribution is the distribution  with the ``most random''
properties -- most random also in the sense that the underlying
process can be realized in the greatest number of different ways.
It is interesting to realize that the state that reveals the most
random distribution if we detect its quanta, reveals a perfectly
determined oscillatory evolution in case we choose to detect its
electric field. Anyway, from the properties of the photo electron
distribution (\ref{El-Poisson}) and the underlying Poisson (or
point) process we can deduce the statistical measures we are
interested in, like photo electron current average over an interval
$\tau$ (which is $\bar{n}\tau$), its standard deviation (which is
$\sqrt{\bar{n}\tau}$), or its spectrum (which for a point process
is the Fourier transform of delta function).

Let us now consider a field, which was emitted by a single mode
laser, say 1~mW at 500~nm wavelength. Because the laser is well
stabilized this field corresponds to a coherent state. According to
above discussion the number of emitted photons per second
$\bar{n}/\tau$  corresponding to 1~mW amounts to $2.5 \cdot
10^{15}$ photons per second, and the standard deviation (RMS shot
noise) of the current is $5 \cdot 10^{7}$ photons per second.
(After photo detection the photons appear as electrons.) In this
quantum detection picture, what makes us saying that the coherent
state represents the quantum mechanical analogue of a classical,
harmonically oscillating field? The quantum mechanical aspect of
the coherent state is its noise, here $5 \cdot 10^{7}$~s$^{-1}$.
The classical aspect of the coherent state becomes visible if we
imagine adding or removing one photon to the coherent state. In
principle we disturb the quantum state by such a manipulation. But
already for coherent states with few photons $\bar{n}$, this
perturbation is smaller than the RMS shot noise level. In above
example the perturbation is in the order of $10^{-15}$ of the
average current, which fluctuates naturally by a factor of
$10^{-6}$. Clearly such a perturbation is very hard, if ever, to
detect. The classical property of the coherent state therefore is
its insensitivity for disturbances on the level of a few quanta.
This quasi classical behavior of the coherent state forms the basis
on which the \textit{semiclassical} approach for describing the
interaction of atoms or molecules (quantum objects) with the
optical field (classical object) is built. For that purpose we keep
only the terms that are significant for a field with a large number
of photons (limit $\hbar \rightarrow 0$. That is, we cut off the
factor $\frac{1}{2}$ in the Hamiltonian (\ref{NGHO}), resulting in
the semiclassical Hamiltonian
\begin{equation}
\OH = \sum_{\K} \hbar \omega_{\K} \cdot \VO_{\K}(t) \AO_{\K}(t)
\quad.
 \label{semiclH}
\end{equation}

Another interesting property is the following: A coherent state can
be generated by a point process. If we attenuate the field with
another point process, the result is still a coherent state,
although one with a reduced $\bar{n}$. Most practical attenuators,
like grey glass, or semi transparent mirrors, are such attenuators.
This is certainly a property we are used to expect for a classical,
harmonically oscillating field, but which is not automatically true
for an arbitrary realization of a quantum state.

\bigskip
Where do we stand now? When we started this discussion we assumed a
source free field. As we have stated at the beginning, we have to
imagine this as the field that subsists after a  source located far
away has stopped to emit. This assumption allowed us to learn about
some essential features of the basic quantum properties of the
electromagnetic field. For example, (\ref{OpE}) illustrates that if
the source were harmonic, the state of the radiated field consists
of a coherent state, and the spectrum an observer records shows a
single sharp line at frequency $\omega$. From a practical point of
view such fields are not very useful. Who ever wants to wait until
some charges in a distant galaxy stopped to wiggle so that we can
receive their coherent field in our laboratory to start the
important spectroscopic investigation we plan to? No, we need
fields we can turn on and off in the way we need them. That means,
we need the sources in our hands, in our laboratory. The sources we
are particularly interested in are molecular sieve microcrystals
that contain fluorescent dye molecules, and under favorable
conditions these molecular-sieve-dye compounds can emit laser
radiation.

So let us proceed the discussion with the simplest source of an
optical field, a single, flourescent molecule. This is not an
unrealistic situation. It is the very simplicity of this
arrangement that makes a single molecule an important device for
exploring the spectroscopic properties of various chemical and
biological environments \cite{BAS97,MOE99}.

\section{Fluorescence in free space}\label{sect-fluofs}

With fluorescence we designate the spontaneous emission of photons
by atoms or mol\-e\-cules. The photon is emitted into a mode of the
electromagnetic field, so augmenting its photon number.
Simultaneously the atom or molecule transits from an electronic
state of energy $E_2$ to one of lower energy $E_1$. The frequency
of the emitted photon then is $\omega = (E_2-E_1)/\hbar$. The
presence of Planck's constant $\hbar$ in the description of this
process indicates that it is a quantum mechanical process. However,
real atoms, even more real molecules, are complicated systems,
which even in their most simple realization, namely hydrogen,
exhibit a non-trivial structure of states. We therefore are forced
to simplify the reality. For the purpose of this discussion we will
restrict ourselves to the simplest model of an atom or molecule,
the \textit{two-level} atom or molecule \cite{ALL75}. In the
following we use the term ``system'' to spare using bulky ``atom or
molecule''.

\bigskip
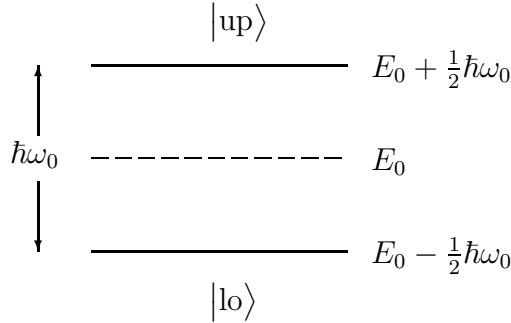
\begin{figure}[!ht]
 \setlength{\unitlength}{0.75em}
\begin{center}
\begin{picture}(27,15)
\thicklines
 \put(5,3){\line(1,0){11}}
 \put(5,11){\line(1,0){11}}
\thinlines
 \multiput(5,7)(1,0){11}{\line(1,0){0.7}}
 \put(2.7,6){\vector(0,-1){3}}
 \put(2.7,8){\vector(0,1){3}}
 \put(1.5,6.8){$\hbar\omega_0$}
 \put(10,0.5){$\ket{\rm lo}$}
 \put(10,12.5){$\ket{\rm up}$}
 \put(17,2.5){$E_0-\frac{1}{2}\hbar\omega_0$}
 \put(17,10.5){$E_0+\frac{1}{2}\hbar\omega_0$}
 \put(17,6.5){$E_0$}
\end{picture}
\end{center}
\caption{\label{2-lev} Energy levels of a quantum system with two
energy states $\ket{\rm lo}$ and $\ket{\rm up}$ separated by the
energy $\hbar\omega_0$. The energy of the states $\ket{\rm lo}$ and
$\ket{\rm up}$ is evaluated relative to the reference energy $E_0$.
}
\end{figure}

\subsection{Two-level system and its variables}

As mentioned above we consider a quantum system with two energy
states $\ket{\rm lo}$ and $\ket{\rm up}$ separated by the energy
$\hbar\omega_0$ as shown in Fig.~\ref{2-lev}. According to the
postulates of quantum mechanics the states $\ket{\rm lo}$ and
$\ket{\rm up}$ are then eigenstates of the (non-interacting)
Hamiltonian $\OH_A$ with eigenvalues
$E_0\pm\frac{1}{2}\hbar\omega_0$:
\begin{equation}
 \label{TL-EVE}
\begin{split}
\OH_A \ket{\rm lo} & = (E_0+\tfrac{1}{2}\hbar\omega_0) \ket{\rm lo}
\\ \OH_A \ket{\rm up} & = (E_0-\tfrac{1}{2}\hbar\omega_0) \ket{\rm up}
 \quad .
\end{split}
\end{equation}
The states $\ket{\rm lo}$ and $\ket{\rm up}$ form an orthonormal
and complete set:
\begin{align}
 \langle \lambda | \lambda^\prime \rangle & = \delta_{\lambda
 \lambda^\prime} \; , \qquad \lambda,\,\lambda^\prime = 1\,({\rm lo}),\,2\,({\rm up})
 \label{2-LEV-ORTHO}\\
 \sum_{\lambda=1}^2 |\lambda\rangle\langle\lambda| & = 1 \quad .
 \label{2-LEV-UNITOP}
\end{align}
In the last section in which we discussed the quantized field, we
introduced the non-Hermitian operators $\AO$ and $\VO$, which
lowered, respectively raised the excitation of the field mode by
one quantum (photon) $\hbar\omega$. Similarly we now introduce the
atomic operators $\VB$ and $\EB$ which lower and raise the energy
of the atom (molecule) by $\hbar\omega_0$. Unlike the field, the
energy of the two-level system is restricted, i.e.\ has a lower
bound $E_1=E_0-\tfrac{1}{2}\hbar\omega_0$ and an upper bound
$E_2=E_0+\tfrac{1}{2}\hbar\omega_0$. Therefore, the effect of $\VB$
on state $\ket{\rm lo}$. as well as of $\EB$ on $\ket{\rm up}$ must
vanish:
\begin{align}
\VB \ket{\rm up} &= \ket{\rm lo} & \EB \ket{\rm up} &= 0
\nonumber\\
\VB \ket{\rm lo} &= 0 & \EB \ket{\rm lo} &= \ket{\rm up} \quad .
\label{B-PROP1}
\end{align}
Repeated application has the following effects
\begin{align}
\VB\EB \ket{\rm up} &= 0 & \EB\VB \ket{\rm up} &= \ket{\rm up}
\nonumber\\
\VB\EB \ket{\rm lo} &= \ket{\rm lo}  & \EB\VB \ket{\rm lo} &= 0
\quad . \label{B-PROP2}
\end{align}
We can see that $\VB\EB$ and $\EB\VB$ have the effect of
\textit{number operators} with eigenvalues 0, 1 for the lower and
upper states respectively, whereas the repeated application of the
same operator always vanishes:
\begin{equation}
\VB^{2} = 0 = \VB^{\dagger 2} \quad .
\label{B-PROP3}
\end{equation}
Those properties can be summarized in the following
\textit{anti-commutation rules}:
\begin{equation}
 \label{ANTI-COMM}
 \begin{split}
 \{\VB,\VB\} = & \, \{\EB,\EB\} = 0 \; , \\
 \{\VB,\EB\} = & \, 1
 \quad ,
 \end{split}
\end{equation}
where for the anti-commutator of $\mbox{\N A}$ and $\mbox{\N B}$ we
used the notation $\{\mbox{\N A},\mbox{\N B}\} \doteq \mbox{\N
A}\mbox{\N B}+\mbox{\N B}\mbox{\N A}$. These anti-commutator
relations are characteristic for fermions, and are analogous to the
commutator relations (\ref{KommVE})--(\ref{KommEE}) of a single
mode of the electromagnetic field (photon field), which is a boson
field.

\subsubsection{Analogy to spin-1/2 system}
All quantum systems characterized by only two possible states  are
mathematically equivalent.  The prototype of such a system is the
spin-$\frac{1}{2}$ particle in a magnetic field. We thus can borrow
from this formalism \cite{RAB37,BLO46}. Be aware that unfortunately
many technical terms keep their original names, although here they
refer to a completely different physical system.

To describe physical observables of the two-level system at hand,
we need Hermitian operators. Borrowing from the mentioned theory we
define the following set of three traceless \textit{Pauli spin
operators}\footnote{Often the following spin operators are defined
that are not traceless: $\HSO{$\sigma$}_+ =\EB,\;\HSO{$\sigma$}_-
=\VB,\;\HSO{$\sigma$}_z =\OR_3$.}
\begin{equation}
 \label{PAULI-OP}
 \begin{split}
 \OR_0 &=\tfrac{1}{2}\cdot\1  \\
 \OR_1 &= \tfrac{1}{2}(\EB + \VB) \\
 \OR_2&=\frac{1}{2\ii}(\EB - \VB) \\
 \OR_3 &=\tfrac{1}{2} (\EB\VB-\VB\EB) \quad .
 \end{split}
\end{equation}
In the original case of a spin-$\frac{1}{2}$ particle in a magnetic
field these four operators describe the dynamics of the
spin-system. In the case of the two-level system discussed here,
however, they are not related to any spin. But because they form a
linearly independent, complete set of Hermitian observables, they
can fully cover the dynamics in the two-dimensional Hilbert space
of two-level atoms or molecules. Therefore, any system operator
$\OO$ can be represented as a series of Pauli spin operators
\begin{equation}
\OO = \sum_{\alpha = 0}^3 g_\alpha \OR_\alpha \quad ,
 \label{REP-RBASE}
\end{equation}
where the coefficients $g_\alpha$ are determined by $\OO$. The
operators (\ref{PAULI-OP}) satisfy the following commutation and
anti-commutation rules\footnote{$\epsilon_{lmn}$ is the fully
antisymmetric Kronecker symbol whose only nonvanishing values are
$\epsilon_{123}=\epsilon_{231}=\epsilon_{312}=-\epsilon_{132}=
-\epsilon_{321}=-\epsilon_{213}=1$ \protect\cite[p.~209]{SIF68}.}
\begin{equation}
 \begin{split}
 \label{COMM-Rs}
 [\OR_l,\OR_m] &= \ii \epsilon_{lmn}\OR_n  \\
 \{\OR_l,\OR_m\} &= \tfrac{1}{2} \delta_{lm}\quad , \qquad
 (l,m,n = 1,2,3) \; ,
 \end{split}
\end{equation}
as well as the relations
\begin{equation}
\begin{split}
 \OR^2_\alpha & = \tfrac{1}{4}\quad , \qquad (\alpha = 0,1,2,3) \\
 \sum_{\alpha =0}^3 \OR_\alpha^2 &= 1 \quad .
 \label{R-COMM_ACOMM}
\end{split}
\end{equation}
In short the following relations will be useful
\begin{align}
 [\VB,\OR_1] &= -\OR_3 \nonumber \\
 [\VB,\OR_2] &= \ii \OR_3 \label{REL-BR}\\
 [\VB,\OR_3] &= \VB \nonumber \quad .
\end{align}

\bigskip
We now construct the representation of the operators in terms of
the two states $\ket{\rm lo}$ and $\ket{\rm up}$ of the two-level
system. The procedure consists in multiplying the operator with the
unit operator (\ref{2-LEV-UNITOP}) from the left side and from the
right side, and observing relations (\ref{B-PROP1}) and
(\ref{B-PROP2}). At the end we obtain
\begin{equation}
 \label{2-LEV-OP1}
 \begin{split}
 \VB &= \ket{\rm lo}\bra{\rm up} \\
 \EB &= \ket{\rm up}\bra{\rm lo} \\
 \VB\EB &= \ket{\rm lo}\bra{\rm lo} \\
 \EB\VB &= \ket{\rm up}\bra{\rm up} \; ,
 \end{split}
\end{equation}
and similarly
\begin{equation}
 \label{2-LEV-OP2}
 \begin{split}
 \OR_3 \ket{\rm up} &= \tfrac{1}{2} \ket{\rm up} \\
 \OR_3 \ket{\rm lo} &= - \tfrac{1}{2} \ket{\rm lo} \; .
 \end{split}
\end{equation}
Inspecting (\ref{2-LEV-OP2}) we see that the states $\ket{\rm lo}$
and $\ket{\rm up}$ are eigenstates of the Hermitian operator
$\OR_3$ that can thus be regarded to measure the amount of
inversion in the 2-level system.

\subsubsection{System energy and dipole moments}
The energy of the two-level system is represented by the
Hamiltonian $\OH_A$ (\ref{TL-EVE}) which by above procedure can be
written as
\begin{equation}
\OH_A = E_0 + \tfrac{1}{2} \hbar\omega_0 \, (\ket{\rm up}\bra{\rm
up} - \ket{\rm lo}\bra{\rm lo}) \; , \label{2_LEV-HAM}
\end{equation}
and after using (\ref{2-LEV-OP1}) and (\ref{PAULI-OP}) we obtain
\begin{equation}
\OH_A = E_0 + \hbar\omega_0 \OR_3 \; .
 \label{HA-R3}
\end{equation}
If, like in our figure (cf.\ Fig.~\ref{2-lev}) the lower state is
the system ground state, then $E_0 = \frac{1}{2}\hbar\omega_0$, and
$\OH_A = \hbar\omega_0(\OR_3 +  \frac{1}{2}) = \hbar\omega_0
\EB\VB$.

\bigskip
In the following we will elucidate the physical significance of the
other Hermitian variables $\OR_1$ and $\OR_2$. Those operators are
closely related to the dipole moment $\Omu$. For atoms, ions or
molecules the dipole moment can be defined as
\begin{equation}
\Omu = \sum_i e \Or_i \; ,
\end{equation}
where $\Or_i$ is the position operator for the $i$-th charge $e$ in
the system. To express the dipole operator in terms of $\VB,
\EB$-operators we apply the unit operator trick
\begin{align}
\Omu &= (\ket{\rm up}\bra{\rm up}+\ket{\rm lo}\bra{\rm lo}) \Omu
(\ket{\rm up}\bra{\rm up}+\ket{\rm lo}\bra{\rm lo}) \nonumber \\
 &= \Vmu_{22}\EB\VB + \Vmu_{11}\VB\EB + \Vmu_{12}\VB + \Vmu_{21}\EB \;
 .
\end{align}
The coefficients $\Vmu_{ij}$ stand for the matrix elements $\langle
i|\Omu |j\rangle $ ($i,j=1\,({\rm lo}),2\,({\rm up})$). We
therefore note that $\Vmu_{11}$ and $\Vmu_{22}$ represent
expectation values for the dipole moment in the lower and upper
system state, respectively. However, we know that the dipole moment
has an odd parity, and therefore those coefficients must vanish.
Because the dipole moment is a Hermitian operator we thus obtain
\begin{equation}
 \Omu = \Vmu_{12}\VB + \Vmu_{12}^{\,\ast}\EB \; .
 \label{Muvonb}
\end{equation}
If the system transition from the upper state $\ket{\rm up}$ to the
lower state $\ket{\rm lo}$ is characterized by $\Delta m = 0$ in
the real system, then $\Vmu_{12}$ is a real valued vector. On the
other hand, for a $\Delta m = \pm 1$-transition (e.g.\ induced by
polarized light), $\Vmu_{12}$ is a complex valued vector.

 \begin{quote}
\textbf{Example:} Hydrogen atom; let us assume the following
correspondence for a $\Delta m = 0$ transition:
\begin{align}
 \mbox{two-level system} && \mbox{hydrogen atom}& \nonumber \\
 \ket{\rm lo}\quad &\longrightarrow & |n=1,l=0,m=0\rangle &\quad \mbox{(s state)} \nonumber \\
 \ket{\rm up}\quad &\longrightarrow & |n=2,l=1,m=0\rangle &\quad \mbox{(p state)} \nonumber
\end{align}
If we take the $z$-axis as the quantization axis and evaluate the
specified hydrogen wave functions, we find \cite{ALL75}:
\begin{equation}
\Vmu_{12} = \bra{\rm lo} \Omu \ket{\rm up} = \frac{128
\sqrt{2}}{243}\, e a_0 \vec{e}_z
 \; ,
\end{equation}
where $a_0$ is the Bohr radius and $\vec{e}_x, \vec{e}_y,
\vec{e}_z$ are unit vectors in direction of the axes. We can see
here that $\Vmu_{12}$ has the properties of a 3-dimensional
Euclidean vector. With (\ref{Muvonb}) and (\ref{PAULI-OP}) we
obtain
\begin{equation}
\Omu = \Vmu_{12} (\VB + \EB) = 2 \Vmu_{12}\OR_1 \; .\label{HYDMU1}
\end{equation}

However, if the transition involves $\Delta m =1$, as in the
correspondence
\begin{align}
 \mbox{two-level system} && \mbox{hydrogen atom}& \nonumber \\
 \ket{\rm lo}\quad &\longrightarrow & |n=1,l=0,m=0\rangle &\quad \mbox{(s state)} \nonumber \\
 \ket{\rm up}\quad &\longrightarrow & |n=2,l=1,m=1\rangle &\quad \mbox{(p state)}
 \nonumber \; ,
\end{align}
then $\Vmu_{12}$ is complex:
\begin{equation}
\Vmu_{12} = \bra{\rm lo} \Omu \ket{\rm up} = \frac{128}{243}\, e
a_0 (\vec{e}_x +\ii \vec{e}_y)
 \; .
\end{equation}
In the case of a complex $\Vmu_{12}$ we can always rewrite
(\ref{Muvonb}) in the following form
\begin{align}
 \Omu &= \Vmu_{12} (\OR_1-\ii\OR_2) + \Vmu_{12}^{\,\ast}(\OR_1+\ii\OR_2)
 \nonumber \\
 &= 2 {\rm Re}(\Vmu_{12}) \OR_1 + 2 {\rm Im}(\Vmu_{12}) \OR_2 \; .
 \label{MU-R}
\end{align}

In this example we have illustrated how the operators $\OR_1,\OR_2$
are related to the dipole moment operator $\Omu$.
 \end{quote}

As next we will derive an expression for the \textit{rate of
change} of the dipole moment operator $\Omu$. (In above example of
the hydrogen atom the rate of change of $\Omu$ would correspond to
the electron velocity $\Ov$ multiplied by $e$.) In general, the
rate of change of an observable is determined by Heisenberg's
equation of motion, which in this case reads as
\begin{equation}
\frac{\dd\Omu}{\dd t} = \frac{1}{\ii\hbar}[\Omu , \OH_A] \; ,
\end{equation}
and which with (\ref{HA-R3}), (\ref{Muvonb}), and the commutation
relations (\ref{COMM-Rs}) or (\ref{REL-BR}) results in
\begin{align}
 \frac{\dd\Omu}{\dd t} &= \frac{1}{\ii\hbar}
 [(\Vmu_{12}\VB+\Vmu_{12}^{\,\ast}\EB)\, , \,
 (\hbar\omega_0\OR_3+E_0)]\nonumber \\
 &= -\ii\omega_0(\Vmu_{12}\VB - \Vmu_{12}^{\,\ast}\EB)\;.
 \label{D-MU-DT}
\end{align}
The operators $\VB$ and $\EB$ are transformed into the interaction
picture according to the usual rule
\begin{equation}
\VB(t)=\exp\left[\frac{\ii}{\hbar}\OH_A(t-t_0)\right]\;\VB(t_0)\;\exp\left[-\frac{\ii}{\hbar}\OH_A(t-t_0)\right]\;
 , \quad t \ge t_0 \; .
\end{equation}
Inserting (\ref{HA-R3}), and observing the operator expansion
theorem, as well as commutation rules (\ref{REL-BR}), we find
\begin{equation}
\begin{split}
 \VB(t) &=\VB(t_0)\,\ee^{-{\rm i }\omega_0(t-t_0)} \\
 \EB(t) &=\EB(t_0)\,\ee^{{\rm i }\omega_0(t-t_0)} \; ,
\end{split}
\end{equation}
and for the dipole operator in the interaction picture
\begin{equation}
\begin{split}
 \Omu(t) &=2 {\rm Re}(\Vmu_{12})\OR_1(t)+2{\rm
 Im}(\Vmu_{12})\OR_2(t)\\
 &=\Vmu_{12}\VB(t_0)\,\ee^{-{\rm i}\omega_0(t-t_0)} +
 \Vmu_{12}^{\,\ast}\EB(t_0)\,\ee^{{\rm i}\omega_0(t-t_0)}\; .
\end{split}
\end{equation}

\bigskip
\subsubsection{Bloch-representation of the state}\label{BLOCH-REP}

We have seen [c.f.\ \eqref{2-LEV-ORTHO} and \eqref{2-LEV-UNITOP}]
that the states $\ket{\rm lo}$ and $\ket{\rm up}$ form an
orthogonal and complete set within the two level model. Hence, any
pure state of the two-level system can be represented as the linear
combination
\begin{equation}
|\psi\rangle = c_1\ket{\rm lo} + c_2 \ket{\rm up} \; ,
\end{equation}
with
\begin{equation}
|c_1|^2+|c_2|^2 =1\; .
\end{equation}
When the system state is not pure and has to be specified in
statistical terms, it is represented by the atomic (or molecular)
density operator $\OrhoA$:
\begin{equation}
\OrhoA = \rho_{11}\ket{\rm lo}\bra{\rm lo} + \rho_{22}\ket{\rm
up}\langle2| + \rho_{12}\ket{\rm lo}\bra{\rm up} +
\rho_{21}\ket{\rm up}\bra{\rm lo}\; ,
\end{equation}
where the coefficients $\rho_{ij}$ stand for the ensemble average
\begin{equation}
\rho_{ij} = \langle c_i\, c_j\rangle\, , \quad i,j = 1.2 \, ,
\end{equation}
which represents a two-dimensional, Hermitian, covariance matrix.
Bloch \cite{BLO46} presented a simple, intuitive, geometrical
interpretation of the state in terms of a real three-dimensional
vector $\vec{r}$ with components $r_1, r_2, r_3$. In the
Schr\"odinger picture they are given by
\begin{equation}
\label{BLOCH-R-RHO-REP}
\begin{split}
 r_1 &= 2\, {\rm Re}(\rho_{12})  \\
 r_2 &= 2\, {\rm Im}(\rho_{12})  \\
 r_3 &= \rho_{22} - \rho_{11} \; ,\\
 \text{where often a forth}&\; \text{component is added:}\\
 r_0 &= \rho_{22} + \rho_{11} = 1 \; .
\end{split}
\end{equation}
The correspondence between the Bloch vector and the density matrix
is a consequence of the properties of the respective symmetry
groups, namely the correspondence between the real orthogonal group
O3 and the special unitary group SU2. Note, that also the Stokes
vector that is often used to represent the polarization state of
light (see e.g.\ \cite{HEC87}) has O3 symmetry.

%

\begin{figure}[!ht]
 \setlength{\unitlength}{0.75em}
\begin{center}
\begin{picture}(25,18)
\thicklines
 \put(12,8){\vector(0,1){8}}
 \put(12,8){\vector(0,-1){8}}
 \put(12,8){\vector(-3,2){3}}
 \put(12,8){\vector(4,3){3.5}}
\thinlines
 \put(12,8){\vector(0,1){10}}
 \put(12,8){\vector(1,0){12}}
 \put(12,8){\vector(-3,-2){11}}
 \linethickness{0.5pt}
 \bezier{25}(12,16)(8.7,16)(6.34,13,66)
 \bezier{25}(6.34,13,66)(4,11.3)(4,8)
 \bezier{25}(4,8)(4,4.7)(6.34,2.34)
 \bezier{25}(6.34,2.34)(8.7,0)(12,0)
 \bezier{25}(12,0)(15.3,0)(17.66,2.34)
 \bezier{25}(17.66,2.34)(20,4.7)(20,8)
 \linethickness{1pt}
 \bezier{200}(20,8)(20,11.3)(17.66,13.66)
 \bezier{200}(17.66,13.66)(15.3,16)(12,16)
 \bezier{200}(12,16)(6,13)(6,4)
 \bezier{200}(6,4)(15,4)(20,8)
 \put(1,1.5){$x$}
 \put(11,17){$z$}
 \put(23,7){$y$}
 \put(12.2,14){\footnotesize fully}
 \put(12.2,12.8){\footnotesize excited $+$1}
 \put(12.2,1){\footnotesize fully excited $-$1}
 \put(15.5,9.7){\footnotesize partly}
 \put(15.5,8.5){\footnotesize excited}
\end{picture}
\end{center}
\caption{\label{BLOCH-VEC}
  Bloch-vector representation of the state
of a two-level quantum system. Pure states are characterized by
vectors of unit length. ``Fully excited'' refers to system
occupying pure state $\ket{\rm up}$, whereas ``fully unexcited'' to
$\ket{\rm lo}$.
  }
\end{figure}
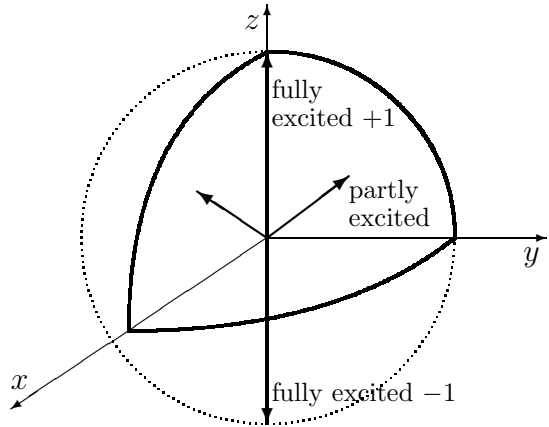
%

In the three-dimensional Bloch-vector representation sketched in
Fig.~\ref{BLOCH-VEC} pure states are characterized by their unit
length. For example the pure state $\ket{\rm lo}$ corresponds to
the down pointing vector $\vec{r}=(0,0,-1)$, whereas $\ket{\rm up}$
to the up pointing vector $\vec{r}=(0,0,1)$. Pure intermediate
states point in various directions, in particular, states with an
equal mixture of upper and lower states ($\rho_{22}=\rho_{11}$) lie
in the horizontal $x,y$-plane.

As mentioned, a pure state is represented by Bloch vectors of unit
length, while for mixed or impure states the length is less than
unity:
\begin{align}
 r=r_1^2 + r_2^2 + r_3^2 &= 4\,|\rho_{12}|^2 +
 (\rho_{22}-\rho_{11})^2 \notag \\
 r &= 1 - 4(\rho_{22}\rho_{11}-|\rho_{12}|^2) \; .
 \label{LENGTH-R}
\end{align}
From the Schwarz inequality follows
\begin{equation}
\rho_{22}\rho_{11}-|\rho_{12}|^2 =
\langle|c_2|^2\rangle\langle|c_1|^2\rangle -|\langle
c_1c_2^\ast\rangle|^2 \ge 0 \; ,
\end{equation}
where equality is realized when the ensemble contracts to a single
realization, thus a pure state. Therefore
\begin{equation}
r_1^2 + r_2^2 + r_3^2 \le 1 \; ,
\end{equation}
with equality only for a pure state. On the other side the Bloch
vector vanishes when the ensemble consists of an equally weighted
mixture of upper and lower states with random phases.

\bigskip
With \eqref{REP-RBASE} we mentioned that the spin operators
$\OR_\alpha$ form a complete set, so allowing to represent any
observables in the two-dimensional Hilbert space of the two-level
system. Thus in this base the density operator $\OrhoA$ has the
expansion
\begin{equation}
\OrhoA = \sum_{\alpha=0}^3 g_\alpha \OR_\alpha \; ,
\end{equation}
where the coefficients $g_\alpha$ are time dependent, whereas the
operators $\OR_\alpha$ are time-in\-de\-pend\-ent. Evaluating
$\langle i|\OrhoA |j\rangle, \;(i,j={\rm lo},\,{\rm up})$ with
\eqref{BLOCH-R-RHO-REP} we can identify the coefficients as
$g_\alpha = r_\alpha$, and therefore
\begin{equation}
\OrhoA = \sum_{\alpha=0}^3 r_\alpha \OR_\alpha \; .
 \label{DENSITY-OP-BLOCH-R}
\end{equation}
This result shows that the Bloch-vector components can be
interpreted as weights in the Schr\"odinger-picture expansion of
the density operator in the spin operator base.

\bigskip
The symmetry of the Bloch-vector representation (cf.\
Fig.~\ref{BLOCH-VEC}) suggests a representation of $\vec{r}$ in
polar coordinates $(r,\theta,\phi)$ rather than Cartesian
coordinates $(r_1,r_2,r_3)$. From \eqref{LENGTH-R} and
\eqref{BLOCH-R-RHO-REP} we deduce
\begin{align}
 \rho_{11} &= \tfrac{1}{2}(1-r_3) = \tfrac{1}{2}(1-r\cos\theta) \notag\\
 \rho_{22} &= \tfrac{1}{2}(1+r_3) = \tfrac{1}{2}(1+r\cos\theta) \\
 \rho_{12} &= \tfrac{1}{2}(r_1+\ii r_2) =
 \tfrac{1}{2}r\,\sin\theta\,{\rm e}^{\rm i\phi} \; ,\notag
\end{align}
resulting in
\begin{equation}\label{BLOCH-POL-COORD}
\begin{split}
 \cos\theta &= \frac{\rho_{22}-\rho_{11}}{r}\;, \; {\rm and}\\
 \tan\phi &= \frac{{\rm Im}(\rho_{12})}{{\rm Re}(\rho_{12})}\; .
\end{split}
\end{equation}

\subsubsection{Spin operator expectation values}
In \eqref{DENSITY-OP-BLOCH-R} we have illustrated the relation
between the spin operators $\OR_\alpha$ and the Bloch-vectors
components $r_\alpha$. In the following we calculate the
expectation values of the spin operators for the system in an
arbitrary quantum state which is characterized by the density
operator $\OrhoA$:
\begin{equation}
\langle\OR_\alpha\rangle = {\rm Tr}\bigl[\OrhoA\OR_\alpha\bigr]\; ,
\; (\alpha=0,\dots,3)\; .
\end{equation}
Inserting \eqref{DENSITY-OP-BLOCH-R} for the density operator in
the Schr\"odinger picture we obtain
\begin{equation}
\langle\OR_\alpha\rangle = \sum_{\beta = 0}^3 {\rm Tr}\bigl(r_\beta
\OR_\beta \OR_\alpha\bigr)\; .
\end{equation}
The product of two different operators $\OR$ gives either $\OR_1$,
$\OR_2$, or $\OR_3$, which all have vanishing trace. The sum
therefore only gets a contribution from term $\alpha=\beta$. Since
the trace of a two-dimensional unit vector is 2, we find with
\eqref{R-COMM_ACOMM}
\begin{equation}
\langle \OR_\alpha\rangle = {\rm Tr}\bigl(\tfrac{1}{4}r_\alpha
\1\bigr) = \tfrac{1}{2} r_\alpha\; , \; \alpha = 0, \dots ,3 \; .
\end{equation}
The components of the Bloch vector can thus also be interpreted in
terms of the expectation values of the spin operators.

We can now calculate the expectation of the energy \eqref{HA-R3} as
\begin{equation}
\langle \OH_A\rangle = E_0 + \tfrac{1}{2}\hbar\omega_0r_3 \;.
\end{equation}
In the same way the expectation value of the dipole moment
\eqref{MU-R} is given by
\begin{equation}
\langle \Omu \rangle = {\rm Re}(\vec{\mu}_{12})r_1 + {\rm
Im}(\vec{\mu}_{12})r_2\;.
\end{equation}
From this equations we can see that the $z$-component of the Bloch
vector is associated with the energy of the two-level system,
whereas the $x$- and $y$-components are related to its dipole
moment. For a two-level system without a permanent dipole moment
the expectation value of the operator $\Omu$ must vanish in the
lower, as well as in the upper state. In those states the operator
$\Omu$ is not well defined and its value fluctuates. From
\eqref{Muvonb} or \eqref{MU-R} we can see that
\begin{equation}
\Omu^2 = |\vec{\mu}_{12}|^2 \1 \;.
\end{equation}
As a consequence, the variance of the dipole moment in the lower
state $\ket{\rm lo}$ and in the upper state $\ket{\rm up}$  is
given by
\begin{equation}
\bra{\rm lo}(\Delta\Omu)^2\ket{\rm lo}= |\vec{\mu}_{12}|^2
=\bra{\rm up}(\Delta\Omu)^2\ket{\rm up} \, ,
\end{equation}
and is maximum in those states. Therefore, even if the expectation
value of the dipole moment of a two-level system vanishes in the
ground and upper state, the system can interact with an
electromagnetic field through the nonvanishing fluctuations of
$\Omu$. In the following we will now discuss these interactions.

\subsection{Interaction of a two-level system with a classical
electromagnetic field}

Let us assume that the classical electromagnetic field is described
by its electric field $\vec{E}$, and that the two-level system is
located at a fixed point in space and its dipole moment is
characterized by the operator $\Omu$. Also, the typical wavelength
of the field is much larger than the length of the dipoles. The
interaction energy $\OH_I$ is then given by the usual expression
for the potential energy of a dipole in an electric field. With
\eqref{MU-R} we can write
\begin{equation}
 \OH_I = -\Omu(t) \cdot \vec{E}=-2[{\rm Re}(\vec{\mu}_{12})\OR_1 +{\rm
 Im}(\vec{\mu}_{12})\OR_2]\cdot \vec{E}
 \label{INT-H-E}
\end{equation}
The interaction of a charge with an electromagnetic field of
moderate power\footnote{field power below the regime where
multiphoton interactions dominate, so that terms in $\OA^2$ and
higher can be neglected} can also be described by the interaction
energy
\begin{equation}
\OH_I = \frac{e}{m}\OP(t)\cdot\vec{A}(\vec{r}_0,t)\;.
 \label{INT-H-A1}
\end{equation}
For the canonical momentum operator $\OP$ we can set
$m\dot{\Omu}/e$. and with \eqref{D-MU-DT} we obtain
\begin{equation}
\OH_I = -\dot{\Omu}(t) \cdot \vec{A}(t) =
\ii\omega_0[\vec{\mu}_{12}\VB(t)-\vec{\mu}_{12}^\ast\EB(t)]\cdot\vec{A}(t)\;.
  \label{INT-H-A2}
\end{equation}
We have given two interaction Hamiltonians, \eqref{INT-H-E} and
\eqref{INT-H-A2}. Depending on the problem at hand one can choose
the more appropriate one, although in most cases they lead to
similar results. The total energy of the system in the field
consists of the sum
\begin{equation}
\OH = \OH_A + \OH_I \; ,
\end{equation}
where $\OH_A$ is diagonal in the states $\ket{\rm lo}$ and
$\ket{\rm up}$, whereas $\OH_I$ is off-diagonal.

In the Schr\"odinger picture the time dependence of the density
operator is given by the equation
\begin{equation}
\frac{\dd \Orho(t)}{\dd t}= \frac{1}{\ii\hbar}\,[(\OH_A+\OH_I)\,
,\; \Orho^{(A)}(t)] \; .
 \label{RHO-EVOL}
\end{equation}

\subsubsection{Bloch equations}

As next we consider \eqref{RHO-EVOL} in matrix element notation.
After inserting the unity operator \eqref{2-LEV-ORTHO} between
$\OH$ and $\Orho^{(A)}(t)$ in the commutator relation on the rhs of
\eqref{RHO-EVOL}, we obtain the following Schr\"odinger picture
equations
\begin{equation}\label{RHO-MATRIX-EVOL}
\begin{split}
 \dot{\rho}_{11} &=\frac{1}{\ii\hbar}\,[\bra{\rm lo}\OH_I(t)\ket{\rm up}\rho_{21}-{\rm
 c.c.}]\\
 \dot{\rho}_{22} &=-\frac{1}{\ii\hbar}\,[\bra{\rm lo}\OH_I(t)\ket{\rm up}\rho_{21}-{\rm
 c.c.}]\\
 \dot{\rho}_{12} &=\frac{1}{\ii\hbar}\,[-\hbar\omega_0\rho_{12}+\langle
 1|\OH_I(t)\ket{\rm up}(\rho_{22}-\rho_{11})]\\
 \dot{\rho}_{21} &=\frac{1}{\ii\hbar}\,[\hbar\omega_0\rho_{21}+\langle
 2|\OH_I(t)\ket{\rm lo}(\rho_{11}-\rho_{22})] \; .
\end{split}
\end{equation}
Using \eqref{BLOCH-R-RHO-REP} we can express
\eqref{RHO-MATRIX-EVOL} in terms of the Bloch vector and the result
is often referred to as \textit{Bloch equations} \cite{BLO46}:
\begin{equation}
\label{BLOCH-EQS}
\begin{split}
 \dot{r}_1 &= \frac{2}{\hbar}\,{\rm Im}(\bra{\rm lo}\OH_I\ket{\rm up})\,
 r_3 - \omega_0r_2 \\
 \dot{r}_2 &= -\frac{2}{\hbar}\,{\rm Re}(\langle
 1|\OH_I\ket{\rm up})\, r_3 + \omega_0r_1 \\
 \dot{r}_3 &= -\frac{2}{\hbar}\,{\rm Im}(\langle
 1|\OH_I\ket{\rm up})\, r_1 + \frac{2}{\hbar}\,{\rm Im}(\langle
 1|\OH_I\ket{\rm up})\, r_2
\end{split}
\end{equation}
We can see that $r_3$ is constant, when there is no external field.
Multiplying the first equation with $r_1$ the second with $r_2$ and
the third with $r_3$, we find
\begin{equation}
\frac{\dd}{\dd t}(r_1^2 + r_2^2 +r_3^2) = 0 \; .
\end{equation}
Thus, in the presence of a classical field the length of the Bloch
vector remains constant. That means a pure state stays pure, and a
mixed state remains mixed. The structure of \eqref{BLOCH-EQS} led
Feynman {\it et al.} \cite{FEY57} to propose an interesting
geometric interpretation. For that they introduce a vector
$\vec{Q}$ with following components
\begin{equation}
\label{FEYNM-Q}
\begin{split}
 Q_1 &= \frac{2}{\hbar}\,{\rm Re}(\bra{\rm lo}\OH_I\ket{\rm up})\\
 Q_2 &= \frac{2}{\hbar}\,{\rm Im}(\bra{\rm lo}\OH_I\ket{\rm up})\\
 Q_3 &= \omega_0 \; .
\end{split}
\end{equation}
With this vector the Bloch equations \eqref{BLOCH-EQS} are
equivalent to
\begin{equation}
 \frac{\dd}{\dd t}\vec{r} = \vec{Q} \times \vec{r}\; .
 \label{FEYNM-QVP}
\end{equation}
In analogy to the mechanics of rigid bodies this equation shows
that the Bloch vector $\vec{r}$ precesses around the vector
$\vec{Q}$ with a frequency that is given by the magnitude of
$\vec{Q}$. When $\vec{Q}$ itself varies with time, the precession
motion may become complicated. We may evaluate the matrix element
$(\bra{\rm lo}\OH_I\ket{\rm up})$ using \eqref{INT-H-E} or
\eqref{INT-H-A2}. For example, with \eqref{INT-H-E} we obtain
\begin{equation}
\bra{\rm lo}\OH_I\ket{\rm up} = - \vec{\mu}_{12} \, \vec{E}(t) \; .
\end{equation}
Let us assume an electric field with frequency $\omega_1$ that is
nearly resonant with the two-level system, so that
$|\omega_1-\omega_0|\ll\omega_0$. Then we can write
\begin{align}
 \vec{E}(t) &= \vec{\varepsilon}\,{\mathcal E}(t) {\rm e}^{{\rm
 i}\omega_1t}+ {\rm c.c.} \nonumber \\
 &= 2|{\mathcal E}(t)|\;\{{\rm Re}(\vec{\varepsilon})\cos[\omega_1t-\phi(t)] +
 {\rm Im}(\vec{\varepsilon})\sin[\omega_1t-\phi(t)] \} \; ,
 \label{E-NEARLY-RES}
\end{align}
where ${\mathcal E}(t) = |{\mathcal E}(t)|{\rm e}^{{\rm i}\phi(t)}$
is a slowly varying, complex amplitude, and $\vec{\varepsilon}$ is
a unit polarization vector. With this we obtain
\begin{equation}
 \bra{\rm lo}\OH_I\ket{\rm up} = - |{\mathcal
 E}(t)|\,\vec{\mu}_{12}\cdot[\vec{\varepsilon}\,
 {\rm e}^{-{\rm i}[\omega_1t-\phi(t)]}+{\rm c.c.}]\;.
\end{equation}
This matrix element can now be substituted in the Bloch equations
\eqref{BLOCH-EQS}.

\subsubsection{The rotating wave approximation}
As an example to illustrate the \textit{rotating wave
approximation}, let us compare now transitions with $\Delta m =1$
and $\Delta m =0$. In the case $\Delta m =1$ the vector $\Vmu_{12}$
has the form
\begin{equation}
 \Vmu_{12} = \frac{\vec{e}_x + \ii
 \vec{e}_y}{\sqrt{2}}\,|\Vmu_{12}|\; ,
 \quad (\vec{e}_x, \vec{e}_y \text{ are unit vectors})\; .
\end{equation}
In addition, assume an external field that is circularly polarized
and propagating in the $z$-direction, thus
\begin{equation}
 \vec{\varepsilon}=\frac{\vec{e}_x + \ii
 \vec{e}_y}{\sqrt{2}}\, .
\end{equation}
Then we obtain for the scalar products
\begin{equation}
 \Vmu_{12}\cdot\vec{\varepsilon} = 0 \; , \quad \text{and} \qquad
 \Vmu_{12}\cdot\vec{\varepsilon}^\ast =|\Vmu_{12}| \; .
\end{equation}
The Bloch equations \eqref{BLOCH-EQS} then become
\begin{equation}
\label{M1-R}
\begin{split}
 \dot{r}_1 &=-\Omega(t) r_3 \sin[\omega_1t-\phi(t)]-\omega_0 r_2  \\
 \dot{r}_2 &= \Omega(t) r_3 \cos[\omega_1t-\phi(t)]+\omega_0r_1\\
 \dot{r}_3 &= -\Omega(t) r_1 \sin[\omega_1t-\phi(t)] -
               \Omega(t) r_2 \cos[\omega_1t-\phi(t)] \; ,
\end{split}
\end{equation}
where we have introduced  the parameter
\begin{equation}
\Omega(t) = \frac{2}{\hbar}\; \Vmu_{12} \cdot
\vec{\varepsilon}^\ast \, |{\mathcal E(t)}| \; ,
 \label{RABIFRQ}
\end{equation}
which is known as the \textit{vacuum Rabi frequency}. The Rabi
frequency in this context is a measure for the coupling strength of
the two-level system with the external field. For the components of
the $\vec{Q}$-vector \eqref{FEYNM-Q} we obtain
\begin{equation}
\label{M1-Q}
\begin{split}
 Q_1 &= - \Omega(t) \cos[\omega_1t-\phi(t)] \\
 Q_2 &= - \Omega(t) \sin[\omega_1t-\phi(t)] \\
 Q_3 &= \omega_0 \;.
\end{split}
\end{equation}

\bigskip
On the other hand, for a $\Delta m =0$ transition $\vec{\mu}_{12}$
is a real vector, and if the incident light is linearly polarized,
then $\vec{\varepsilon}$ is also real. In this case the Bloch
equations \eqref{BLOCH-EQS} read as
\begin{equation}
\label{M0-R}
\begin{split}
 \dot{r}_1 &=-\omega_0 r_2\\
 \dot{r}_2 &= 2\Omega(t) r_3 \cos[\omega_1t-\phi(t)]+\omega_0r_1\\
 \dot{r}_3 &= -2\Omega(t) r_2 \cos[\omega_1t-\phi(t)] \; ,
\end{split}
\end{equation}
and following \eqref{FEYNM-QVP} we obtain for the corresponding
$\vec{Q}$-vector
\begin{equation}
\label{M0-Q}
\begin{split}
 Q_1 &= -2 \Omega(t) \cos[\omega_1t-\phi(t)] \\
 Q_2 &= 0 \\
 Q_3 &= \omega_0 \;.
\end{split}
\end{equation}
Looking at the two cases of the field interaction, the first with a
$\Delta m = 1$-system, the second with a $\Delta m = 0$-system,
\eqref{M1-R} and \eqref{M1-Q} seem to be rather different from
\eqref{M0-R} and \eqref{M0-Q}. However, Allen and Eberley
\cite[sect.~2.4]{ALL75} have shown that they only differ by some
anti-resonant terms. Comparing the expressions for $\vec{Q}$ we can
see that \eqref{M0-Q} can be decomposed into a sum of two vectors,
of which the first vector is \eqref{M1-Q}, and the second,
auxiliary vector consists of components $(-\Omega(t)
\cos[\omega_1t-\phi(t)],\; + \Omega(t) \sin[\omega_1t-\phi(t)],\;0
)$. This auxiliary vector rotates around the $z$-axis with
frequency $\omega_1-\dot{\phi}(t)$, whereas the Bloch vector
$\vec{r}$ rotates in the opposite direction with frequency
$\omega_0$. Their relative rotation frequency thus amounts to
$\omega_0+\omega_1-\dot{\phi}(t)$. Thus, when integrating the
equations of motion over a characteristic time interval, say
$\Delta t > 1/\omega_0$, the contributions of the fast precessing
auxiliary vector $\vec{Q}$ are small. To a good approximation we
can therefore neglect the auxiliary vector $\vec{Q}$. If we drop
the auxiliary $\vec{Q}$, then the $\Delta m = 0$ interaction with
linearly polarized light is described by the same set of equations
as the $\Delta m = 1$ interaction with circularly polarized light.
This is known as the \textit{rotating wave approximation}.

\subsubsection{Bloch equations in a rotating frame}
Let us go back to \eqref{M1-R}. This set of equations describes a
rotation around the $z$-axis at an optical frequency. We will now
introduce a rotating reference frame, in which the motion of the
Bloch vector $\Vr$ is slower. At first sight, Eq.~\eqref{M1-R}
seems to justify  the atomic (or system) frequency $\omega_0$ as a
suitable choice. But after we consider that the atomic frequency in
a spectroscopic sample varies from atom to atom, we realize that
the frequency of the applied field $\omega_1$ is better suited if
we want to refer the time evolution of all the differing atoms to
the same reference frame. The transformation from the stationary
frame $\vec{r}=(r_1.r_2,r_3)$ to the rotating frame
$\vec{r}^{\;\prime}=(r_1^\prime.r_2^\prime,r_3^\prime)$ is given by
\begin{equation}
\Vec{r}^{\;\prime} =\matrize{\Theta} \vec{r} \; ,
 \label{ROT-TRAFO}
\end{equation}
where $\matrize{\Theta}$ is the $3 \times 3$ orthonormal rotation
matrix
\begin{equation}
\matrize{\Theta} =
 \begin{pmatrix}
 \cos\omega_1t & \sin\omega_1t & 0 \\
 -\sin\omega_1t & \cos\omega_1t & 0 \\
 0 & 0 & 1
 \end{pmatrix}
 \label{THETA}
\end{equation}

Let us now transform Bloch equations \eqref{M1-R} to the rotating
(primed) frame. For that we rewrite \eqref{M1-R} in matrix form as
\begin{equation}
\Dot{\Vec{r}} = \matrize{C}\vec{r} \;,
\end{equation}
where $\matrize{C}$ is the $3 \times 3$ coefficient matrix.
Inserting \eqref{ROT-TRAFO} we obtain
\begin{align}
\frac{\dd \vec{r}^{\;\prime}}{\dd t} &=
  \dot{\matrize{\Theta}}\; \vec{r} +
  \matrize{\Theta}\:\Dot{\Vec{r}}\notag\\
 &=\dot{\matrize{\Theta}}\matrize{\Theta}^{-1}\matrize{\Theta}\;\vec{r} +
  \matrize{\Theta}\matrize{C}\matrize{\Theta}^{-1}\matrize{\Theta}\;\vec{r} \\
 &=(\dot{\matrize{\Theta}}\matrize{\Theta}^{-1} +
  \matrize{\Theta}\matrize{C}\matrize{\Theta}^{-1})\;\vec{r}^{\;\prime}\notag
\end{align}
After we insert the elements of $\matrize{C}$ from \eqref{M1-R} and
\eqref{THETA} we get the Bloch equations in the rotating frame
\begin{equation}
\begin{split}
\dot{r}_1^\prime &= (\omega_1-\omega_0) r_2^\prime + \Omega\sin
 \phi \; r_3^\prime \\
\dot{r}_2^\prime &= (\omega_0-\omega_1) r_1^\prime + \Omega\cos
 \phi \; r_3^\prime \\
\dot{r}_3^\prime &=  - \Omega\sin
 \phi \; r_3^\prime -  \Omega\cos
 \phi \; r_2^\prime \; ,
\end{split}
\end{equation}
which with $\vec{Q}=(-\Omega\cos\phi\, ,\;\Omega\sin\phi\,
,\;\omega_0-\omega_1)$ can be expressed as
\begin{equation}
\frac{\dd \vec{r}^{\;\prime}}{\dd t} = \vec{Q}^{\;\prime} \times
\vec{r}^{\;\prime}
\end{equation}
We can see that the Bloch vector $\vec{r}^{\;\prime}$ precesses
around $\vec{Q}^{\;\prime}$ with the frequency depending on
$|\Omega^2+(\omega_1-\omega_0)^2|^{1/2}$ and the orientation of
$\vec{Q}^{\;\prime}$. If the initial direction of the Bloch vector
$\vec{r}^{\;\prime}$ is approximately parallel to
$\vec{Q^{\;\prime}}$, then $\vec{r}^{\;\prime}$ precesses around
$\vec{Q^{\;\prime}}$ on a cone with small angle, and this cone
tends to follow slow variations of the direction of
$\vec{Q^{\;\prime}}$. This is called \textit{adiabatic following},
and can be used to prepare the quantum states of atoms.

\subsubsection{The Rabi solution}
In the following we discuss the interaction with a sinusoidal
exciting field (well stabilized laser), so that in
\eqref{E-NEARLY-RES} the complex amplitude ${\mathcal E}$ is
constant and the phase $\phi$ can be made zero by a proper choice
of the time origin. Historically the solution of the interaction of
a two-level system with such a field was given by Rabi \cite{RAB37}
when he studied  a spin $\frac{1}{2}$ system in a magnetic field.
For our discussion here, let us assume that at $t=0$ the system
starts in the lower state $\ket{\rm lo}$, then $r_3(0)=-1$ and
$r_1(0)=r_2(0)=0$, and the Bloch equations in the rotating frame
are given by
\begin{equation}
\label{RABI-SOLUTION}
\begin{split}
r_1^\prime (t) &=
 \frac{(\omega_0-\omega_1)\Omega}{\Omega^2+(\omega_0-\omega_1)^2}
 \,\{1-\cos[\Omega^2+(\omega_0-\omega_1)^2]^{1/2}\,t \} \\
r_2^\prime(t) &=
 \frac{-\Omega}{[\Omega^2+(\omega_0-\omega_1)^2]^{1/2}}
 \,\sin[\Omega^2+(\omega_0-\omega_1)^2]^{1/2}\,t \\
r_3^\prime(t) &=
 -\frac{(\omega_0-\omega_1)^2+\Omega^2\cos[\Omega^2+
 (\omega_0-\omega_1)^2]^{1/2}\,t}{\Omega^2+(\omega_0-\omega_1)^2}\;
 .
\end{split}
\end{equation}
These equations characterize an intricate motion.  In
\eqref{BLOCH-R-RHO-REP} we have seen that the $r_3$ component of
the Bloch vector is related to the population inversion of the
two-level system. Eq.~\eqref{RABI-SOLUTION} reveals an oscillation
of population characterized by $r_3^\prime(t)$ around
$(\omega_0-\omega_1)^2/[\Omega^2+(\omega_0-\omega_1)^2]$ with
frequency $[\Omega^2+(\omega_0-\omega_1)^2]^{1/2}$ and amplitude
$\Omega^2/\Omega^2+(\omega_0-\omega_1)^2$. This phenomenon is known
as \textit{Rabi oscillation} or \textit{optical nutation}, and can
be observed in systems, which are well isolated from disturbances,
like in a low pressure gas \cite{WAL77,DAG78}, that means in
systems, in which damping time constants are longer than the time
scale of system motion.

\subsection{Interaction of a two-level system with a quantum
field}\label{Sect-2-lev-int}

According to the postulates of quantum mechanics, a two-level
system initially in its excited state $\ket{\rm up}$ will stay
there for all times in the absence of an interaction. Experience
shows, however, that normally the system will return to its ground
state $\ket{\rm lo}$ after a certain time. In the last sections we
discussed how the interaction with a classical field can change the
occupation of states. On the other hand, spontaneous emission is a
quantum mechanical process, and its description requires the
quantization of both, the atoms and the field. A result of this
theory is that the rate of spontaneous emission is proportional to
the mode density of the surrounding environment.

To illustrate this theory we start with the interaction of a
two-level system with a single mode of the electromagnetic field.
The total Hamiltonian thus consists of a component describing the
two-level system $\HSO{H}_A$, the component of the electromagnetic
field $\HSO{H}_F$, and the interaction Hamiltonian $\HSO{H}_I$
\begin{equation}
\HSO{H}_{\rm tot} = \HSO{H}_A + \HSO{H}_F + \HSO{H}_I \; ,
 \label{TOTHAM}
\end{equation}
where the two-level system Hamiltonian $\HSO{H}_A$ is given by
\eqref{2_LEV-HAM}, and where for the purpose of this short overview
we insert the field Hamiltonian $\HSO{H}_F$ in its semiclassical
form \eqref{semiclH} restricted to one mode $\K$. The interaction
is described by a Hamiltonian analogous to \eqref{INT-H-E}, in
which the field is expressed as Hilbert space operator
\begin{equation}
 \HSO{H}_I = -\Omu \cdot \HSO{E} \; .
\label{QINTHAM}
\end{equation}
Thus, the quantum properties of the field enter at two places,
first through $\HSO{H}_F$ \eqref{semiclH}, and second through
$\HSO{H}_I$. In $\HSO{H}_F$
\begin{equation}
\HSO{H}_F =  \hbar \omega_{\K} \cdot \VO_{\K}(t) \AO_{\K}(t)
 \label{SEMICLHAM}
\end{equation}
the field is represented as the product of annihilation and
creation operators. Even though in this semiclassical Hamiltonian
the zero point energy term $\frac{1}{2}\hbar\omega_{\K}$ of the
fully quantum field Hamiltonian \eqref{NGHO} is missing, a part of
the quantum reality is still represented in the noncommutativity of
$\VO_{\K}$ and $\AO_{\K}$; cf.~\eqref{KommVE}. On the other hand,
in \eqref{QINTHAM} the quantum aspect is introduced by the electric
field, which appears as the Hilbert space operator
\eqref{E-FIELDOP}
\begin{equation}\label{TRAV-WAV-EL-FIELD-OP}
\HSO{E} = {\T \ii \vec{\varepsilon} \sqrt{\frac{\hbar \omega}{2
\epsilon_0 L^3}}} \; [\AO_{\K}\,{\rm e}^{{\rm
i}\vec{k}\cdot\vec{r}}
 \;-\;  \VO_{\K}\,{\rm e}^{-{\rm i}\vec{k}\cdot\vec{r}}] \; .
 \end{equation}
In this notation $\oE$ describes a \textit{travelling wave}
mode\footnote{Depending on the geometry of the problem, a
\textit{standing wave} mode can be more suitable
\protect\cite{SHO90}. The corresponding operator is then given by:
$\HSO{E} = \vec{\varepsilon} \sqrt{\frac{\hbar \omega}{\epsilon_0
L^3}}\,\sin\vec{k}\vec{r}\;(\AO+\VO)$}, and $\vec{\varepsilon}$ is
the polarization vector. In the rotating wave approximation the
interaction Hamiltonian \eqref{QINTHAM} can be written as
[cf.~\eqref{2-LEV-OP1}]
\begin{equation}
\begin{split}
 \label{QINTHAM-RWA}
 \HSO{H}_I &= \frac{\hbar
\Omega}{2}\,(\VO_{\K}\,\ket{\rm lo}\bra{\rm up} + {\rm h.c.})
\\
&= \frac{\hbar \Omega}{2}\,(\VO_{\K}\,\HSO{b} + {\rm h.c.})\;,
\end{split}
\end{equation}
where the coupling coefficient $\Omega$ represents the vacuum Rabi
frequency \eqref{RABIFRQ}, which in general depends on the location
$\vec{r}$ of the two-level system. The total Hamiltonian thus is
given by
\begin{equation}
\HSO{H}_{\rm tot} = E_0 + \hbar\omega_0 \OR_3 +  \hbar \omega_{\K}
\cdot \VO_{\K}\AO_{\K} + \frac{\hbar \Omega}{2}\,(\VO_{\K}\,\HSO{b}
+ {\rm h.c.}) \; .
 \label{TOTHAMILTON}
\end{equation}

\bigskip
In \eqref{QINTHAM-RWA} we have represented the interaction
Hamiltonian in a form which permits an intuitive interpretation:
the two-level system can
\begin{itemize}
\item absorb a photon from the field and make a transition from state $\ket{\rm
lo}$ to $\ket{\rm up}$, or
\item emit a photon to the field and make a transition from $\ket{\rm
up}$ to $\ket{\rm lo}$.
\end{itemize}
A state $\ket{\Psi}$ of the combined quantum system consisting of
the two-level system in state $\ket{\lambda}$ and the field say in
a number state $\ket{n}$ can be represented symbolically as
$\ket{\Psi}=\ket{\lambda,n}$. Within the framework of the model
described by \eqref{TOTHAMILTON} a state $\ket{\Psi}=\ket{{\rm
up},n}$ can only couple with state $\ket{\Psi^\prime}=\ket{{\rm
lo},n+1}$. Consequently, we are allowd to consider the interaction
for each manifold of levels ${\ket{{\rm up},n},\ket{{\rm lo},n+1}}$
independently. Obviously, in each manifold the number of
excitations amounts to $n+1$ and is conserved. Technically, this
means that for each manifold the problem is reduced to solving a
two-level problem. The full dynamics is then obtained by summing
over the dynamics within the appropriate manifold.

\bigskip
To illustrate the basic concept of spontaneous emission, let us
assume that the two-level system is initially in its excited state
$\ket{\rm up}$ and interacts with only one field mode which is in
the vacuum state $\ket{0}$,
\begin{equation}
 \ket{\Psi(t=0)} = \ket{{\rm up},0}\; .
 \label{2-LEV-SYS-INI-COND}
\end{equation}
As stated above, the two-level-field system remains in the
one-quantum excitation manifold for all time (given the validity of
\eqref{TOTHAMILTON}), and its state at time $t$ is given by a
superposition of $\ket{{\rm up},0}$ and $\ket{{\rm lo},1}$. At
resonance $\omega_0 = \omega_{_K}$, the state is described by
\begin{equation}
\ket{\Psi(t)} = \cos\left(\frac{\Omega t}{2}\right) \ket{{\rm
up},0} - \ii \sin\left(\frac{\Omega t}{2}\right)\ket{{\rm lo},1}\;
.
\end{equation}
The probability to find the two-level system in its ground state
$\ket{\rm lo}$ after time $t$ is
\begin{equation}
P_{\rm lo} = \abs{\bra{{\rm lo},1}\Psi(t)\rangle}^2 =
\sin^2\left(\frac{\Omega t}{2}\right) \; .
\end{equation}
This result shows, that the initially unoccupied ground state
$\ket{\rm lo}$ spontaneously becomes occupied, even though the
field initially was in a vacuum state. This is due to the effects
of the quantum nature of the field, which we pointed out at the
beginning of this section \ref{Sect-2-lev-int}. Note, however, that
if the two-level system is initially in its ground state $\ket{\rm
lo}$ and the field in vacuum state $\ket{0}$, then the two-level
system will remain there for all times.

On the other hand, above result for $P_{\rm lo}$ exhibits an
oscillatory behavior at Rabi frequency $\Omega$, which is in
contrast to experience where the decay is irreversible. In the
simple model described by \eqref{TOTHAMILTON} the two-level system
interacts with only one field mode, which also is undamped.
Therefore the two-level system can reabsorb the photon from the
field, so returning to its initial state. Later on we will discuss
how recent experimental progress has allowed to observe this
oscillatory regime. But for the moment, we will discuss how
spontaneous emission becomes irreversible, when the photon is
emitted into a multimode vacuum field.

\subsection{The Fermi golden rule}\label{Fermigoldenr}

A realistic model of a two-level system in free space involves the
interaction with a multimode field. Instead of \eqref{TOTHAMILTON}
we therefore consider a Hamiltonian of the following form
\begin{equation}
\HSO{H}_{\rm tot} = E_0 + \hbar\omega_0 \OR_3 + \sum_{\K} \hbar
\omega_{\K} \cdot \VO_{\K}\AO_{\K} + \sum_{\K} \frac{\hbar
\Omega_{\K}}{2}\,(\VO_{\K}\,\HSO{b} + {\rm h.c.}) \; ,
 \label{MULTIM-TOTHAMILTON}
\end{equation}
where the last term describing the interaction is the multimode
interaction Hamiltonian $\HSO{H}_I$; cf.\ the single mode
Hamiltonian \eqref{QINTHAM-RWA}.

Similar as above in the single mode case we assume that all the
modes of the field are initially in their vacuum state and that the
two-level system is in its excited state, so that we can write
\begin{equation}
\ket{\Psi(t=0)} = \ket{{\rm up},\{0\}} \; ,
\end{equation}
where $\ket{\{0\}}$ labels the multimode vacuum, and for each mode
$\K$ we have $\HSO{a}_{\K}\ket{\{0\}}=0$. The state of the combined
system can then formally be described by the superposition of
states
\begin{equation}
\ket{\Psi(t)} = a(t)\,{\rm e}^{-{\rm i}\omega_0 t}\,\ket{{\rm
up},0} + \sum_{\K} b_{\K}(t)\;{\rm e}^{-{\rm i}\omega_{\K}t}\,
\ket{{\rm up}, 1_{\K}}\; ,
 \label{MULTI-MODE-SUP}
\end{equation}
where the coefficients at $t=0$ are set to fulfill the initial
conditions, thus $a(t=0)=1,\; b_{\K}(t=0)=0$. For convenience we
have extracted the fast time dependence, and we have reset the
origin of the two-level system energy so that $\hbar\omega_{\rm
lo}=0$ and $\hbar\omega_{\rm up}=\hbar\omega_0$. The field state
$\ket{1_{\K}}$ can be represented by
\begin{equation}
\ket{1_{\K}} = \HSO{a}^\dagger_{\K}\,\ket{0}\; .
\end{equation}
The equations of motion for the coefficients $a(t)$ and $b(t)$ are
readily obtained as
\begin{align}
 \frac{\dd a(t)}{\dd t} &= - \frac{\ii}{2}\sum_{\K}\Omega_{\K}
  \ee^{-{\rm i}(\omega_{\K}-\omega_0)t}b_{\K}(t)
   \label{EQMOT-A} \\
 \frac{\dd b_{\K}(t)}{\dd t} &= - \frac{\ii}{2}\,\Omega^\ast_{\K} \ee^{{\rm
   i}(\omega_{\K}-\omega_0)t}a(t)
  \label{EQMOT-B}
\end{align}
If we restrict ourselves to times $t$ close enough to $t=0$ so that
the initial coefficients do not change significantly, we can
approximate $a(t)$ in \eqref{EQMOT-B} by its initial value
$a(t=0)=1$ (this approximation is often named first order
perturbation theory). With this we can readily integrate
\eqref{EQMOT-B} and we obtain
\begin{equation}
\abs{b_{\K}(t)}^2 = \frac{\abs{\Omega_{\K}}^2}{4} \times
\frac{\sin^2\left[\tfrac{1}{2} (\omega_{\K}-\omega_0)t
\right]}{\tfrac{1}{4}(\omega_{\K}-\omega_0)^2} \; .
 \label{B-SQUARE}
\end{equation}
The probability to find the two-level system in its excited state
after time $t$ is then given by
\begin{equation}
P_{\rm up}= 1 - \sum_{\K} \abs{b_{\K}(t)}^2 \; ,
 \label{UPPERSTATE-PROB}
\end{equation}
where the sum collects the contributions of every mode which the
two-level system can couple with. Here we can see for the first
time, how the structure of the mode space affects the spontaneous
emission: As the sum only gets positive contributions it clearly
increases when the number of participating modes increases, so
reducing the probability $P_{\rm up}$. Since the frequency
bandwidth over which the dipole moment couples with the field is
limited, so actually the number of modes per frequency interval is
the significant parameter governing the size of the sum. And this
parameter depends on the geometry of the space in which the
fluorescent system can radiate into. This is the observation on
which attempts to modify the spontaneous emission rate will hook
on. We will come back to this in a moment.

When the two-level system radiates into a space which is densely
populated with modes, such as free space for example, then it
couples to a whole continuum of modes over which we have to extend
the summation. Mathematically this means replacing the sum in
\eqref{UPPERSTATE-PROB} over the modes $\K$ by an integral. We
discussed this problem in part \ref{class-HAM-sourcefree}, where we
obtained the transformation rule \eqref{SUM-TO-OMEGAINT}. If we
equate $\zeta$ in \eqref{SUM-TO-OMEGAINT} with \eqref{B-SQUARE}
where we insert \eqref{RABIFRQ} for the vacuum Rabi frequency with
\eqref{E-FIELD-PER-PHOTON} as electric field, we can work out the
integrals and we obtain
\begin{equation}
P_{\rm up}= 1 - \frac{1}{6 \epsilon_0 \pi^2 \hbar c^3}
 \int_{\K} \dd \omega_{\K}\; \omega_{\K}^3 \abs{\mu_{12}}^2\;
\frac{\sin^2\left[\tfrac{1}{2} (\omega_{\K}-\omega_0)t \right]}{[
\tfrac{1}{2}(\omega_{\K}-\omega_0)]^2} \; .
\end{equation}

Now, as time $t$ moves on, the term with ${\sin^2\left[\tfrac{1}{2}
(\omega_{\K}-\omega_0)t \right]}/{[
\tfrac{1}{2}(\omega_{\K}-\omega_0)]^2}$ will be significantly over
zero only for modes with frequency $\omega_{\K}\approx\omega_0$.
This term therefore secures energy conservation in the interaction;
in fact mathematically
\begin{equation}
\lim_{t\to\infty}\frac{\sin^2\left[\tfrac{1}{2}
(\omega_{\K}-\omega_0)t \right]}{[
\tfrac{1}{2}(\omega_{\K}-\omega_0)]^2} = 2\pi t \;
\delta(\omega_{\K}-\omega_0)\; . \label{ENERGY-CONS}
\end{equation}
Inserting this we find for the decay rate of the excited state in a
free space vacuum environment (which is also known as the
\textit{Einstein A-coefficient})
\begin{equation}
\gamma_{\rm fs}= \frac{\dd P_{\rm up}}{\dd t}= -
\frac{\omega_0^3}{c^3}
 \;\frac{\abs{\vec{\mu}_{12}}^2}{3\pi\epsilon_0\hbar}
 =-\abs{k_0}^3 \;\frac{\abs{\vec{\mu}_{12}}^2}{3\pi\epsilon_0\hbar} \; .
 \label{GAMMA-GOLDEN-R}
\end{equation}
This is an example of a special case of Fermi's golden rule, which
predicts that for times large enough that energy conservation
holds, but short enough that first-order perturbation theory
applies, the excited upper state irreversibly decays according to
\eqref{GAMMA-GOLDEN-R}. In summary, when all the approximations
described above are considered the Fermi golden rule for a general
transition can be expressed as
\begin{equation}
\text{Fermi golden rule:}\qquad\gamma =
 \frac{2\pi}{\hbar^2}\sum_f\abs{\bra{f}\HSO{H}_I\ket{i}}^2\;
 \delta(\omega_i-\omega_f)\; ,
 \label{FERMI-GOLDEN-RULE}
\end{equation}
where $\gamma$ characterizes the decay rate from the initial state
$\ket{i}$ to the final state $\ket{f}$.

\medskip
Let us go back to the initial first-order perturbation theory
assumptions, in which we agreed to restrict ourselves to short
times $t$ so that for the coefficients $a$ and $b$ we can write
$a(t=0)=1,\; b_{\K}(t=0)=0$. Of course, when this holds, then the
upper state is still nearly fully excited, and $P_{\rm up}\approx
1$. Thus we still stay within the limits of our initial assumptions
when we write
\begin{equation}
 \frac{\dd P_{\rm up}}{\dd t} \approx - \gamma_{\rm fs}\;P_{\rm up}\; ,
 \label{EXP-DECAY-CONJ}
\end{equation}
which of course corresponds to an exponential decay with decay rate
$\gamma_{\rm fs}$.

In the next paragraph we will show that this is true even for
longer times than we have considered here.

\subsection{The Weisskopf-Wigner theory}\label{WignerWeisskopf}

An other approach to solve the dynamical equations for the
coefficients $a$ and $b$ \eqref{EQMOT-A}, \eqref{EQMOT-B} was
introduced by Weisskopf and Wigner \cite{WEI30}. They start by
formally integrating \eqref{EQMOT-B}, and inserting this result in
\eqref{EQMOT-A}, which will produce
\begin{equation}
 \frac{\dd a(t)}{\dd t} = -
 \sum_{\K}\frac{\abs{\Omega_{\K}}^2}{4}\;
 \int_0^t \dd t^\prime \ee^{-{\rm
 i}(\omega_{\K}-\omega_0)(t-t^\prime)}a(t^\prime) \; .
\end{equation}
As above and for the same reason we replace the sum over the modes
by an integral, which results to
\begin{equation}
 \frac{\dd a(t)}{\dd t}  = - \frac{1}{6 \epsilon_0 \pi^2\hbar c^3}
 \int_{\K} \dd \omega_{\K}\; \omega_{\K}^3 \abs{\mu_{12}}^2\;
 \int_0^t \dd t^\prime \ee^{-{\rm
 i}(\omega_{\K}-\omega_0)(t-t^\prime)}a(t^\prime)   \; .
 \label{EQMOT-A-WW}
\end{equation}
The construction of the dependent superposition of states
\eqref{MULTI-MODE-SUP} was such that compared with all other time
dependent functions the variation in $a$ and $b$ is slow, and
therefore we assume that the variation of $a(t)$ in
\eqref{EQMOT-A-WW} is much slower than in the exponential as well,
and can thus be pulled out of the time integration. At the end we
may check, if this assumption is consistent with the result.
Because of
\begin{equation}
 \lim_{t\to\infty}\int_0^t \dd t^\prime \ee^{-{\rm
 i}(\omega_{\K}-\omega_0)(t-t^\prime)} =
 \pi\, \delta(\omega_{\K}-\omega_0) - {\cal P}\left[
 \frac{\ii}{\omega_{\K}-\omega_0}\right] \; ,
 \label{DELTA-REPRES}
\end{equation}
for larger times we have a similar situation as in
\eqref{ENERGY-CONS}, where here now the exponential assures energy
conservation. Neglecting the frequency shift due to the principal
value term (which is analogous to the Lamb shift) we find
\begin{align}
\frac{\dd a(t)}{\dd t}& = -\frac{\gamma_{\rm fs}}{2}\, a(t)\; ,\\
\text{or} \qquad \qquad & \notag\\
 \frac{\dd P_{\rm up}}{\dd t} &= - \gamma_{\rm fs}\;P_{\rm
 up}\; ,
 \label{WEISSKOPF-WIGNER}
\end{align}
which is the same as we guessed in \eqref{EXP-DECAY-CONJ}.

\subsection{Reservoir theory and master equation}\label{MasterEquation}

In the last two paragraphs we have seen how the irreversibility of
spontaneous emission emerged when the source was coupled with a
quantized field, more exactly, its vacuum modes. Notwithstanding,
the Hamiltonians are perfectly energy conserving, so the apparent
nonreversible dynamics of the system comes somewhat surprising and
its physical reason is rather obscure. With the reservoir theory
approach we will discuss as next, we will be able to gain a more
intuitive understanding of the physical processes which lead to the
nonreversible decay of the excited level of a two-level system
\cite[p.\ 374ff]{MEY91}. In reservoir theory we shift the
perspective from the particular two-level system coupled with the
field to a more general situation, which is \textit{a small system
coupled to a large system}. We characterize the small system by its
Hamiltonian $\HSO{H}_s$, the large system by $\HSO{H}_r$, and their
coupling by the interaction Hamiltonian $\HSO{V}$. Thus:
$\HSO{H}=\HSO{H}_s+\HSO{H}_r+\HSO{V}$. For our particular case the
small system can be identified with the two-level system and the
large system with the continuum of modes of the field. In addition
we assume that the large system always stays in thermal equilibrium
at some temperature $T$. This means it acts as a \textit{thermal
reservoir}. A thermal reservoir is usually described by an
equilibrium (i.e.\ time independent) density operator of the form
\begin{equation}
\Orho_r = \frac{1}{Z} \, \exp{(-\HSO{H}_r/k_BT)} \; ,
\end{equation}
where with $k_B$ we denote Boltzman's constant, and the
\textit{partition function} $Z$ is given by the trace over the
reservoir ${\rm Tr}_r$
\begin{equation}
Z= {\rm Tr}_r\bigl[\exp{(-\HSO{H}_r/k_BT)}\bigr] \; .
\end{equation}

Now, we are interested in the dynamics of the small system only. In
that case we can find the dynamics in the evolution of the
\textit{reduced density operator} $\Orho_s$
\begin{equation}
\Orho_s = {\rm Tr}_r \bigl(\Orho_{sr}\bigr)\; ,
\end{equation}
where $\Orho_{sr}$ is the density operator associated with the full
system, i.e.\ small system \textit{and} reservoir. Thus the reduced
density operator $\Orho_s$ is the trace over the reservoir of the
total density operator. If we know the reduced density operator at
any time $t$ we can calculate the expectation value of any system
operator. The equation of motion for $\Orho_{sr}$ is called a
\textit{master equation}, and this is what we will derive in the
following. In order to focus directly to the relevant dynamic time
scale of the system, we switch to the interaction picture, where
all the free evolution is eliminated.  The interaction between the
small system and the reservoir is described by the Schr\"odinger
picture interaction Hamiltonian $\HSO{V}$, for which the
interaction picture representation $\HSO{V}_I(t-t_0)$ is obtained
by the unitary transformation
\begin{equation}
\HSO{V}_I(t-t_0) = \exp{[\ii\HSO{H}_0(t-t_0)/\hbar]}\;\HSO{V}\;
\exp{[-\ii\HSO{H}_0(t-t_0)/\hbar]} \; , \label{IP-INTHAM}
\end{equation}
and where we have set $\HSO{H}_0=\HSO{H}_s+\HSO{H}_r$. Similarly we
can relate the full system density operator in the interaction
picture $\HSO{P}_{sr}$ to the Schr\"odinger picture density
operator $\Orho_{sr}$  by the unitary transformation
\begin{equation}
\Orho_{sr}(t) = \exp{[-\ii\HSO{H}_0(t-t_0)/\hbar]}\;\HSO{P}_{sr}\;
\exp{[\ii\HSO{H}_0(t-t_0)/\hbar]} \; .
\end{equation}
Observing the Schr\"odinger picture rule $\dot{\Orho}=-\frac{{\rm
i}}{\hbar}[\HSO{H},\Orho] $ we obtain the following time derivative
\begin{equation}
\frac{\partial \Orho_{sr}}{\partial t} =\frac{\ii}{\hbar}\;
 \exp{[-\ii\HSO{H}_0(t-t_0)/\hbar]}\;
 \left\{
 [\HSO{H}_0 , \HSO{P}_{sr}(t)]+
 \frac{\partial \HSO{P}_{sr}}{\partial t}
 \right\} \;
 \exp{[\ii\HSO{H}_0(t-t_0)/\hbar]} \; ,
\end{equation}
in which the motion of the density operator in the Schr\"odinger
picture is related to its motion in the interaction picture. From
this we obtain the interaction picture equation of motion
\begin{equation}
 \frac{\partial \HSO{P}_{sr}}{\partial t} =
 -\frac{\ii}{\hbar}\;\left[\HSO{V}_I(t-t_0), \HSO{P}_{sr}(t)  \right]
 \; .
\end{equation}

We may assume that at $t=t_0$ the small system and the reservoir do
not exhibit any correlations. We can therefore approximately solve
this equation to second order in perturbation theory, and obtain
\begin{equation}
\begin{split}
 \label{IP-TOT-RHO}
 \HSO{P}_{sr}(t) =  \HSO{P}_{sr}(t_0) &-
 \frac{\ii}{\hbar}\;\int_{t_0}^t\dd t^\prime
 \left[\HSO{V}_I(t-t_0), \HSO{P}_{sr}(t_0)  \right] - \\
 &- \frac{1}{\hbar^2}\;\int_{t_0}^t\dd t^\prime \int_{t_0}^{t^\prime}\dd t^{\prime\prime}
 \left[\HSO{V}_I(t^\prime-t_0),[\HSO{V}_I(t^{\prime\prime}-t_0),\HSO{P}_{sr}(t_0)]
 \right] + \dots \quad .
\end{split}
\end{equation}
We trace out the reservoir and obtain the reduced density operator
in the interaction picture $\Orho(t)={\rm Tr}_r
\bigl[\HSO{P}_{sr}(t)\bigr]$ for which we can write
\begin{equation}
\Orho(t) = \exp{[\ii\HSO{H}_s(t-t_0)/\hbar]}\;\Orho_s\;
\exp{[-\ii\HSO{H}_s(t-t_0)/\hbar]}
\end{equation}

When the time interval $\tau=t-t_0$ is long compared to the
relaxation time (memory time) of the reservoir $\tau_c$, but short
compared to times in which the small system variables show
significant changes (for example $\gamma_{\rm fs}^{-1}$ in
spontaneous emission), we can define a \textit{coarse-grained}
equation of motion (time derivative) by
\begin{equation}
\dot{\Orho}(t) \approx \frac{\Orho(t)-\Orho(t-\tau)}{\tau}\; .
\label{DEF-COARSE-GRAINED-DERIV}
\end{equation}
Applying this to \eqref{IP-TOT-RHO} we obtain after some algebra
\begin{equation}
 \label{IP-DOT-REDRHO}
\begin{split}
 \dot{\Orho}(t) = &
 -\frac{\ii}{\hbar\tau} \int_0^\tau \dd\tau^\prime\;
 {\rm Tr} \bigl\{ \HSO{V}_I(\tau^\prime)\HSO{P}_{sr}(t) \bigr\} - \\
 & -\frac{1}{\hbar^2\tau} \int_0^\tau\dd\tau^\prime
 \int_0^{\tau^\prime}\dd\tau^{\prime\prime}
 {\rm Tr} \bigl\{\HSO{V}_I(\tau^{\prime})\HSO{V}_I(\tau^{\prime\prime})
  \HSO{P}_{sr}(t) - \HSO{V}_I(\tau^{\prime}) \HSO{P}_{sr}(t)
  \HSO{V}_I(\tau^{\prime\prime}) \bigr\} + {\rm adj.}\; .
\end{split}
\end{equation}
It can be shown that with the properties of the reservoir given
here the first term on the rhs of \eqref{IP-DOT-REDRHO} vanishes,
and that the second term is composed of a sum of two-time
correlation functions of reservoir operators only \cite[p.\
380]{MEY91}. In fact, with the dipole coupling Hamiltonian
\eqref{MULTIM-TOTHAMILTON} and taking into account the transform
\eqref{IP-INTHAM} we have
\begin{equation}
\HSO{V}_I(t) = \hbar \, \HSO{b}^\dagger \HSO{F}(t) + {\rm adj.} \;
,
\end{equation}
where the operator
\begin{equation}
\HSO{F}(t) = \sum_{k}\frac{\Omega_{k}}{2}\;\HSO{a}_k\;\ee^{{\rm
i}(\omega_0-\omega_{k})t}
\end{equation}
acts only in the reservoir Hilbert space. As is better visible from
their Fourier transformed form, such operators are usually
associated with noise sources and are thus called \textit{noise
operators}. The trace over the reservoir involves first order
correlation functions of the forms
$\langle\HSO{F}(t^\prime)\HSO{F}(t^{\prime\prime})\rangle\, , \;
\langle\HSO{F}^\dagger(t^\prime)\HSO{F}^\dagger(t^{\prime\prime})\rangle
\, ,
\;\langle\HSO{F}(t^\prime)\HSO{F}^\dagger(t^{\prime\prime})\rangle
\, , \;
\langle\HSO{F}^\dagger(t^\prime)\HSO{F}(t^{\prime\prime})\rangle
\;$.  For example,
\begin{equation}
 \langle\HSO{F}(t^\prime)\HSO{F}^\dagger(t^{\prime\prime})\rangle_r
 = \sum_{k,k^\prime}\frac{\Omega_k}{2}\,\frac{\Omega_{k^\prime}^\ast}{2}
 \; \langle\HSO{a}_k\HSO{a}^\dagger_{k^\prime}\rangle_r\;
 \ee^{{\rm i} \omega_0(t^\prime-t^{\prime\prime})}
 \ee^{{\rm
 i}(\omega_kt^\prime-\omega_{k^\prime}t^{\prime\prime})}\; ,
 \label{F-NOISE-CORR}
\end{equation}
where $\langle\dots\rangle_r$ denotes the average over the
reservoir. As we have
\begin{equation}
 \langle\HSO{a}_k\HSO{a}_{k^\prime}^\dagger\rangle_r =
 (\bar{n}_k+1)\,\delta_{kk^\prime}\; ,
\end{equation}
where $\bar{n}_k$ denotes the average number of thermal photons in
mode $k$ ($\bar{n}_k=0$ at zero temperature), the correlation
function \eqref{F-NOISE-CORR} reduces to
\begin{equation}
 \langle\HSO{F}(t^\prime)\HSO{F}^\dagger(t^{\prime\prime})\rangle_r
 = \sum_{k}\frac{\abs{\Omega_k}^2}{4}\,(\bar{n}_k+1)\;
 \ee^{{\rm i}(\omega_0-\omega_k)(t^\prime-t^{\prime\prime})} \; .
 \label{NOISE-CORR-RES}
\end{equation}
For the other correlation functions appearing in
\eqref{IP-DOT-REDRHO} analogous expressions can be obtained. As we
assumed a reservoir in thermal equilibrium, thus with stationary
statistical properties, the reservoir correlations depend only on
the time difference $\Delta t = t^\prime - t^{\prime\prime}$, and
we obtain for an example term appearing in \eqref{IP-DOT-REDRHO}
\begin{equation}
 \int_0^\tau\!\!\dd \tau^\prime\!
 \int_0^{\tau^\prime}\!\!\!\dd\tau^{\prime\prime}\,
 \langle\HSO{F}(\tau^\prime)\HSO{F}^\dagger(\tau^{\prime\prime})\rangle_r
 = \int_0^\tau\!\!\dd\tau^\prime\! \int_0^{\tau^\prime}\!\!\!\dd\Delta t
 \sum_{k}\frac{\abs{\Omega_k}^2}{4}\,(\bar{n}_k+1)\;
 \ee^{{\rm i}(\omega_0-\omega_k)\Delta t} \; .
 \label{NOISE-INT-1}
\end{equation}
The time dependence of \eqref{NOISE-INT-1} is governed by the first
order correlations existing in the reservoir, for which
\eqref{NOISE-CORR-RES} is an example. Given the reservoir
properties introduced in the discussion of
\eqref{DEF-COARSE-GRAINED-DERIV} (reservoir correlation time much
shorter than characteristic small system time constants) we can
extend the upper integration limit $\tau^\prime$ in
\eqref{NOISE-INT-1} to infinity, and we obtain
\begin{align}
 \int_0^\tau\!\!\dd \tau^\prime\!
 \int_0^{\tau^\prime}\!\!\!\dd\tau^{\prime\prime}\,
 \langle\HSO{F}(\tau^\prime)\HSO{F}^\dagger(\tau^{\prime\prime})\rangle_r
 &=
 \int_0^\tau\!\!\dd\tau^\prime\! \int_0^\infty\!\!\!\dd\Delta t
 \sum_{k}\frac{\abs{\Omega_k}^2}{4}\,(\bar{n}_k+1)\;
 \ee^{{\rm i}(\omega_0-\omega_k)\Delta t} \notag \\
 &=
 \frac{\tau}{6\epsilon_0\pi^2\hbar c^3}\,
 \int \!\!
 \dd\omega_k\;\omega_k^3\, \abs{\mu_{12}}^2 (\bar{n}_k+1)
 \int_0^\infty\!\!\!\dd t \; \ee^{{\rm i}(\omega_0-\omega_k)t}\; ,
 \label{NOISE-INT-2}
\end{align}
where we have replaced the sum over the modes by an integral in the
same way as we discussed in paragraph \ref{Fermigoldenr} and
\ref{WignerWeisskopf}. Obviously this equation has a similar
structure as \eqref{EQMOT-A-WW}. After some algebra, as inserting
the delta function representation \eqref{DELTA-REPRES} and ignoring
the associated principal part frequency shifts, combining the
different contributions of \eqref{IP-TOT-RHO}, and explicitly
focusing on the two-level system as the small system, we obtain the
interaction picture master equation
\begin{equation}
\begin{split}
 \dot{\Orho}_A =  - \frac{\gamma_{\rm fs}}{2}(\bar{n}+1)\;
 &[\HSO{b}^\dagger\HSO{b}\Orho_A(t) -
 \HSO{b}\Orho_A(t)\HSO{b}^\dagger]-\\
  -\frac{\gamma_{\rm fs}}{2}\bar{n}\;
 &[\Orho_A(t)\HSO{b}\HSO{b}^\dagger-
 \HSO{b}^\dagger\Orho_A(t)\HSO{b}] + {\rm adj.}\quad .
 \label{INT-MASTER-EQ-2LEVSYS}
\end{split}
\end{equation}
In contrast to \eqref{EQMOT-A-WW} of the Weisskopf-Wigner theory,
\eqref{NOISE-INT-2} gives us some insight into the processes
leading to the decay of the excited state. Together with
\eqref{INT-MASTER-EQ-2LEVSYS}, \eqref{NOISE-INT-2} relates the
decay with properties of the reservoir expressed as first order
correlation functions. The approximation of these correlation
functions with delta functions is known as the Markov
approximation. The decay appears thus as a result of the delta
function like memory time of the reservoir which instantaneously
looses track of the interaction with the two level system. So the
way of the excitation back to the two level system is forgotten, so
to speak. Combining the Markov approximation with second-order
perturbation theory to derive the master equation is usually
labeled Born-Markov approximation.

The master equation \eqref{INT-MASTER-EQ-2LEVSYS} describes the
decay of a two-level system which is coupled to an electromagnetic
field characterized as a reservoir at temperature $T$. In this
equation $\bar{n}$ represents the number of thermal quanta of the
reservoir at the two-level transition energy $\hbar\omega_0$. For a
reservoir at zero temperature, the population $P_{\rm up}$ of the
excited state of the two-level system is given by
\begin{equation}
 P_{\rm up}(t) =
 {\rm Tr}\bigl[\,\ket{\rm up}\bra{\rm up}\,\Orho_A(t)\,\bigr]
 = {\rm Tr} \bigl[\,(\HSO{R}_3 +\frac{1}{2})\,\Orho_A(t)\,\bigr]\; ,
\end{equation}
and exhibits the equation of motion
\begin{equation}
 \frac{\dd P_{\rm up}(t)}{\dd t} =
 {\rm Tr}\bigl[\,\ket{\rm up}\bra{\rm up}\,\dot{\Orho}_A(t)\,\bigr]
 = -\gamma_{\rm fs} P_{\rm up}(t) \; ,
\end{equation}
which is the Weisskopf-Wigner result \eqref{WEISSKOPF-WIGNER}.

\bigskip
Let us  look again at the master equation of the two-level system
\eqref{INT-MASTER-EQ-2LEVSYS}. In the Schr\"odinger picture and at
temperature $T=0$ this equation has the form
\begin{equation}
 \dot{\Orho}_A(t) =-\frac{\ii}{\hbar}[\HSO{H}_A,\Orho_A]-
 \frac{\gamma_{\rm fs}}{2}[\HSO{b}^\dagger\HSO{b}\Orho_A(t)+
 \Orho_A(t)\HSO{b}^\dagger\HSO{b}-
 2\HSO{b}\Orho_A(t)\HSO{b}^\dagger ] \; ,
 \label{INT-MASTER-EQ-2LEVSYS-ZEROTEMP}
\end{equation}
where $\HSO{H}_A$ denotes the two-level system Hamiltonian
\eqref{2_LEV-HAM}. If we formally define a non-Hermitian effective
Hamiltonian $\HSO{H}_{\rm eff}$ as
\begin{equation}
\HSO{H}_{\rm eff} = \HSO{H}_A - \ii\hbar\frac{\gamma_{\rm
fs}}{2}\,\HSO{b}^\dagger\HSO{b} \; ,
\end{equation}
then introducing this in \eqref{INT-MASTER-EQ-2LEVSYS-ZEROTEMP}
leads to
\begin{equation}
 \dot{\Orho}_A(t) =
 -\frac{\ii}{\hbar}
 [\HSO{H}_{\rm eff}\Orho_A(t) - \Orho_A(t)\HSO{H}_{\rm eff}^\dagger]+
 \gamma_{\rm fs}\HSO{b}\Orho_A(t)\HSO{b}^\dagger \; .
 \label{INT-MASTER-EQ-2LEVSYS-ZEROTEMP-EFF}
\end{equation}
Going back to \eqref{MULTI-MODE-SUP} we define the nonnormalized
excited two-level system state as
\begin{equation}
\ket{\phi_{\rm up}(t)} = a(t)\,\ket{\rm up}\; .
\label{UNNORMALIZED-EXC-STATE_1}
\end{equation}
With this the master equation
\eqref{INT-MASTER-EQ-2LEVSYS-ZEROTEMP} can be replaced by the
following \textit{effective Schr\"odinger equation}
\begin{equation}
 \ii\hbar\;\frac{\dd }{\dd t} \ket{\phi_{\rm up}(t)}=
 \HSO{H}_{\rm eff} \ket{\phi_{\rm up}(t)} \; .
 \label{EFF-SCHROED-EQ-1}
\end{equation}
Obviously such an equation is technically easier to deal with than
the corresponding master equation, because here state vectors take
over the role of density operators. Recently similar effective
Schr\"odinger equations were introduced to develop Monte Carlo wave
function simulation techniques \cite{GIS92,MOE93}.

\subsubsection{Harmonic oscillator coupled to a reservoir}

In section \ref{CanonicalQieldQuantization} we have shown, that one
mode of the electromagnetic field is mathematically equivalent to a
harmonic oscillator to which in the semiclassical approximation we
can associate the Hamiltonian $\HSO{H}_F = \hbar \omega_{k} \;
\VO_{k} \AO_{k}$ (cf.~\eqref{semiclH}), where the wavenumber $k$
labels the particular mode. The advantage of reservoir theory is
that the particular structure of the reservoir does not affect the
result, as long as the reservoir stays in thermal equilibrium at
all times, and that perturbations of its state decorrelate
immediately in a Markovian sense. We can therefore imagine the
reservoir as a large collection of harmonic oscillators which are
in thermal equilibrium at temperature $T$. For this situation,
namely the harmonic oscillator associated with the field mode,
which is coupled to a reservoir of harmonic oscillators, we can
work out the calculations in analogy to section
\ref{MasterEquation}, and we obtain the following master equation
\begin{equation}
\begin{split}\label{DECAY-HARMON-OSC}
 \dot{\Orho}_F = &\frac{\ii}{\hbar}\; [\HSO{H}_F,\Orho_F] -
 \frac{\kappa}{2}\,(\bar{n}+1)\,
 \bigl[\HSO{a}^\dagger\HSO{a}\Orho_F(t) -
 \HSO{a}\Orho_F(t)\HSO{a}^\dagger\bigr]\\
 &-\frac{\kappa}{2}\,\bar{n}\,
 \bigl[\Orho_F(t)\HSO{a}\HSO{a}^\dagger -
 \HSO{a}^\dagger\Orho_F(t)\HSO{a}\bigr]\quad + \quad{\rm adj.} \; ,
\end{split}
\end{equation}
where $\bar{n}$ represents the number of thermal excitations of the
harmonic oscillator at frequency $\omega_k$. This equation shows
that the decay of the mean number of quanta
$\langle\HSO{a}^\dagger\HSO{a}\rangle$ of the harmonic oscillator
occurs with the rate $\kappa$ at zero temperature, whereas the
expectation value of the annihilation operator $\HSO{a}$ (which is
proportional to the positive frequency part of the electric field
mode) decays with a rate of $\kappa / 2$.

\section{Spontaneous emission in an optical resonator}\label{Sect-fluoor}

Up to now we have shown that spontaneous emission is a consequence
of the coupling of the atomic or molecular system to its
surrounding electromagnetic field which in the last section we
assumed to be the ``universe'' composed of a continuum of plane
wave modes. Thus, spontaneous emission is not a generic property of
the atom or molecule, but depends on the mode structure of the
surrounding environment. Purcell \cite{PUR46} found out that the
spontaneous emission rate can be enhanced by placing an atom into a
resonator in which a mode is tuned to the atomic transition
frequency, and Kleppner \cite{KLE81} described the opposite case,
inhibition of spontaneous emission. Spontaneous emission needs not
be irreversible. Irreversibility is a consequence of the coupling
with a Markovian Reservoir, i.e.\ a reservoir with a very short
memory. For vacuum modes which can not be approximated in this way,
spontaneous emission can exhibit considerable different properties,
as in the case of long memory times, where the excitation can be
periodically exchanged between the atom or molecule and the field
\cite{RAI89,BER92}.

In the microstructures discussed in section \ref{Sect-realiz} the
atoms or molecules are coupled to one, or a few, resonator modes.
However, in section \ref{microresonators} we show that in
wavelength size microresonators it is inevitable that the resonator
modes couple with the continuum of modes of the external world as
well. This, if the resonator cavity is not completely closed around
the atom (or molecule), it can couple to the continuum of vacuum
modes. In an optical resonator we therefore have to consider at
least tree coupling coefficients: (1) The Rabi frequency $\Omega$
\eqref{RABIFRQ} which characterizes the atom (or molecule) coupling
with the resonator modes, (2) the decay rate $\kappa$ of the
resonator mode, and (3) the rate $\gamma_r$ of spontaneous emission
into the vacuum modes to which the system couples with. In
wavelength scale resonators $\gamma_r$ differs from the spontaneous
emission rate into free space $\gamma_{\rm fs}$ because of the
reduced mode density in the resonator. If the system couples with
more than one field mode or when more than two levels are involved,
then we have to consider the associated coefficients as well.

%
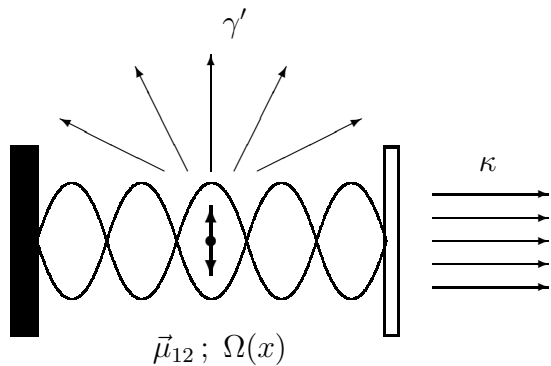
\begin{figure}[!ht]
 \setlength{\unitlength}{0.75em}
\begin{center}
\begin{picture}(25,17)
\thicklines
 \put(1,2){\framebox(1,8)}
 \put(17,2){\framebox(0.5,8)}
 \multiput(1,2)(0.1,0){11}{\line(0,1){8}}
 \put(9.5,6){\vector(0,1){1.5}}
 \put(9.5,6){\vector(0,-1){1.5}}
 \put(9.5,6){\circle*{0.5}}
\thinlines
 \put(19,4){\vector(1,0){5}}
 \put(19,5){\vector(1,0){5}}
 \put(19,6){\vector(1,0){5}}
 \put(19,7){\vector(1,0){5}}
 \put(19,8){\vector(1,0){5}}
 \put(9.5,9){\vector(0,1){5}}
 \put(10.5,9){\vector(1,2){2,24}}
 \put(11.5,9){\vector(2,1){4,47}}
 \put(8.5,9){\vector(-1,2){2,24}}
 \put(7.5,9){\vector(-2,1){4,47}}
 \linethickness{0.5pt}
 \bezier{200}(2,6)(3.5,11)(5,6)
 \bezier{200}(5,6)(6.5,11)(8,6)
 \bezier{200}(8,6)(9.5,11)(11,6)
 \bezier{200}(11,6)(12.5,11)(14,6)
 \bezier{200}(14,6)(15.5,11)(17,6)
 \bezier{200}(2,6)(3.5,1)(5,6)
 \bezier{200}(5,6)(6.5,1)(8,6)
 \bezier{200}(8,6)(9.5,1)(11,6)
 \bezier{200}(11,6)(12.5,1)(14,6)
 \bezier{200}(14,6)(15.5,1)(17,6)
 \put(10,15){$\gamma^\prime$}
 \put(21,9){$\kappa$}
 \put(7,1){$\vec{\mu}_{12}\, ; \;\Omega(x)$}
\end{picture}
\end{center}
\caption{\label{F-P-Resonator}
 Fabry-P\'erot resonator with a standing wave mode of frequency
 $\omega_r$ which is excited by the two-level system with dipole
 moment $\vec{\mu}_{12}$. The coupling of the two-level system to the
 resonator mode $\omega_r$ is characterized by the
 coefficient $\Omega(x)$ which depends on the position of
 $\vec{\mu}_{12}$ with respect to the standing wave field
 distribution. The coupling coefficient $\gamma^\prime$ describes
 the spontaneous emission rate into exterior vacuum modes (because
 of the mirrors, the solid angle which determines the coupling to exterior modes
 differs from the free space case), and $\kappa$ describes the
 decay rate of the field mode due to diffraction losses, mirror
 absorption and transmission.
  }
\end{figure}
%
To illustrate this concept we consider a linear resonator build
with two ideal mirrors spaced by the distance $L$ (Fabry-P\'erot
interferometer). The space between the mirrors defines the
resonator. If we neglect diffraction losses at the mirrors, the
modes of the electromagnetic field in this resonator of volume $V$
consist of a discrete set of standing waves with frequency
$\omega_n=c k_n= c\pi n/L$, where $n$ is an integer. In section
\ref{microresonators} we will show that also transversal modes
exist, which we neglect in this example. If the cavity is small
enough the mode frequencies are well separated, so that only one of
them couples with the transition of the two level system. We label
this mode frequency with $\omega_r$ and with $\omega_0$ we denote
the frequency of the two-level transition (nearly resonant mode).
In reality the mirrors are not perfect and the field is diffracted,
so that the mode is damped. We characterize the damping by the
damping rate $\kappa$. Damping of the field mode is tantamount to
widening the mode frequency to a width proportional to $\kappa$. In
analogy to the free space mode density
\eqref{FREE-SPACE-MODE-DENSITY} we can associate an
\textit{effective mode density} $\rho_r$ with the damped mode
situation in which the two-level system now radiates into. If we
assume that the mode width is less than the separation of the mode
frequency, the effective mode density can be approximated by a
Lorentzian as
\begin{equation}
 \rho_r(\omega)=\frac{\kappa}{2\pi V}\cdot
 \frac{1}{
 \bigl(\tfrac{\kappa}{2}\bigr)^2 +
 (\omega_r-\omega)^2
 } \; .
 \label{RESONATOR-MODE-DENSITY}
\end{equation}
Usually a resonator is characterized by its \textit{quality factor}
$Q$ \cite[p.~430]{SIE86} which is proportional to the number of
resonator round trips of a photon, and which is related to the
damping rate $\kappa$ by
\begin{equation}
Q= \frac{\omega_r}{\kappa} \; . \label{DEF-QUALITY}
\end{equation}
We can calculate the spontaneous emission rate in the resonator in
a similar way as before in sections \ref{Fermigoldenr},
\ref{WignerWeisskopf}, or \ref{MasterEquation}, except that we have
to replace the free space mode density
\eqref{FREE-SPACE-MODE-DENSITY} by \eqref{RESONATOR-MODE-DENSITY}
which represents the resonator geometry. For a resonator tuned near
the atomic transition $\omega_0$ we obtain
\begin{equation}
\text{in resonance:}\qquad\gamma_r \approx \gamma_{\rm fs} \times
Q\; \frac{\lambda_0^3}{V} \;. \label{APPR-ENH-DECAY}
\end{equation}
It is interesting to observe what happens, when the resonator is
tuned out of the resonance. Note that the radiating dipole is still
in the resonator, and not in free space, only  the oscillating
dipole has no mode to radiate into. It can be shown that for
coupling to the off-resonance mode the decay rate can be
approximated by
\begin{equation}
\text{off-resonance:}\qquad\gamma_r \approx \gamma_{\rm fs} \times
Q^{-1}\; \frac{\lambda_0^3}{V} \;.  \label{APPR-INHIB-DECAY}
\end{equation}
Although the situation is not utterly realistic, these equations
illustrate the effect of enhanced or inhibited spontaneous
emission. The equations are very crude approximations in which we
have neglected a number of things, for example that the two-level
system interacts with free modes associated with the open sides of
the resonator volume. In the following we will fix some of the
approximations.

\subsection{Master equation of a two-level system in a
resonator}\label{Sect-Mast-Eq-Res}

As above we assume a two-level system in a resonator which exhibits
sufficiently small dimensions so that the mode frequency separation
$c/2L$ is large compared to the characteristic coupling parameters
of the two-level system
$\Omega,\;\frac{1}{\kappa},\;\frac{1}{\gamma_r}$. In this way the
two level system interacts only with one resonator mode. Let us
also assume that this resonator mode is tuned to near resonance
with the two-level transition: $\omega_r\approx\omega_0$. In
analogy to section \ref{MasterEquation} we now identify the small
system with the two-level system combined with the resonator mode
it couples to. Corresponding to this definition of the small system
we define the small system Hamiltonian $\HSO{H}_s$ as
\begin{equation}
\HSO{H}_s= \HSO{H}_A + \HSO{H}_F + \HSO{H}_I \; ,
\end{equation}
which corresponds to $\HSO{H}_{\rm tot}$ of \eqref{TOTHAM}. The
suitable mode set for describing the field in a resonator are the
standing waves modes. For standing waves the coupling constant
$\Omega$ becomes a function of the position $x$ inside the
resonator, $\Omega(x)= \Omega\cos (kx)$
(cf.~Fig.~\ref{F-P-Resonator}), where $k=\omega_r/c$ and
\begin{equation}
\Omega=2\sqrt{\frac{\hbar\omega_r}{\epsilon_0 V}}
\;\;(\vec{\mu}_{12}\cdot\vec{\varepsilon})\; ,
\label{RABI-FRQ-GEN-SYS}
\end{equation}
with $\vec{\varepsilon}$ denoting a unit vector in the direction of
the electric field polarization; cf.\ \eqref{RABIFRQ} and footnote
on p.~\pageref{TRAV-WAV-EL-FIELD-OP}. Transforming into a reference
frame which rotates with the resonator mode frequency $\omega_r$
the Hamiltonian becomes
\begin{equation}
\HSO{H}_s = -\hbar \delta \HSO{R}_3 + \HSO{H}_I \; ,
\end{equation}
with
\begin{equation}
 \HSO{H}_I = \frac{\hbar\Omega(x)}{2}\;
 \bigl( \HSO{a}^\dagger\HSO{b}+\HSO{b}^\dagger\HSO{a} \bigr)
 \; ,
\end{equation}
and where $\delta = \omega_r-\omega_0$ denotes the detuning.

As we have mentioned before, and illustrated in
Fig.~\ref{F-P-Resonator}, the small system
two-level-system-cavity-mode is subject to two dissipative
processes. The first one consists in the coupling with a cone of
free space modes through the open sides of the resonator structure.
The second dissipative process is the loss of the standing wave
resonator mode through the mirrors to the exterior world, which is
characterized by the mode decay rate $\kappa$. Those two mechanisms
are statistically independent and can therefore be combined by a
simple addition, which in a reservoir theoretical approach looks as
follows: The small system consisting of the two-level system and
the single resonator mode is coupled to two thermal reservoirs that
represent the cone of electromagnetic modes at the open resonator
sides, and the mirror losses. The thermal reservoirs are both
modeled as a continuum set of harmonic oscillators. In accordance
with the Born-Markov approximation the coupling of the small system
to the two reservoirs is described by a master equation in which
the dissipative (non-Hermitian) part consists of the sum of
\eqref{INT-MASTER-EQ-2LEVSYS} and \eqref{DECAY-HARMON-OSC}, where
we now have to take into account that the two-level system only
interacts with free space modes contained in a cone defined by the
openings in the resonator walls. As a consequence we replace the
free space spontaneous emission rate $\gamma_{\rm fs}$ in
\eqref{INT-MASTER-EQ-2LEVSYS} by the smaller rate $\gamma^\prime$.
We then can derive the small system master equation, which is
similar to \eqref{INT-MASTER-EQ-2LEVSYS-ZEROTEMP-EFF}, and which at
zero temperature looks like
\begin{equation}
 \dot{\Orho}_s(t) =
 -\frac{\ii}{\hbar}
 \bigl[\HSO{H}_{\rm eff}\Orho_s - \Orho_s\HSO{H}_{\rm eff}^\dagger\bigr] +
  \kappa\HSO{a}\Orho_s\HSO{a}^\dagger +
 \gamma^\prime\HSO{b}\Orho_s\HSO{b}^\dagger \; .
 \label{MASTER-EQ-HARM-OSC-RES}
\end{equation}
where
\begin{equation}
\HSO{H}_{\rm eff} = \HSO{H}_s + \HSO{H}_{\rm loss}\; ,
\label{EFF-HAM}
\end{equation}
and
\begin{equation}
 \HSO{H}_{\rm loss} =
 -\ii\hbar \,\frac{\gamma^\prime}{2}\;\HSO{b}^\dagger\HSO{b}-
 \ii\hbar \, \frac{\kappa}{2}\;\HSO{a}^\dagger\HSO{a} \; .
\end{equation}

As in section \ref{Sect-2-lev-int} [cf.\
\eqref{2-LEV-SYS-INI-COND}] we assume that initially at $t=0$ the
state of the system is characterized by an excited two-level system
and an empty resonator mode
\begin{equation}
\ket{\Psi(0)} = \ket{{\rm up},0} \; .
\end{equation}
As next we have to identify the possible states the system can
occupy. The energy conserving part of the small system evolution
consists in the exchange of quanta between the two-level system and
the resonator mode. As we found out in section
\ref{Sect-2-lev-int}, the total number of quanta in the energy
conserving process is constant -- in this case here the total
number is one quantum, or excitation respectively. On the other
hand, the dissipative processes -- that are the
coupling-to-the-reservoir processes -- are characterized by the
irreversible loss of the small system excitation. In summary we
therefore distinguish three relevant states: two states
corresponding to full excitation (one quantum states), $\ket{{\rm
up},0}$ and $\ket{{\rm lo},1}$, and the state corresponding to the
total loss of excitation $\ket{{\rm lo},0}$ (zero quantum state).
The evolution of the total quantum system takes place in the fully
excited state subspace and in the zero quantum state subspace. In
analogy to the way we derived \eqref{UNNORMALIZED-EXC-STATE_1} and
\eqref{EFF-SCHROED-EQ-1} we can define the following
\textit{unnormalized} single excitation state
\begin{equation}
 \ket{\Psi(t)} =
 c_{\rm up}(t)\ee^{\frac{\delta t}{2}}\ket{{\rm up},0}
 +
 c_{\rm lo}(t)\ee^{\frac{\delta t}{2}}\ket{{\rm lo},1} \;
 \label{SINGLE-EX-WAVFUNC}
\end{equation}
where $\delta = \omega_r-\omega_0$ denotes the detuning. From
\eqref{MASTER-EQ-HARM-OSC-RES} we can derive the effective
Schr\"odinger equation describing the system evolution
\begin{equation}
\ii\hbar\;\frac{\dd}{\dd t}\;\ket{\Psi(t)}=\HSO{H}_{\rm
eff}\ket{\Psi(t)}\; , \label{EFF-SCHROED-EQ-2}
\end{equation}
where $\HSO{H}_{\rm eff}$ is given by \eqref{EFF-HAM}. Inserting
\eqref{SINGLE-EX-WAVFUNC} we obtain the equations of motion for the
coefficients $c_{\rm up}(t)$ and $c_{\rm lo}(t)$
\begin{align}
 \frac{\dd c_{\rm up}(t)}{\dd t} & =
 -\frac{\gamma^\prime}{2}\, c_{\rm up}(t)
 -\ii \frac{\Omega}{2}\, c_{\rm lo}(t)
 \label{DOT-C-UP} \\
 \frac{\dd c_{\rm lo}(t)}{\dd t} &=
 -(\ii \delta+\kappa/2)c_{\rm lo}(t)
 -\ii \frac{\Omega}{2}c_{\rm up}(t) \; ,
 \label{DOT-C-LO}
\end{align}
which obviously is a system of coupled first order differential
equations.

In the following we will discuss two important limit cases: the
\textit{bad cavity limit} or \textit{weak coupling regime} where
$\kappa,\gamma^\prime > \Omega$, and the \textit{good cavity limit}
or \textit{strong coupling regime}, where $\kappa,\gamma^\prime <
\Omega$.

\subsection{Bad cavity limit (weak coupling)}\label{Weak-coupling}

In the last section we have derived the evolution of the system,
which was described in \eqref{EFF-SCHROED-EQ-2} with
\eqref{DOT-C-UP} and \eqref{DOT-C-LO}. Eq.~\eqref{DOT-C-LO} can be
formally integrated:
\begin{equation}
 c_{\rm lo}(t) = -\ii\,\frac{\Omega}{2}
 \int_0^t\!\!\! \dd t^\prime\; c_{\rm up}(t^\prime)
 \ee^{-({\rm i}\delta+\kappa/2)(t-t^\prime)} \; .
\end{equation}
For partially open resonators the weak coupling regime is
characterized by $\kappa,\gamma^\prime > \Omega$. Inspecting
\eqref{DOT-C-UP} we can see that $c_{\rm up}(t)$ is slowly changing
if $\Omega/2$ and $\gamma^\prime$ are smaller than
$\abs{\delta}+\abs{\kappa}/2$, what corresponds to weak coupling.
In this case we can thus pull $c_{\rm up}(t)$ in front of the
integral, which for $t \gg 1/\kappa$ gives
\begin{equation}
 c_{\rm lo}(t) = \frac{-\ii\,\Omega}{2(\ii\delta+\kappa/2)}\; c_{\rm up}(t)
 \; .
\end{equation}
Substituting this in \eqref{DOT-C-UP} we obtain
\begin{equation}
 \frac{\dd c_{\rm up}(t)}{\dd t} =
 - \left[\frac{\gamma^\prime}{2}+\frac{\Omega^2}{4}\cdot
 \frac{\kappa/2-\ii\delta}{\delta^2+\kappa^2/4}
 \right]\; c_{\rm up}(t) \; .
\end{equation}
Obviously this differential equation is solved by an exponential
ansatz, and the occupation probability $P_{\rm up}$ for the excited
level decays as
\begin{equation}
P_{\rm up} = \abs{c_{\rm up}(t)}^2 \propto \ee^{-\gamma t}\; ,
\end{equation}
with
\begin{equation}
 \gamma = \gamma^\prime + \gamma_0 \; ,
\end{equation}
and
\begin{equation}
 \gamma_0 = \frac{\Omega^2}{2\kappa}\cdot\frac{1}{1+2(2\delta/\kappa)^2}\; ,
 \label{WEAK-COUPL-DAMP-COEFF}
\end{equation}
where again $\delta = \omega_r-\omega_0$ denotes the detuning. The
decay rate $\gamma$ is thus composed of a sum of terms of which the
first can be attributed to the contribution of the vacuum field,
and the second to a resonator contribution. In the limit where the
resonator is completely open (free space)
$\kappa\rightarrow\infty$, and as a result
$\gamma^\prime\rightarrow\gamma$. Thus the decay rate $\gamma$
reduces to the Weisskopf-Wigner coefficient $\gamma_{\rm fs}$.

\subsubsection{Enhanced spontaneous emission}
Let us imagine a completely closed resonator that surrounds the
two-level system. Then the loss mechanism which in
Fig.~\ref{F-P-Resonator} is characterized by the coefficient
$\gamma^\prime$ is absent, thus $\gamma^\prime = 0$. Let us also
assume resonance between the two-level system and the field mode,
so that the detuning $\delta = 0$, and still assume weak coupling
$\kappa > \Omega$. Then the decay coefficient $\gamma_0$ becomes
maximal, $\gamma_0=\Omega^2/2\kappa$. With \eqref{GAMMA-GOLDEN-R},
\eqref{DEF-QUALITY}, \eqref{RABI-FRQ-GEN-SYS}, and $\lambda_0=2\pi
c/\omega_0$ we obtain
\begin{equation}
 \gamma_{\rm max} = \frac{3
 Q}{4\pi^2}\cdot\frac{\lambda_0^3}{V}\cdot\gamma_{\rm fs} \; .
 \label{ENHANCED-SPONT-EMISS-RATE}
\end{equation}
We can see that $\gamma_{\rm max}$ has the same wavelength
dependence as the approximate result \eqref{APPR-ENH-DECAY}. By
taking into account the coupling of the dipole moment with the
field polarization \eqref{RABI-FRQ-GEN-SYS} we now also obtained
the correct geometrical dependence. We can see that for resonators
of good quality $Q$ with sizes approaching $V\approx\lambda^3$ a
substantial \textit{enhancement} of the free space spontaneous
emission rate $\gamma_{\rm fs}$ is expected.

\subsubsection{Inhibited spontaneous emission}

Let us consider the same two-level system in the same resonator
which, however, now is tuned far off the two-level system resonance
$\omega_0$ so that the detuning can be set to
$\abs{\delta}=\omega_0$. For a good resonator with $Q\gg 1$
\eqref{WEAK-COUPL-DAMP-COEFF} then becomes
\begin{equation}
 \gamma_{\rm inh}\approx\gamma_{\rm max}\cdot \frac{1}{4Q} =
 \frac{3}{16 \pi^2Q}\cdot\frac{\lambda_0^3}{V} \cdot
 \gamma_{\rm fs}\;.
 \label{INHIB-SPONT-EMISS-RATE}
\end{equation}
The spontaneous emission of a two-level system in a
one-mode-resonator with a large quality factor $Q$ can therefore be
nearly completely suppressed by detuning.

\bigskip
We now have derived three different rates, \eqref{GAMMA-GOLDEN-R},
\eqref{ENHANCED-SPONT-EMISS-RATE}, and
\eqref{INHIB-SPONT-EMISS-RATE}, by which an excited two-level
system can decay. One might ask, how the two-level system decides
which one to choose \cite{PAR87,COO87}. Obviously the choice
depends on the electromagnetic boundary conditions defining the
mode structure which surrounds the two-level system, and the
question therefore is, how does the two-level system explore its
mode environment. To answer this question we have to consider the
very initial processes of the decay, which thus involve time scales
short compared to the inverse decay rate. Unfortunately this is a
range, which is not consistent with some approximations we have
made in our previous discussion, especially it is not consistent
with the assumptions for using the coarse grained derivative
\eqref{DEF-COARSE-GRAINED-DERIV}. Thus, to answer this question we
need a more precise theory. Nevertheless, let us try a qualitative
argument: Let us assume the two-level system spontaneously starts
to emit, radiating a wave packet into the mode. If the emission
occurs into a free space mode, nothing comes back. However, if the
emission occurs into a high-$Q$ resonator mode, then the emitted
packet is reflected at the resonator boundary and returns to the
emitting dipole, i.e.\ the two-level system, where it is picked up
by the dipole. Coded in its phase, the reflected packet carries the
information about the geometry of the mode, as well as the state of
the emitting two-level system at earlier times. Thus, the question
now is, so to speak, what the phase is of the reflection arriving
at the generator driving the dipole. If the reflection is received
with a phase opposite to the actual emitting phase, then further
emission is inhibited, whereas in-phase reception enhances the
emission. We can see that a valid theory for the early emission
stages must involve the phases of the quantum systems involved. On
the other side the phase is not a Hermitian observable, therefore
we must accept that the early stages of the emission are hard to
divulge.

\subsection{Good cavity limit (strong coupling)}\label{Strong-coupling}

Strong coupling of the two-level system with the resonator mode is
observed when the photon which is emitted by the two-level system
into the resonator mode lives long enough so that it can be
reabsorbed by the two-level system. In this case we have to
consider the exact solution of the system of coupled differential
equations \eqref{DOT-C-UP}--\eqref{DOT-C-LO}. The general solution
of a system of two linear, first order differential equations is
given by the exponential ansatz
\begin{equation}
 c_{\rm up}(t) = c_{\rm up1}\; \ee^{\alpha_1t} +
 c_{\rm up2}\; \ee^{\alpha_2t} \; ,
\end{equation}
where the constants $c_{\rm up1},\;c_{\rm up2}$ are  chosen to fit
the initial conditions, for example $c_{\rm up}(t=0)=1$, and the
exponential are given by
\begin{equation}
 \alpha_{1,2} = -\frac{1}{2}\;
 \left(\frac{\gamma^\prime}{2}+\frac{\kappa}{2}+\ii\delta\right)
 \pm \frac{1}{2}\left[
 \left(\frac{\gamma^\prime}{2}+\frac{\kappa}{2}+\ii\delta\right)^2
 -\Omega^2 \right]^{\frac{1}{2}} \; .
\end{equation}
For strong coupling we have $\Omega/2 \gg
\gamma^\prime,\,\kappa,\,\delta$, and then the exponents reduce to
\begin{equation}
 \alpha_{1,2} = -\frac{1}{2}\;
 \left(\frac{\gamma^\prime}{2}+\frac{\kappa}{2}+\ii\delta\right)
 \pm \ii \frac{\Omega}{2}  \; .
\end{equation}
In this expression the imaginary part is larger than the real part,
so that the time evolution of the upper level probability $P_{\rm
up }$ is characterized by oscillations at the vacuum Rabi frequency
$\Omega$, which are slowly damped. The emission spectrum of course
is no longer a simple Lorentzian, but a  doublet of Lorentzian
lines, each with a width of $(\gamma^\prime+\kappa)/4$, split by
the vacuum Rabi frequency $\Omega$ \cite{CAR89}.

\subsubsection{Dressed states}

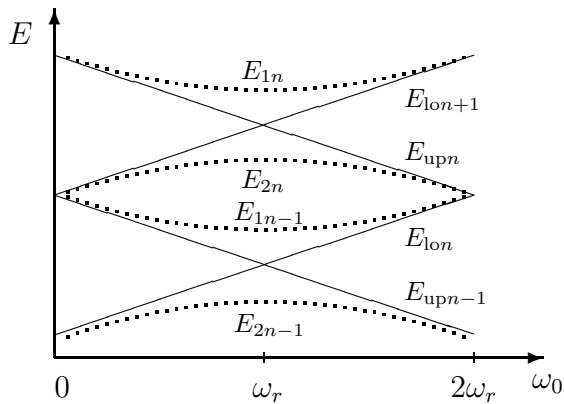
\begin{figure}[!ht]
 \setlength{\unitlength}{0.75em}
\begin{center}
\begin{picture}(25,18)
\thicklines
 \put(3,2.5){\vector(0,1){15}}
 \put(3,2.5){\vector(1,0){21}}
\thinlines
 \put(3,3.5){\line(3,1){18}}
 \put(3,9.5){\line(3,1){18}}
 \put(3,9.5){\line(3,-1){18}}
 \put(3,15.5){\line(3,-1){18}}
 \put(12,2.3){\line(0,1){0.4}}
 \put(21,2.3){\line(0,1){0.4}}
 \linethickness{1pt}
 \bezier{40}(3.5,3.4)(12,6.4)(20.5,3.4)
 \bezier{40}(3.5,9.4)(12,6.6)(20.5,9.4)
 \bezier{40}(3.5,9.6)(12,12.4)(20.5,9.6)
 \bezier{40}(3.5,15.4)(12,12.6)(20.5,15.4)
 \put(1,16){$E$}
 \put(23.5,1.2){$\omega_0$}
 \put(11.5,0.8){$\omega_r$}
 \put(20,0.8){$2\omega_r$}
 \put(3,0.8){0}
 \put(10.7,3.6){\footnotesize $E_{2n-1}$}
 \put(10.7,8.4){\footnotesize $E_{1n-1}$}
 \put(11,9.8){\footnotesize $E_{2n}$}
 \put(11,14.5){\footnotesize $E_{1n}$}
 \put(18,5){\footnotesize $E_{{\rm up}n-1}$}
 \put(18,7.3){\footnotesize $E_{{\rm lo}n}$}
 \put(18,11){\footnotesize $E_{{\rm up}n}$}
 \put(18,13.3){\footnotesize $E_{{\rm lo}n+1}$}
\end{picture}
\end{center}
\caption{\label{Fig-DressedStates}
 Eigenenergy of two manifolds of dressed states (dots) and the
 corresponding bare states (lines) as a function
 of the two-level system transition frequency $\omega_0$. The
 anticrossings occur at resonance $\omega_0=\omega_r$.
  }
\end{figure}
%
We can generalize the situation if we merge the small system
consisting of the two-level system and the resonator mode into a
single quantum system. As we have done before, we associate the
Hamiltonian \eqref{TOTHAM} to this system.\footnote{In the
literature the Hamiltonian \protect\eqref{TOTHAM} is often referred
to as \protect\textit{Jaynes-Cummings Hamiltonian}
\protect\cite{JAY63}.} As we have discussed before, this
Hamiltonian $\HSO{H}$ conserves the total number of small system
excitations, more explicit, $\HSO{H}$ only couples state $\ket{{\rm
up},n}$ with $\ket{{\rm lo},n+1}$. Thus in a system characterized
by $\HSO{H}$ transitions only occur inside the $(n+1)$-quanta
manifold $\{ \ket{{\rm up},n},\,\ket{{\rm lo},n+1}\}$. It is
therefore possible to decompose $\HSO{H}$ into the sum
\begin{equation}
 \HSO{H} = \sum_n \HSO{H}_n \;
\end{equation}
where $\HSO{H}_n$ only acts in the $(n+1)$-quanta subspace. In this
subspace the states $\ket{{\rm up},n}$ and $\ket{{\rm lo},n+1}$
form a basis in which $\HSO{H}_n$ can be represented by a
$2\times2$ matrix. The eigenvalues of $\HSO{H}_n$ are obtained by
diagonalization as
\begin{equation}
\begin{split}
 E_{2n} &= \hbar \; \bigl( n+\tfrac{1}{2}\bigl) \omega_r -
 \hbar\Omega^\prime_n \\
 E_{1n} &= \hbar \; \bigl( n+\tfrac{1}{2}\bigl) \omega_r +
 \hbar\Omega^\prime_n \;
\end{split}
\label{DRESSED-STATES-EIGENENERGY}
\end{equation}
where with $\Omega^\prime_n$ we denote the $n$-photon Rabi
frequency
\begin{equation}
\begin{split}
 \Omega^\prime_n &= \sqrt{\delta^2+\Omega^2(n+1)}\\
 &= \sqrt{(\omega_r-\omega_0)^2+\Omega^2(n+1)}\;,
\end{split}
\label{N-PHOTON-RABI-FREQ}
\end{equation}
and where the eigenstates are given as
\begin{equation}
\begin{split}
 \label{DRESSED-STATES}
 \ket{2n} &= -\sin\theta_n\ket{{\rm up},n}+\cos\theta_n\ket{{\rm
 lo},n+1}\\
 \ket{1n} &= \quad\cos\theta_n\ket{{\rm up},n}+\sin\theta_n\ket{{\rm
 lo},n+1}\; ,
\end{split}
\end{equation}
in which
\begin{equation}
 \tan 2\theta_n = \frac{\Omega\sqrt{n+1}}{\delta} \; .
\end{equation}
The states \eqref{DRESSED-STATES} are called the \textit{dressed
states} of the two-level system, which refers to the picture of a
two-level state dressed by the strongly coupled resonator mode. In
this terminology the states $\ket{{\rm up},n}$ and  $\ket{{\rm
lo},n+1}$ are called \textit{bare states}. Note that the
zero-quantum manifold has a single eigenstate  $\ket{{\rm up},0}$
with eigenvalue $E_0=0$.
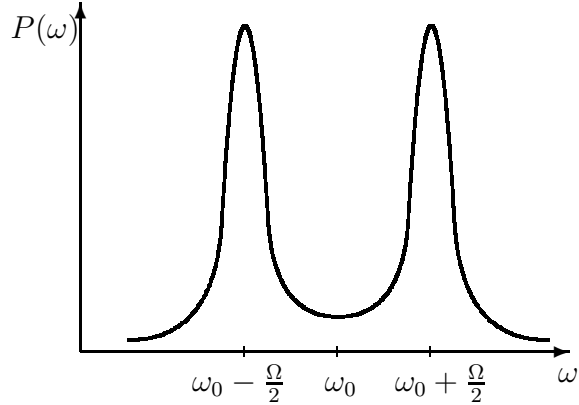
\begin{figure}[!ht]
 \setlength{\unitlength}{0.75em}
\begin{center}
\begin{picture}(25,18)
\thicklines
 \put(3,2.5){\vector(0,1){15}}
 \put(3,2.5){\vector(1,0){21}}
\thinlines
 \put(10,2.3){\line(0,1){0.4}}
 \put(14,2.3){\line(0,1){0.4}}
 \put(18,2.3){\line(0,1){0.4}}
 \linethickness{1pt}
 \bezier{100}(5,3)(8.7,3)(9,8)
 \bezier{300}(9,8)(10,25)(11,8)
 \bezier{100}(11,8)(11.3,4)(14,4)
 \bezier{100}(14,4)(16.7,4)(17,8)
 \bezier{300}(17,8)(18,25)(19,8)
 \bezier{100}(19,8)(19.3,3)(23,3)
 \put(0,16){$P(\omega)$}
 \put(23.5,1.2){$\omega$}
 \put(7.8,0.8){$\omega_0-\frac{\Omega}{2}$}
 \put(13.5,0.8){$\omega_0$}
 \put(16.5,0.8){$\omega_0+\frac{\Omega}{2}$}
\end{picture}
\end{center}
\caption{\label{Fig-doublet}
 Spectrum of the spontaneous emission doublet in the one-quantum
 manifold.
  }
\end{figure}
%
The dressed state energy spectrum given by
\eqref{DRESSED-STATES-EIGENENERGY} and \eqref{N-PHOTON-RABI-FREQ},
and the corresponding bare state eigenenergies are shown in
Fig.~\ref{Fig-DressedStates} as a function of the resonator
detuning. The energy of the bare states $\ket{{\rm up},n}$ and
$\ket{{\rm lo},n+1}$ cross at resonance. However, for the dressed
state eigenenergy this degeneracy is removed by the the interaction
of the two-level system with the field, causing dressed states of
the same manifold to repel each other, an effect that is often
called \textit{anticrossing}.

Spontaneous emission of the two-level system occurs in the
one-quantum manifold, where two transitions are allowed,
$\ket{1,0}\rightarrow\ket{{\rm lo},0}$ and
$\ket{2,0}\rightarrow\ket{{\rm lo},0}$, corresponding to the
frequencies $-\delta/2+\Omega^\prime_0$ and
$-\delta/2-\Omega^\prime_0$. At resonance $\delta=0$ the separation
becomes equal to the vacuum Rabi frequency $\Omega$; cf.\
Fig.~\ref{Fig-doublet}.

\section{Stimulated emission in an optical resonator and thresholdless lasing}
\label{sect-stimem}

In section \ref{Weak-coupling} we discussed the modification of the
two-level system decay when a surrounding resonator reduces the
density of field states. Equation \eqref{WEAK-COUPL-DAMP-COEFF} was
derived for a two-level system that interacts either with an empty,
or with a one photon, damped resonator mode. As a consequence,
three system states had to be considered: $\ket{{\rm up},0}$,
$\ket{{\rm lo},1}$, and $\ket{{\rm lo},0}$. In
\eqref{WEAK-COUPL-DAMP-COEFF} we distinguished two contributions,
the first is attributed to the mode density, and the second, which
was characterized by the vacuum Rabi frequency $\Omega$, to the
field. \textit{Stimulated emission} occurs when the excited
two-level system interacts with a mode which contains several
photons. In analogy to section \ref{Sect-Mast-Eq-Res}, and
observing the approximations for the Fermi golden rule,
cf.\eqref{FERMI-GOLDEN-RULE}, we can derive the following result
for the decay rate of a two-level system in a damped resonator,
interacting with an $n$-photon field
\begin{equation}
 \gamma_{\rm st} = \frac{\Omega^{{\,\prime\;}^2}_n}{2\kappa}\cdot
 \frac{1}{1+2(2\delta/\kappa)^2}\; ,
 \label{WEAK-COUPL-STIM-EMISSION}
\end{equation}
where with $\Omega^{\,\prime\;}_n$ we denote the $n$-photon Rabi
frequency  $\Omega^{\,\prime}_n = \sqrt{\delta^2+\Omega^2(n+1)}$;
cf.~\eqref{N-PHOTON-RABI-FREQ}. We can see that for a resonator
tuned to resonance ($\delta=0$) the stimulated emission decay rate
scales with the number of photons $n$ interacting with the
two-level system. Equation \eqref{WEAK-COUPL-STIM-EMISSION} can be
written as
\begin{equation}
 \gamma_{\rm st} = \gamma_0 \; (n+1) \; ,
\end{equation}
where $\gamma_0$ stands for the spontaneous emission rate which is
modified by the resonator, cf.~\eqref{WEAK-COUPL-DAMP-COEFF}. For
fluorescent dyes we can assume that the lower laser state is
rapidly depopulated, and that nonradiative processes, inversion
saturation and collective spontaneous emission are absent. With
this assumptions the number of photons in the resonator mode can be
expressed as
\begin{equation}
 \dot{n}=\gamma_0(n+1)N-\kappa n \; ,
 \label{RATEEQ-PHOT}
\end{equation}
where $N$ denotes the inversion $N=N_{\rm up}-N_{\rm lo}$, that is
the difference between the population of the upper laser level
$N_{\rm up}$ and the lower laser level $N_{\rm lo}$.  The rate by
which the inversion $N$ is changed depends on the rate with which
the upper level is populated, that is the pumprate  $W$, and by the
rate of depopulation given by the photon emission rate, thus
\begin{equation}
 \dot{N}= W-\gamma_0(n+1)N
 \label{RATEEQ-INV}
\end{equation}
The steady state solution of the system of differential equations
\eqref{RATEEQ-PHOT} and \eqref{RATEEQ-INV} has the simple form
\begin{align}
 n &= \frac{W}{\kappa} \label{RATEEQ-OUTPUT}\\
  \text{and}\qquad & \notag \\
 N &= \frac{W\kappa}{\gamma_0(W+\kappa)}\; .\label{RATEEQ-INVERSION}
\end{align}
From \eqref{RATEEQ-OUTPUT} we can see that the light output $n$
increases for any pump rate $W$ in a linear way, thus the laser
operates without threshold. This is a phenomenon called
\textit{thresholdless lasing} or \textit{lasing without inversion}.
This is the consequence of the exclusive coupling of the laser
active two-level systems to one single resonator mode. On the other
hand, the inversion $N$ approaches the limit of $N_t=\kappa /
\gamma_0$ when the pump rate $W$ increases. This limit can be
interpreted as the inversion at which the photon emission changes
its character from spontaneous emission to stimulated emission.
This is the transition which traditionally is associated with the
onset of lasing. Note that this phenomenon occurs when a collection
of two-level systems interacts with a single field mode, which is
thus different from the interaction of only one two-level system
with a single mode, i.e.\ a \textit{one atom laser} \cite{MES85}.
In summary, the conditions for inversionless lasing as expressed in
\eqref{RATEEQ-OUTPUT} and \eqref{RATEEQ-INVERSION} are that the
collection of two-level systems interacts only with a (damped
$\kappa)$ single mode field. In particular there are no open
resonator side walls through which the two-level system can couple
to the mode continuum (characterized by $\gamma^\prime$); cf.
Fig.~\ref{F-P-Resonator}. Of course there is a continuous
transition in the threshold behavior from the open resonator case
shown in Fig.~\ref{F-P-Resonator} to the closed resonator discussed
here \cite{YOK92} and thresholdless lasing was observed in a dye
solution placed a plane-plane resonator in which the mirrors were
spaced by half an emission wavelength \cite{MAR88}. In section
\ref{Sect-realiz} we will discuss threshold reduction in dye loaded
molecular sieve microlasers.

\section{The mode structure of microresonators}\label{microresonators}

\subsection{The concept of dielectric constant, dielectric
interfaces}

Maxwell's equations are the foundations on which both classical and
quantum optics are built. The possibility of creating a double-slit
diffraction pattern with a single photon  - a famous manifestation
of the wave-particle dualism - is a result of the fact that photons
occupy the modes of the classical electromagnetic field, cf.\ the
definition of the bosonic operators in terms of normal modes in
section \ref{sect-mode}. These modes can be discrete in frequency,
as they are in lossless closed resonators, or have a continuous
spectrum as is the case in scattering experiments such as the
diffraction example \cite{BLO90}; there, the whole information
about the diffraction process is encoded in the modes themselves
because they must satisfy the boundary conditions on the
diffraction slits.

Boundary conditions are a way of introducing rapid spatial
variations of a medium into the macroscopic Maxwell equations
without having to give up the concept of a dielectric constant or
permittivity, by which the microscopic properties of the medium are
taken into account in a very efficient mean-field manner
\cite{JAC98}. Even when details of the quantum-mechanical
light-matter interaction are of interest, such a mean-field
approach is a good starting point in order to define an appropriate
modal basis set in which to expand the relevant matrix elements.
This applies to dielectrics of infinite extent without losses
\cite{GLA91} or with losses \cite{SEL98}, but especially when
fluorescent or laser emission in the presence of a microcavity are
concerned.

The dielectric ``constant'' is moreover a function of field
strength, if the polarizability -- which describes the microscopic
matter-field interaction -- depends nonlinearly on electric field.
When such nonlinearity occurs in \textit{combination} with boundary
conditions, the relative importance of the two depends on the size
scales and field strengths involved. At low pump levels, phenomena
such as vortex formation which has long been known in nonlinear
media \cite{HOH93}, can cross over to \textit{linear} vortices
\cite{WEI92}, an example of which are the whispering-gallery modes
which will be discussed in section \ref{sec:wgmodes}. In small
cavities, boundary effects become dominant, and nonlinear effects
can be understood as interactions between modes of the linear
cavity \cite{WEI92}.

As we discussed above (cf.\ section \ref{Sect-fluoor}),  a well
known microcavity effect is due to Purcell \cite{PUR46} who argued
based on Fermi's Golden Rule that the Einstein A coefficient for
spontaneous emission can be enhanced in a microcavity due to its
highly peaked density of modes. For this perturbative approach to
the quantum electrodynamic problem of spontaneous emission, the
modes into which photons are emitted are determined completely by
Maxwell's equations with the boundary conditions defining the
cavity. In the strong-coupling regime, this ceases to be correct
when the radiant matter and light field are mixed in comparable
proportion in the eigenstates of the total system, as can occur,
e.g. in cavity polaritons in a quantum well microstructure
\cite{HOU94}, or for the Rabi oscillations of an atom in close
proximity to a dielectric microsphere \cite{KLI99}. Nevertheless,
the first step in all these cases is to determine the modes of the
electromagnetic field for the resonator goemetry at hand.

The conclusion from these preliminary remarks is that especially in
microcavities, an understanding of the  effects of the resonator
geometry on the field distribution, neglecting nonlinearities of
the medium, is of central importance. Even in the conceptually
simple problem that remains, we shall see how the well-understood
fundamental equations of electrodynamics lead to solutions that are
at present only partially understood, in the sense of predicting
their dependence on system parameters, or even giving conditions
for the existence of certain solutions -- an important example
again being the whispering-gallery modes.

\subsection{Fields at dielectric interfaces}

\subsubsection{Matching conditions}

We assume that all fields have the stationary time dependence
$\ee^{-i\omega t}$. Then Maxwell's equations become
\begin{eqnarray}
 \vec{\nabla}\times\vec{E}&=&
 -\frac{1}{c}\,\frac{\partial\vec{H}}{\partial t} =
 \ii k\vec{H} \label{max1}\\
 \vec{\nabla}\times\vec{H}&=&
 \frac{1}{c}\,\frac{\partial\vec{D}}{\partial t} =
 -\ii k n^2 \vec{E} \label{max2}\;,
\end{eqnarray}
where the wave number is given by
\begin{equation}
 \label{eq:tdepend}
 k=\omega/c.
\end{equation}
Since we want to illustrate the effects of boundaries, let us make
the further simplification of considering the refractive index $n$
to be piecewise constant, but not necessarily real. Combining these
equations, we obtain the wave equations
\begin{eqnarray}
 \vec{\nabla}\times\vec{\nabla}\times\vec{E}&=&(nk)^2\vec{E}\label{wave1}\; ,\\
 \vec{\nabla}\times\vec{\nabla}\times\vec{H}&=&(nk)^2\vec{H}\label{wave2}\;.
\end{eqnarray}
Since charge density can only appear at the surface of the
dielectric, we have $\vec{\nabla}\cdot\vec{E}=0$ in each domain of
constant $n$. In these regions, (\ref{wave1}) therefore becomes
\begin{equation}
 -\nabla^2\vec{E}=(nk)^2\vec{E} \;.
 \label{wave}
\end{equation}
The conditions to be satisfied at a dielectric interface are
deduced from the requirement that no current flow is possible along
the interface \cite{SEN95}, implying that the tangential components
of $\vec{E}$ and $\vec{H}$ must be continuous. At vertices or
edges, however, the tangent is undefined. In this case we must
invoke the additional requirement that the energy contained in any
volume element of the fields should not diverge near such
singularities of the surface \cite{SEN95}.

These conditions do not constitute true boundary conditions in the
traditional sense, but are instead \textit{matching conditions}.
The field on one side of the interface is determined by the field
on the other side, and only if we already know the latter, can the
former be obtained by solving a boundary-value problem. For some
simple problems, it is in fact easy to eliminate the field on one
side of the interface, if one knows its form \textit{a priory}. The
simplest example is a plane wave in air, impinging with wavevector
$\vec{k}$ on a planar interface with a lossless dielectric of
refractive index $n$. The knowledge that the transmitted wave in
the medium is again a plane wave allows us to straightforwardly
obtain Snell's law of refraction and Fresnel's formulas for the
reflectivity. The latter depend on polarization, which in this case
can be chosen either perpendicular to (TE) or in the plane of
incidence (TM). With this we then can obtain decoupled scalar wave
equations. The essential difference between TE and TM polarizations
is that the latter exhibits the Brewster angle $\chi_B$ at which
perfect transmission occurs. Independently of polarization, Snell's
law relates the incident angle $\chi_0$ (which we measure with
respect to the surface normal), to the transmitted angle $\chi$ of
the plane wave in the medium by
\begin{equation}
 \sin\chi=\frac{\sin\chi_0}{n} \; .
\end{equation}
Note that for large $n$, transmitted waves in the dielectric are
allowed to propagate only in a progressively narrower interval
around the perpendicular direction $\chi\approx 0$. Conversely, the
critical angle $\chi_c=\arcsin (1/n)$ therefore defines the
``escape cone'' for plane waves inside the material. Total internal
reflection prevents all waves with $\chi>\chi_c$ from escaping to
the optically thinner medium, i.e.\ the condition for confinement
is
\begin{equation}
 \label{eq:totalrefl}
 \sin\chi>\frac{1}{n}\;.
\end{equation}

\subsubsection{Impedance boundary conditions}

The fact that high index contrast leads to a narrow escape cone is
at the heart of a large body of literature summarized in Ref.\
\cite{SEN95}.  The aim is to replace the continuity requirements
for the fields at a dielectric interface by approximate boundary
conditions called \textit{(normal) impedance conditions}. There, an
approximate knowledge of the field on the high-index side allows
one to eliminate it from the problem. Yet, the method becomes
unreliable at low refractive-index contrast and at non-planar
interfaces; in particular, the Brewster effect is not correctly
reproduced unless additional corrections are introduced which,
however, do not permit a simple physical interpretation, because
they involve derivatives of the fields that are of higher order
than the original continuity conditions for the fields themselves.
We shall therefore make no use of impedance boundary conditions,
except to point out that they reduce to the familiar Dirichlet
boundary conditions (vanishing fields on the surface) if the index
contrast becomes infinite, as would be the case in an ideal metal.

\subsection{Scattering resonances versus cavity modes in the
dielectric cylinder}\label{sec:wgmodes}

One way to probe the interaction of dielectric bodies with light is
by elastic scattering. Atmospheric phenomena such as the rainbow,
the halo, or the glory arise from light scattering \cite{NUS92},
and in fact atmospheric science relies on scattering experiments as
diagnostic tools. When scattering experiments are carried out with
high spectral resolution, a ripple structure is observed in
scattering cross sections, which cannot be understood in a purely
ray-optics framework. These ripples are resonances which occur when
the incident light couples to long-lived cavity modes. In order to
illustrate the relationship between resonances and modes, we
consider here the example of a dielectric rod with a circular cross
section.

Dielectric cylinders are of great practical interest because they
are the archetypical model for an optical fiber. In the context of
this review, the approximate cylindrical symmetry of the molecular
sieve microcrystals makes it desirable to establish some
fundamentals of cylindrical systems. The modes of an optical fiber
can be divided into two classes: guiding modes and ``leaky'' modes.
As is well known, guided modes are responsible for the ability of
silica fibers to carry long-distance communication signals,
corresponding in the ray picture to zigzagging trajectories
traversing the fiber core in an almost planar motion -- the plane
of propagation coincides with the cylinder axis. The confinement
mechanism in the ray picture is \textit{total internal reflection}
at the interface to a lower-index medium, either the cladding of
the fiber or -- as we will assume for simplicity -- the abrupt
interface with air. However, over short distances a significant
power transport can take place through the leaky modes as well,
which must be taken into account in any comprehensive treatment of
optical waveguides \cite{SNY91}. It is also known \cite{SNY91} that
one has to distinguish between \textit{tunneling} leaky modes which
are confined by frustrated total internal reflection, and
\textit{refracting} leaky modes whose attenuation is even larger
because they correspond to rays that violate the condition for
total internal reflection. Leaky modes correspond to rays spiraling
down the fiber in a helical trajectory, and are therefore also
called ``spiral modes'' \cite{POO98}.

Of these three types of modes -- guided, tunneling, and refracting
-- for propagation along a fiber, only the leaky ones remain if the
radiation is incident with propagation vector
\textit{perpendicular} to the fiber axis. This is the situation we
shall focus on, because the applications we have in mind are not
intended for power transport but for power \textit{storage}, i.e.\
resonators. The corresponding ray trajectories are then confined to
a plane in which they perform a circulating motion. An example
where this mode structure has in fact been observed directly is a
laterally structured cylindrical VCSEL \cite{AHN99} where such a
mode supports lasing action at an unexpectedly low pump power.
Because the circulating mode structure is analogous to the acoustic
``whispering-gallery'' phenomenon in which sound waves cling to the
curved walls of certain buildings, modes with a ring-shaped
intensity distribution are commonly termed ``whispering-gallery
(WG) modes''.

The range of material parameters relevant to the design of
dielectric resonator structures is increasing continuously, as
novel materials enter device applications. For glasses, refractive
indices between $n\approx 1.4$ and $2$ are available (glass for
application in fiber amplifiers has $n\approx 1.8$). Semiconductor
materials extend this interval to even larger indices, while
organic compounds border on the lower end of the index range. The
fundamental wave equations for elastic light scattering in the case
of no propagation along the axis of the dielectric cylinder are
much simpler than for arbitrary oblique incidence, and consequently
treatments of plane wave scattering from dielectric cylinders at
normal incidence can be found in many textbooks
\cite{HUL81,KER69,BAR90}. Because of the ever increasing range of
applications, we review them here, emphasizing, however, what in
the literature is usually missing, namely the aspect of the
relation between resonances and cavity modes. One important piece
of notation that we introduce here is conspicuously absent from the
classic texts: the formulation of the light scattering problem in
terms of S-matrix theory, as it is used in quantum mechanical
scattering. Investigations of leaky cavities may well profit from a
more unified scattering-theoretical terminology.

\subsubsection{Metastable well in the effective potential}

To understand the origin of long-lived whispering gallery mode
resonances we examine Maxwell's equations for an infinite
dielectric cylinder. Following an argument used by Johnson
\cite{JOH93} for the dielectric sphere, we can rewrite (\ref{wave})
in a form similar to the Schr{\"o}dinger equation of quantum
mechanics,
\begin{equation}
 \label{wavequant}
 -\nabla^2\vec{E} + k^2(1-n^2)\vec{E} = k^2\vec{E} \; .
\end{equation}
This shows that dielectric regions ($n>1$) correspond to an
attractive potential well in the quantum analogy, except that here
the potential is itself multiplied by the eigenvalue $k^2$.

The reason why extremely long-lived resonances are created when the
scatterer has rotational symmetry is that after separation of
variables in cylindrical or spherical coordinates,
(\ref{wavequant}) gives rise to a radial equation with a repulsive
potential term due to the angular momentum barrier, as well as the
attractive term just noted.  For the case of a dielectric cylinder
studied here the resulting equation reads
\begin{equation}
 \label{rwave}
 -\left[\frac{d^2}{dr^2}+\frac{1}{r}\frac{d}{dr}\right]\,
 \vec{E}(r)+V_{\rm eff}(r)\,\vec{E}(r) =
 k^2\,\vec{E}(r)\; ,
\end{equation}
where the effective potential is
\begin{equation}
 \label{effpot2}
 V_{\rm eff}(r)=k^2\left(1-n^2\right)+\frac{m^2}{r^2} + k_z^2\; .
\end{equation}
The additional centrifugal potential term has appeared as a
consequence of conservation of the z-component of angular momentum,
and the offset $k_z^2$ results from the conservation of the
z-component of linear momentum.  As inidcated above, we focus on
planar propagation perpendicular to the $z$-axis of the cylinder,
so that $k_z=0$, and the incident wave has its $k$-vector in the
$x-y$ plane.  The problem then becomes effectively two-dimensional
and we will use the two-dimensional polar coordinates $r=\sqrt{x^2
+ y^2},\phi$.

\begin{figure} [hbt]
\begin{center}
 \resizebox{0.48\textwidth}{!}{%
 \includegraphics{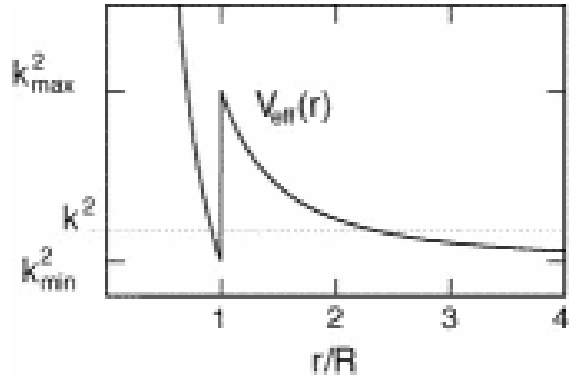}
 }
\end{center}
\caption{Effective potential picture for whispering gallery
resonances of the cylinder; $k_{min} = m/(nR)$ and $k_{max} = m/R$.
\label{effpotentialfig} }
\end{figure}

The radial ``potential'' which results from the sum of the
attractive well due to the dielectric and the repulsive angular
momentum barrier is shown in Fig.~\ref{effpotentialfig}, cf.\ also
Ref.\ \cite{NOE96}.
%
%
At nonzero angular momentum one sees that a metastable well is
formed within the dielectric.  The electric field inside is
separated from the propagating field outside by a tunnel barrier,
and one may expect solutions of the wave equation in which the
intensity inside the well is exponentially larger than outside the
well.  In order to complete the formal analogy to the
one-dimensional Schr{\"o}dinger equation we make a substitution of
variables to eliminate the first derivative term in the radial
equation (\ref{rwave}). This can be achieved by introducing a new
coordinate $\xi := \ln(kr)$; the resulting equation is
\begin{equation}
 \label{xiwave}
 \left[\frac{d^2}{d\xi^2}+ q^2(\xi)\right]\vec{E}(r) = \vec{0}\; ,
\end{equation}
where $q(\xi)=(n^2 \exp(2\xi) - m^2)^{1/2}$, and $q(\xi)$ is the
effective wave vector of the rescaled problem. In order that there
are any resonant solutions, $kR$ must be sufficiently large so that
$q$ is real for some value of $r$ within the dielectric.  The
largest value of the argument of the square root within the
dielectric occurs when $r=R$, so we deduce the condition
\begin{equation}
 k_{\rm min}=\frac{m}{nR}\; .
\end{equation}
On the other hand, if infinitesimally outside the dielectric  where
$n=1$, $q$ is real, then there is no region of exponential decay,
and no sharp resonances exist. The maximum value of $k$ such that
$q$ is imaginary just outside the dielectric is given by
\begin{equation}
 k_{\rm max} = \frac{m}{R}\; .
\end{equation}
Consequently, we expect narrow resonances of angular momentum $m$,
broadened only by tunnelling decay (evanescent leakage), for wave
vectors satisfying
\begin{equation}
 \frac{m}{nR} < k <\frac{m}{R} \label{kcond}
\end{equation}
or equivalently
\begin{equation}
 \frac{1}{n} < \frac{m}{nkR} < 1 \; .
 \label{kcond2}
\end{equation}
We will label these metastable well states by a discrete index
$\nu$ which corresponds to the number of radial nodes. If the
refractive index is changed, the barrier top in
Fig.~\ref{effpotentialfig} remains the same, but the well grows
deeper so that the lowest allowed $k$ will give rise to a narrower
resonance. Later we will show that essentially all of the
metastable states satisfying (\ref{kcond}) are discrete in the
sense that their spacing in $k$ is larger than their decay width
through the barrier. In this way they will give rise to isolated
resonances in the scattering functions which are well set apart
from the background.

It is worth quoting here the main features of the numerically
observed resonances in plane-wave scattering off a dielectric
cylinder at normal incidence, as given in Ref.\ \cite{BAR90} (but
using our notation):
\begin{quote}
\ldots their widths decrease as $m$ [angular momentum] increases
for a given $\nu$ [radial node number], and their widths increase
as $\nu$ increases for a given $m$. Also, as the index of
refraction [$n$] is increased, the positions of the resonances
shift to lower $kR$, and their widths become narrower. Resonances
having relatively narrow widths occur in the range $x$ to $n\,x$.
\end{quote}
All of the above phenomena can already be understood qualitatively
by inspecting the effective potential picture.

The inequality (\ref{kcond2}) can be brought into a very suggestive
form which can be justified by considering the short-wavelength
limit in the WKB approximation \cite{NOE96}:
\begin{equation}
 \sin\chi := \frac{m}{nkR} \; ,
\end{equation}
then (\ref{kcond2}) leads to
\begin{equation}
 \label{eq:sinchi}
 \frac{1}{n}<sin\chi < 1 \;.
\end{equation}
The right inequality is trivial, but the left relation is precisely
the condition for \textit{ total internal reflection} which must be
satisfied by the angle of incidence $\chi$ on the high-index side
at an interface to air, cf.~(\ref{eq:totalrefl}). At this point we
have provided no reason why $\chi$, as defined here, should be the
classical angle of incidence. Let us try a non-rigorous argument:
If we interpret the ray trajectories as classical particles with
momentum $\hbar (nk)$ inside the resonator, and angular momentum
$L=\hbar m$, then the classical relation between angular and linear
momentum would imply $\vec{L}=\vec{r}\times\vec{p} \leadsto
L=R\,(\hbar nk)\,\sin\chi\leadsto m=R\,nk\,\sin\chi$, where
$\sin\chi$ arises from the vector product of linear momentum and
radius vector ${\bf r}$, which in turn has magnitude $R$ at every
reflection from the boundary. At such a reflection, $\chi$ is then
just the angle of incidence.

The introduction of $\hbar$ into the present optical context is
somewhat arbitrary. The question is, does the optical field in the
cavity truly acquire angular momentum. We know that photons, as
massless fundamental particles, carry a unit spin angular momentum,
but no \textit{orbital} angular momentum in the physical sense.
Therefore, when we refer to $m$ as an angular momentum quantum
number, we invoke a formal analogy that arises between the
mathematical structure of bound photonic states and bound quantum
systems: There are, in our case, two degrees of freedom which are
associated with a finite region of space, giving rise to a discrete
frequency spectrum. The modes are then labeled by as many discrete
indices, or ``quantum'' numbers as there are confined degrees of
freedom. For a rotationally symmetric problem, the indices, or
quantum numbers, are simply the number of radial and azimuthal
nodes of the wave function. The azimuthal quantum number is what we
call $m$. Its relation to the angle of incidence is rigorously
derivable from the eikonal (or WKB) limit \cite{NOE96}.

There is an additional complication to this argument if the cavity
is not closed. Due to the openness of the system, it is then not
clear how to define the photons in a resonant mode. However, the
cavity can approach the closed limit as $n\to\infty$. Before we
discuss this problem, we will elucidate the concept of modes in the
leaky cavity setting, where discrete resonances appear but have a
finite width in $k$. We will return to this issue under the heading
of  \textit{quasibound states} in section \ref{sec:qbstates}.

For this moment let us nevertheless dwell on \eqref{eq:sinchi},
which we can use to interpret some results. For example, it
indicates that states at the bottom of the well correspond to a ray
motion tangential to the boundary, whereas states at the top
correspond to rays colliding with the boundary at exactly the
critical angle. As next we turn to a more detailed discussion of
the actual wave solutions for the dielectric cylinder where the
incident wave propagation is normal to the axis.

\subsubsection{Matching conditions for TM polarization}\label{sec:matchcond}

Our general considerations so far have not addressed the boundary
conditions at the interface between dielectric and vacuum (or air)
-- conditions which also depend on the polarization of $\vec{E}$.
The arguments given above for the existence of narrow resonances
are, however, independent of the boundary conditions, because we
rely only on the fact that a tunnel barrier is formed where the
field decays. Thus there should exist resonances corresponding to
two different polarization states. Henceforth we will focus on the
transverse magnetic polarization for which the boundary conditions
lead to the simplest matching conditions for the electric field
inside and outside the dielectric. For this polarization state we
assume an incident plane wave of the form
$\vec{E}=E(x,y)\,\vec{e}_z$, where the unit vector $\vec{e}_z$
points parallel to the cylinder axis (and hence the magnetic field
is transverse to the cylinder axis). From now on we will focus on
the scalar function $E(x,y)=E(r,\phi)$ which is the amplitude of
the electric field in $z$-direction. Since a pure dielectric is an
insulator, there is no current flowing on its surface in response
to the incident field and $\vec{H}$ is continuous everywhere
(neglecting any variation in the magnetic permeability). From
\eqref{max1}, we derive the azimuthal component of the magnetic
field, which in polar coordinates has the form
\begin{equation}
 \label{derivcont}
 H_{\phi} = \frac{\ii}{k}\left[\nabla\times E\vec{e}_z\right]_{\phi}
 = - \frac{\ii}{k}\,\frac{\partial E}{\partial r}\; .
\end{equation}
Therefore the radial derivative of $E(r,\phi)$ is continuous at the
dielectric surface.  Since the tangential component of $E$ is
always continuous at such an interface, we arrive at the simple
matching conditions that both $E$ and ${\partial E}/\partial r$ are
continuous at the cylinder surface.  TE polarization ($\vec{H}$
along $\vec{e}_z$) does not result in these familiar requirements
\cite{HUL81,KER69}, however one can still obtain a scalar wave
equation, which can be treated with a generalization of the methods
used below. Most conveniently, one writes the wave equation for the
magnetic field $H$, which is again continuous, but exhibits a jump
of magnitude $n^2$ in the normal derivative at the interface. The
effective potential picture with purely tunneling escape for the
whispering gallery modes remains unchanged by this, since our
arguments above did not involve polarization. Note that for TE
polarization the electric field is in the plane of incidence.
Transmission out of the dielectric becomes unity at the Brewster
angle which neglecting the finite curvature of the interface is
given by
\begin{equation}
 \sin\chi_B=\frac{1}{\sqrt{1+n^2}}\; .
\end{equation}
The whispering gallery mode criterion $\sin\chi>1/n$ is clearly not
modified by this effect since $\sin\chi_B<1/n$.

An important point is that the same matching conditions remain
valid even when the cross section of the cylinder is
\textit{non-circular}, because for TM polarization $\vec{E}$
remains tangential to the surface. As long as we consider only
convex surfaces, the continuity of $\partial E/\partial r$ in polar
coordinates implies the continuity of the normal derivative of $E$.
These boundary conditions are identical to those for quantum
scattering from a potential well. Hence, we are allowed to use the
terms familiar from quantum theory, such as tunneling, as we did in
the previous sections. However, as was already noted with
\eqref{wavequant}, it should be kept in mind that the analogy is
incomplete, because the dispersion relation in quantum mechanics is
$\omega=\hbar\,k^2/(2m)$ and not $\omega=n\,c\,k$ as in optics. The
results presented here can therefore not be obtained by simply
copying known quantum mechanical calculations. An important
difference between the quantum and optical wave equations can be
read off \eqref{effpot2}, where the effective potential well is
seen to become progressively \textit{deeper} with increasing $k$.
As a consequence, above barrier reflection remains significant even
for large $k$, whereas it would become negligible in an attractive
circular quantum well at high energies. This effect is just the
well known classical Fresnel reflection at a dielectric interface,
which independently of the wavelength has the value
$R=(1-n)^2/(1+n)^2$ at normal incidence, i.e.\ $\sin\chi=m/(nkR)\to
0$.

\subsubsection{Resonances in elastic scattering}

The wave equation for TM (or TE) polarization as defined above is
of the scalar form
\begin{equation}
 \label{eq:wavescatscalar}
 \nabla^2\psi+n^2k^2\,\psi = 0\; ,
\end{equation}
where $\psi$ denotes the electric field for the TM case, and the
refractive index $n$ is a constant different from $1$ inside the
dielectric. The wavenumber $k$ is not position dependent because it
enters the problem simply through the definition $\omega=c\,k$ for
the harmonic time dependence. All the scattering functions for the
elastic scattering of light (i.e.\ scattering at the fixed
frequency $\omega$) by a dielectric cylinder can be calculated if
we know the scattering states
\begin{equation}
 \label{general}
 E_m(r) =
\begin{cases}
 A_m\,J_m(nkr) & (r<R)\\
 H^{(2)}_m(kr)+S_m\,H^{(1)}_m(kr) &(r>R),
\end{cases}
\end{equation}
for all $m$ ($J_m$ and $H_m$ denote the Bessel and Hankel
functions, respectively). These are solutions to the radial
equation\eqref{rwave}, assuming that the refractive index is unity
outside the cylinder and $n(r) = 1 + (n-1) \Theta (R-r)$ inside.
The amplitudes $A_m,S_m$ are determined by the matching conditions
for $E$ and its radial derivative $E'$,
\begin{eqnarray}
 A_m\,J_m(nkR)   &=& H^{(2)}_m(kR)+S_m\,H^{(1)}_m(kR) \nonumber \\
 A_m\,nJ'_m(nkR) &=& H^{(2)\prime}_m(kR)+S_m\,H^{(1)\prime}_m(kR)\;
 ,
 \label{inhomogenarrayeqn}
\end{eqnarray}
where, the primes denote differentiation. e.g.\ $J'(x)=dJ/dx$. This
can be solved for the scattering amplitude $S_m$ as a function of
the size parameter $x\equiv kR$. In terms of quantum mechanical
scattering theory, $S_m$ is a diagonal element of the scattering or
S-matrix which for the rotationally invariant scatterer is diagonal
in the basis of angular momentum states $m$. Using these solutions,
one can for example calculate the scattering properties for an
incident plane wave. A plane wave can be decomposed into a
combination of Bessel functions \cite{COU68} as
\begin{equation}
\ee^{-{\rm i}kx} = \sum\limits_{m=-\infty}^{\infty}(-\ii)^m
\ee^{{\rm i}m\phi}J_m(kr).
\end{equation}
This then determines the coefficients with which the partial waves
\eqref{general} are superimposed to get the full solution. The
conventional representation of this solution is \cite{BAR90}
\begin{equation}
E(r,\phi)=
 \begin{cases}
 \sum\limits_{m=-\infty}^{\infty}e^{im\phi}\,(-i)^md_m\,J_m(nkr) & (r<R) \\
 \sum\limits_{m=-\infty}^{\infty}e^{im\phi}\,(-i)^m\left[
   J_m(kr)-b_m\,H^{(1)}_m(kr)\right] & (r>R)
 \end{cases}
 \label{conventnomencleqn}
\end{equation}
Using $J_m=\frac{1}{2}(H^{(1)}_m+H^{(2)}_m)$ in the incoming wave,
this becomes for $r>R$
\begin{equation}
 E(r,\phi)=\frac{1}{2}\,\sum\limits_{m=-\infty}^{\infty} \ee^{{\rm i}m\phi}
\,(-\ii)^m\left[H^{(2)}_m(kr)+(1-2\,b_m)\,H^{(1)}_m(kr)\right],
 \label{conventnomencleqn1}
\end{equation}
so that we can extract the relationship between $S$- and the
$b$-coefficients,
\begin{equation}
 S_m=1-2\,b_m \; .
\end{equation}
Hence the $b_m$ are analogous to the quantum mechanical transition
matrix $T$, defined by $S=1-2\pi \ii\,T$. The S-matrix satisfies an
additional condition as a consequence of flux conservation, namely
\cite{SIF68}
\begin{equation}
 S^{\dagger}S=1.
\end{equation}
This unitarity relation also holds for deformed resonators, where
$S$ is not diagonal. It provides an independent equation with which
to check the accuracy of the wavefunction matching. For the
circular cylinder, unitarity just implies that all $S_m$ have
modulus one. For numerical calcultions borrowing this formalism
from quantum-mechanics is therefore an added advantage.

\begin{figure}[!hbt]
\begin{center}
 \resizebox{0.48\textwidth}{!}{%
 \includegraphics{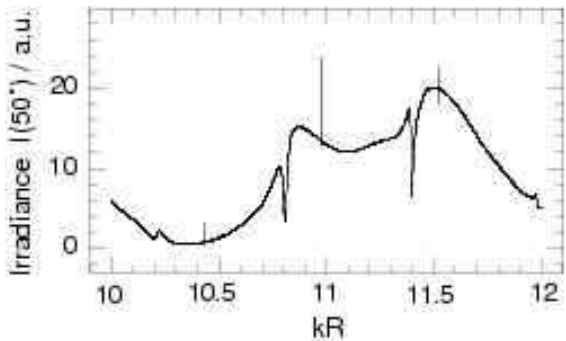}
 }
\end{center}
\caption{\label{circleresosfig} Irradiance $I$ which is scattered
off a circular cylinder with refractive index $n=2$ at
$\phi=50^{\circ}$ with respect to the direction of the incoming
wave which is assumed to be plane and TM polarized.}
\end{figure}

If a plane wave is incident on the cylinder, the scattered
intensity has the angular dependence \cite{HUL81,KER69}
\begin{equation}
I(\phi)\propto \left|\sum\limits_{m=-\infty}^{\infty}b_m\,\ee^{{\rm
i}m\phi}\right|^2 \; ,
\end{equation}
where $\phi$ is the angle with the beam. From the matching
conditions \eqref{inhomogenarrayeqn}, one finds that the expansion
coefficients can be written in the form
\begin{eqnarray}
 b_m &=& \frac{1}{1+i\,\beta_m} \; ,\label{bmaxcondeqn}\\
 \beta_m &=& \frac{n\,{\rm J}_{m-1}(nx)\,{\rm Y}_m(x)-
 {\rm J}_m(nx)\,{\rm Y}_{m-1}(x)} {n\,{\rm J}_{m-1}(nx)\,
 {\rm J}_m(x)-{\rm J}_m(nx)\,{\rm J}_{m-1}(x)} \; ,\label{betadefeqn}
\end{eqnarray}
where we have used the abbreviation
\begin{equation}
 x\equiv kR \; ,
\end{equation}
and the function ${\rm Y}_m(x)$ is the Bessel function of the
second kind. This provides an explicit formula for the scattering
intensity $I(\phi)$, an example of which is plotted in Fig.\
\ref{circleresosfig}. As is seen in the figure, the scattered
intensity shows rapid resonant variations. The resonances are not
simply Lorentzian peaks, but are of the Fano shape \cite{FAN61}
with varying values of the Fano asymmetry parameter.  This
asymmetric shape is due to interference between the nonresonant
scattered waves and the resonant scattering. As long as the
resonance is sufficiently isolated the resonance widths can be
extracted by fitting a Lorentzian to the Fano function.  It can
immediately be observed that the resonances have widely varying
widths corresponding to how close they are to the top of the
barrier in the effective potential well.

The decomposition \eqref{bmaxcondeqn} is analogous to the case of a
scattering sphere \cite{JOH93}, and in the same way one observes
that intensity maxima occur near the value of $k$ at which one of
the $b_m$ has a maximum (if there is no interference with other
nonresonant waves then it will be at exactly that value of $k$).
This in turn occurs when $\beta_m=0$. Thus, we set
\eqref{betadefeqn} to zero and obtain an implicit equation for the
resonance positions $x=kR$ in each angular momentum channel $m$:
\begin{equation}
 \label{resozerocondeqn}
 n\,{\rm J}_{m-1}(nx)\,{\rm Y}_m(x)={\rm J}_m(nx)\,{\rm Y}_{m-1}(x).
\end{equation}
It should be noted that the denominator of \eqref{bmaxcondeqn} is
nonzero only for the regime of below barrier resonances, $x<$.
Otherwise, additional resonances arise whenever $\beta_m$ becomes
infinite, i.e.\ when
\begin{equation}
 \label{resozerocondeqn1}
 n\,{\rm J}_{m-1}(nx)\,{\rm J}_m(x)={\rm J}_m(nx)\,{\rm J}_{m-1}(x).
\end{equation}
Above-barrier resonances thus occur if either
\eqref{resozerocondeqn} or \eqref{resozerocondeqn1} is satisfied.
Yet, these resonances will typically be so broad that they do not
give rise to isolated peaks in the scattering functions.

\subsubsection{Quasibound states at complex wavenumber}

Having obtained the condition for when a resonance is excited in
elastic scattering, we now turn to a complementary approach in
which no incident wave is present. Resonances can also be excited
without any incoming wave at that frequency, e.g.\ if the light is
generated inside, such as in a laser. Also the specific lineshape
of a resonance in elastic scattering depends on the nature of the
incident wave and is not a property of the scatterer alone. It is
thus useful and important to consider an alternative definition of
the resonant state which is independent of the manner of its
excitation.  We will first introduce the concept of the quasibound
state. We will discuss this concept using the example of the
dielectric cylinder we have have considered above, and then turn to
the more general significance of such states.

Solutions of the wave equation with \textit{no incoming wave}  are
termed quasibound states (or in the recent literature on resonances
of dielectric spheres, quasinormal modes \cite{CHI96}, hinting at
the fact that they are not an orthonormal set in the usual sense).
Due to the unitarity of the $S$-matrix for real $k$, there can be
no such solutions for real values of $k$.  Instead quasibound
states are directly connected with the complex \textit{poles} of
the $S$-matrix (or in our simple case the matrix element $S_{m}$),
the real part of $k$ giving the resonance frequency, and the
imaginary part giving the resonance width.  This is because the
condition for having a solution to the matching equations with no
incoming wave is just the condition for a pole of the $S$-matrix
\cite{NOE94}:  If in \eqref{general} we give the incoming wave
($H^{(2)}_{m}$ an arbitrary amplitude $A_{i}$, then the outgoing
wave will have amplitude
\begin{equation}
 A_{o}=S_{m}\,A_{i}\; .
\end{equation}
If we now set $A_{i}\to 0$, there can be a finite outgoing
radiation if at the same time $S_{m}$ has a divergence, requiring
us to tune $k$ to the complex value where $S_{m}$ has a pole.

Thus the complex wavenumber of the quasi-bound state is determined
by
\begin{equation}
 \label{generalquasieqn}
 {\tilde E}_m(r) =
 \begin{cases}
 A_m\,J_m(nkr) & (r<R)\\
 {\tilde S}_m\,H^{(1)}_m(kr) & (r>R) \; ,
 \end{cases}
\end{equation}
and the accompanying condition for continuity of the derivatives.
These matching conditions for $\tilde{E}$ can be satisfied only for
a discrete set of complex $k$ at which there exists a nontrivial
solution to the homogenous part of the linear system
\eqref{inhomogenarrayeqn}. The latter occurs only at the zeroes of
the determinant
\begin{equation}
 D=\left|\begin{array}{cc} J_m(nkR)&-H^{(1)}_m(kR)\\
 nJ'_m(nkR)&-H^{(1)\prime}_m(kR)\end{array}\right| \; ,
\end{equation}
which leads to
\begin{equation}
 \label{derivatformcondeqn}
 nJ'_m(nkR)\,H^{(1)}_m(kR) = J_m(nkR)\,H^{(1)\prime}_m(kR)\; .
\end{equation}
Using the recursion relations for the Bessel functions to eliminate
the derivatives, we obtain the resonance condition
\begin{eqnarray}
&&n\left[
J_{m-1}(nkR)-\frac{m}{nkR}J_m(nkR)\right]\,H^{(1)}_m(kR)\\&&=
J_m(nkR)\,\left[H^{(1)}_{m-1}(kR)-\frac{m}{kR}H^{(1)}_m(kR)\right]
\nonumber\\ &\Rightarrow&n J_{m-1}(nkR)\,H^{(1)}_m(kR)=
J_m(nkR)\,H^{(1)}_{m-1}(kR)\; .
 \label{rescond}
\end{eqnarray}
This can be solved numerically to find the real and imaginary parts
of $kR$ at which a metastable state occurs. Equation
\eqref{rescond} is a complex equation for a complex variable. If we
restrict $kR$ to be real, then the real and imaginary parts of the
equation are precisely \eqref{resozerocondeqn} and
\eqref{resozerocondeqn1} which we obtained for the scattering
resonances. This shows the direct relationship between the
quasibound states and the scattering resonances: For real $kR$,
\eqref{rescond} cannot be exactly fulfilled, but the closest one
can get to satisfying it with real $kR$ is given by the scattering
resonance conditions.

\begin{figure}[!ht]
\begin{center}
 \resizebox{0.35\textwidth}{!}{%
 \includegraphics{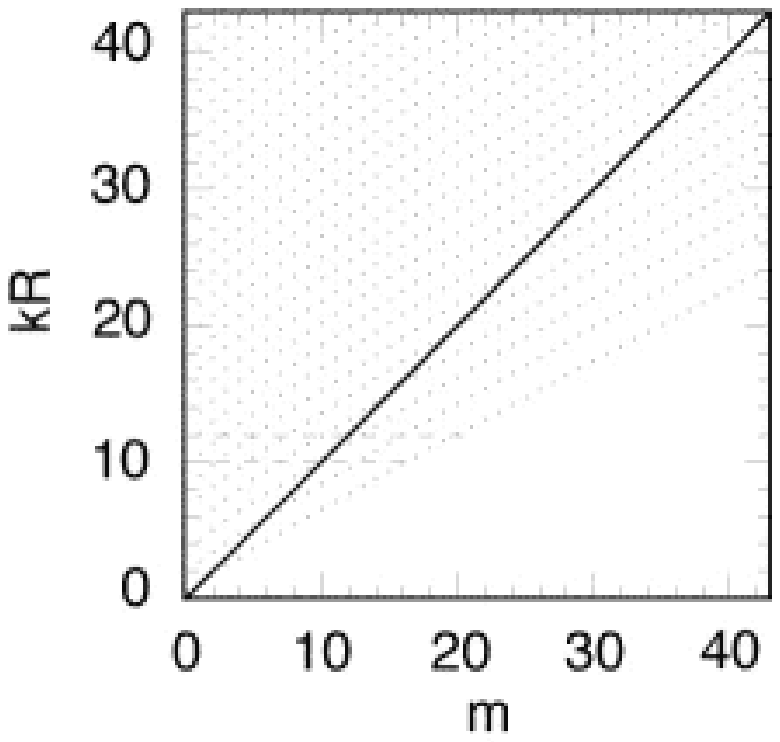}
 }
\end{center}
\vspace*{-1.2cm}
  \caption{
\label{circlegridfig} Resonance positions in the circle of
refractive index $n=2$ with TM polarization for the lowest $44$
angular momentum numbers $m$. Each dot corresponds to one
resonance, and the solid line $kR=m$ represents the dividing line
between broad and narrow resonances, with narrow widths expected
below it. The horizontal dashed lines enclose a $kR$ interval in
which we count a total of $9$ resonances below the critical line.
Compare this observation with the scattered intensity in Fig.\
\protect{\ref{circleresosfig}}. The three narrowest resonances
there correspond to the three lowest $kR$ points at
$m=17,\,18,\,19$. The closer the points are to the critical line,
the broader the resonances get.}
\end{figure}

The exact results obtained numerically for the real parts of the
quasibound state wavevectors are summarized in Fig.\
\ref{circlegridfig}. For each $m$, there exists an infinite
sequence of discrete points in this grid, corresponding to an
increasing number of radial nodes in the effective potential well.
The quasibound states are hence labeled by angular momentum and
radial quantum numbers (which we call $m$ and $\nu$), as in the
closed circle.  We have drawn in Fig.\ \ref{circlegridfig} a line
of slope $m/kR=1$ corresponding to the condition $\sin \chi
=\frac{1}{n}$; resonances to the right of that line are true
whispering gallery modes and we expect their widths to be much less
than the resonance spacing, with the narrowest resonances
corresonding to the points with the largest distance from this
line.

Note that the spacing in $k$ of resonances for a given value of $m$
is roughly constant above the line $m/kR=1$, whereas the spacing of
resonances below the line increases and hence is not independent of
$kR$ for a given $m$. The latter are the whispering gallery mode
resonances of primary interest to us.

The uniform spacing of the above barrier resonances can be
demonstrated by using the large argument expansions of the Bessel
functions in \eqref{rescond}. This is justified at large enough
size parameter $nkR\gg m$, noting that $nkR>m$ always holds. One
obtains
\begin{eqnarray}
&&n\cos\left(nkR-\frac{\pi}{2}m-\frac{3\pi}{4}\right)\,
H_m^{(1)}(kR)\nonumber\\ &=& \label{largeargeqn}
\cos\left(nkR-\frac{\pi}{2}m-\frac{\pi}{4}\right)\,
H_{m-1}^{(1)}(kR).
\end{eqnarray}
Noting that $\cos(\alpha-\pi/2)=\sin\alpha$, this simplifies to
\begin{equation}
 \label{besselratioeqn}
 \tan\left(nkR-\frac{\pi}{2}m-\frac{\pi}{4}\right)=
\frac{1}{n}\;\frac{ H_{m}^{(1)}(kR)}{ H_{m-1}^{(1)}(kR)}\; .
\end{equation}
An important limiting case is $n\to \infty$, for which the large
argument expansion always applies if $m$ and $kR$ are fixed. Then
the righthand side becomes negligible, so that the resonance
position is given by
\begin{equation}
 \label{neumannlimiteqn}
 kR=\frac{\pi}{n}\left[ \frac{m}{2}+\nu+\frac{1}{4}\right]\: ,
\end{equation}
where the nonnegative integer $\nu$ is the radial quantum number.
This is just the limit of a closed system with Neumann boundary
conditions, $J'_m(kR)=0$, in the limit $nkR\gg m$, and it clearly
leads to a purely real $kR$, i.e.\ to truly bound states.

On the other hand large $n$ is not always necessary to get an
approximate result: one obtains the same expression for the
resonance position \textit{independently} of $n$, provided the
quotient of the Hankel functions is purely imaginary. This is
precisely the case for \textit{above-barrier} resonances, where
$kR>m$, and hence we can use the large argument expansion for the
Hankel functions, too. The righthand side in \eqref{besselratioeqn}
then becomes equal to $-\ii/n$, and one can solve exactly for
\begin{equation}
nkR=\frac{\pi}{2}m+\frac{\pi}{4}
+\frac{i}{2}\left[\ln\frac{1-1/n}{1+1/n}-i 2\pi \nu\right].
\end{equation}
Thus the real and imaginary parts of $kR = x + iy$ can be given
analytically in the limit $kR>m$:
\begin{eqnarray}
 \label{repart}
 x &=& \frac{\pi}{n}\left[ \frac{m}{2}+\nu+\frac{1}{4}\right]\\
 y &=& \frac{1}{2n}\ln\frac{1-1/n}{1+1/n}\; .
\label{impart}
\end{eqnarray}
Since the righthand side in \eqref{besselratioeqn} has no real part
when $kR\ll m$, it does not affect the condition for the real part
of $kR$. The approximations made here are thus valid either for
large $n$ or for above barrier resonances.

Of greatest interest to us, however, are the long-lived resonances
with $kR>m$. For this case, the ratio of Hankel functions will
acquire an $m$- and $k$-dependent real and imaginary part, which
will emerge in both resonance positions and widths. Approximate
explicit expressions for resonance widths and positions can be
obtained, but are not generally valid for the entire range of
interesting parameters $n$, $kR$, $m$ \cite{NOE97} -- therefore, we
can for now be content with the simple limiting cases considered
here. They are already of practical use because a numerical
solution of the implicit equation \eqref{rescond} for the real and
imaginary part of $k$ at given $m$ and $n$ is made more efficient
by using the approximate resonance positions from \eqref{repart} as
an initial guess in finding the roots of \eqref{rescond}.

\subsection{Quasibound states in lasing and fluorescence}\label{sec:qbstates}

The previous considerations have brought us to the quasibound state
concept via the resonances in elastic light scattering. Even if we
allow the cross section of our dielectric cylinder to be deformed,
TE and TM polarization can still be  decoupled (see the remarks in
section \ref{sec:matchcond}), provided we continue to restrict
ourselves to propagation normal to the cylinder axis. After we
clarified the role a complex wavenumber $k$ can play in the
original wave equation, we deduce a rather general statement about
quasibound states. To arrive at the time-independent wave equation
\eqref{wave}, we assumed a monochromatic time variation $\ee^{-{\rm
i}ck t}$, with $k=\omega/c$ according to \eqref{eq:tdepend}. With
the quasibound state boundary condition (no incoming wave), this
same wave equation admits solutions only at discrete
\textit{complex} $k = \omega/c - \ii\,\gamma/c$, where we have
split the frequency into its real and imaginay part. Returning to
the ansatz for the time dependence, this implies that the
quasibound state \textit{decays} in time as
$\exp[-\ii\omega\,t-\gamma\,t]$.

This decay is of course a consequence that the system is open and
radiates energy away to infinity. Consequently, the corresponding
``radiation boundary condition'' is composed only of outgoing waves
which are present far away from the cavity. This is the type of
boundary condition that occurs in many emission problems, in
particular in lasing. To see how these ``leaky modes'', which we
found in the treatment of the passive cavity, arise in lasing, we
have to take a new point of view: Consider a scalar wave equation
of the form
\begin{equation}
\label{eq:wavescalar} \nabla^2\psi+{\tilde n}^2k^2\,\psi =0,
\end{equation}
as it arises for each of the coupled polarization directions in the
cylinder. This is similar to \eqref{wave} where we made the ansatz
of a steady-state time dependence with real $k$. If we do indeed
require $k$ to be \textit{real}, then in order to find a solution
satisfying the radiation condition, we permit the
\textit{refractive index} to be complex inside the resonator, which
we denote by $\tilde{n}= n - \ii\,n'$, where $n$ is the real part.
The imaginary part allows to introduce an \textit{amplifying
medium}. Consider the simple example of a plane wave
$\exp[\ii\tilde{n}kx]$ which will clearly grow with $x$ in the
direction of propagation. This is the ``energy source'', but note
that the detailed mechanism of this energy production, for example
the pumping mechanism, is not defined in this way. Nevertheless,
outside the cavity, we again assume air with  $\tilde{n}=1$ with
matching conditions that are appropriate for the given
polarization. Finally, let us assume TM polarization as before in
the scattering problem. The solutions for the exterior and interior
field are $\psi_{\rm ext}$ and  $\psi_{\rm int}$.

Let us now recast \eqref{eq:wavescalar} as
\begin{equation}
\label{eq:wavescalarqb} \nabla^2\psi+n^2\tilde{k}^2\,\psi =0\; ,
\end{equation}
where $n$ is the real part of $\tilde{n}$ as defined above, and
$\tilde{k} = k-\ii\,k\,n'/n$ is the complex wavenumber inside the
cavity, which reduces to $\tilde{k}= k$ outside. This recast
equation is almost of the form of \eqref{eq:wavescatscalar}, except
that $k$ has different values inside and outside the cavity. If
instead,  we also had $\tilde{k}= k-\ii\,k\,n'/n$ outside, the
solutions of \eqref{eq:wavescalarqb} were exactly the quasibound
states of the passive resonator (i.e.\ solutions of the Helmholtz
equation at complex $k$) if the deacy rate was defined as as
\begin{equation}
\gamma=ck\,n'/n.
\end{equation}
We now may argue that adding or dropping the imaginary part of
$\tilde{k}$ outside the resonator makes only a small difference if
$\gamma$ is small:
In fact, as can be checked (cf.\ \cite{NOE97}) by inspecting the
large-argument asymptotics of the Hankel function $H^{(2)}$, the
field of a quasibound state at distances larger than $\approx
c/(2\gamma)$ from the cavity grows exponentially due to
retardation, but within this physical range $\psi_{\rm ext}$
vanishes as $\gamma\to 0$. Therefore one can write $\psi_{\rm
ext}({\bf r})\approx \gamma \zeta({\bf r})$. If we expand
$\psi_{\rm int}(\gamma)$ and $\zeta(\gamma)$ in a Taylor series in
$\gamma$, then to linear order the $\gamma$-dependence of $\zeta$,
but not that of $\psi_{\rm int}$, can be dropped in the full
solution. Therefore, the stationary state of the active medium
described by \eqref{eq:wavescalar}) or \eqref{eq:wavescalarqb}, as
well as the metastable decaying state obtained by replacing the
real valued outside wavenumber $k$ by $k-\ii\,k\,n'/n$ are
identical to first order in $\gamma$ within an area of order
$\gamma^{-2}$.

In conclusion, we have seen that long-lived quasibound states
appear not only as resonances in elastic light scattering but also
as stationary states in emission problems at low pump rate
(characterized by $\gamma=ck\,n'/n$). When asking for the intrinsic
properties of such long-lived states, it is often a good starting
point to consider first the limiting case of no losses, i.e.\ the
closed cavity. In the next section, we discuss the familiar concept
of Gaussian beams as freely propagating waves, and how they can be
used to describe resonator modes. Although since the work of Fox
and Lee \cite{FOX61} this is the standard approach to describe
laser resonators \cite{SIE86}, this approach has some severe
limitations for microresonators.

\subsection{Paraxial approximation and the parabolic equation}

In free space, the spectrum of electromagnetic waves is not only
continuous but highly degenerate, as can be seen, for example, by
choosing plane-wave solutions of the Helmholtz equation and noting
that at each frequency an infinite number of propagation directions
can be chosen. There exists a class of \textit{approximate} free
space solutions of the Helmholtz equation, the \textit{Gaussian
beams}. Gaussian beams are obtained by introducing the
slowly-varying envelope approximation in the wave equation. This is
a special case of a frequently encountered approximation which is
labelled with different names, for example adiabatic approximation,
paraxial approximation, or Born-Oppenheimer approximation. In the
theory of partial differential equations one also finds the name
``parabolic-equation'' method, because the main step is to neglect
the second derivative of a slowly-varying ``envelope'' function,
which results in an equation of the type of the time-dependent
Schr{\"o}dinger equation for the latter. In optics, this
approximation leads to the Fresnel propagator describing
diffraction in the limit of small angles with the optical axis.

\subsubsection{Gaussian beams and the short-wavelength limit}

Gaussian beams play a prominent role in traditional laser optics,
because they arise naturally as solutions in parallel-mirror
resonator structures, and they also represent diffraction-limited
propagation. However, they are not in general acceptable functions
to describe the modes of a microcavity. In contrast to free space,
it is not even clear \textit{a priori} how to define the equivalent
of the paraxial condition in a general microresonator structure,
such as for example a circular disk. To  clarify this point, we
will discuss the paraxial approximation as next.

Behind the paraxial approximation is the picture of a ``light
beam'' propagating in $z$-direction. Let us assume that the beam is
described by the envelope function
\begin{equation}
\label{eq:envelopeansatz}
 \psi(x,y,z)\equiv u(x,y;z)\,\ee^{{\rm i}kz}
\end{equation}
which we insert into the Helmholtz equation
\begin{equation}
\label{eq:helmholtz}
 \nabla^2\psi+k^2\psi=0\; .
\end{equation}
As aresult we obtain
\begin{equation}
\label{eq:helmholtz2}
 u_{xx}+u_{yy}+u_{zz}+2\ii\, k u_{z}=0\;.
\end{equation}
With this step we have formally reduced the wavenumber exponent by
one, without loss of information.

In the paraxial approximation we take \eqref{eq:helmholtz2}, but
neglect the second order derivative
$u_{zz}\equiv\partial^2u/\partial z^2$. In what situation is this
appropriate? The name ``paraxial'' comes from a comparison between
length scales over which wave solution extends along $z$ and in the
transverse directions, respectively. Let $L$ be the relevant length
in the $z$ direction. Now let us introduce rescaled coordinates
which measure the transverse space in dimensionless form as
\cite{BAB72}
\begin{equation}
 x'=x\,\sqrt{k/L},\quad y'=y\,\sqrt{k/L},\; ,
\end{equation}
while for the longitudinal dimension we set
\begin{equation} z'=z/L\; .
\end{equation}
With this \eqref{eq:helmholtz2} appears as
\begin{equation}\label{eq:helmholtz3}
\frac{k}{L}\left(u_{x'x'}+u_{y'y'}\right)+\frac{1}{L^2}\,u_{z'z'}+
2ik\,\frac{1}{L}\,u_{z'}=0\; .
\end{equation}
With multiplication by $L/k$ we obtain
\begin{equation}\label{eq:helmholtz4}
u_{x'x'}+u_{y'y'}+\frac{1}{kL}\,u_{z'z'}+ 2iu_{z'}=0\; .
\end{equation}
Thus, the second order $z^\prime$ derivative can be neglected, if
$1/kL\ll 1$. In cases in which this is fulfilled the paraxial
approximation is appropriate.

As $1/kL = \lambda/2\pi L \ll 1$ implies $\lambda\ll L$, the
paraxial approximation corresponds in fact to a
\textit{short-wavelength approximation}. The approximation consists
in the assumption that the envelope $u$ is slowly-varying compared
to the wave fronts along $z$. The resulting equation in the
original coordinates reads as
\begin{equation}
\label{eq:parabolic} -\left(\frac{\partial^2}{\partial
x^2}+\frac{\partial^2}{\partial y^2}\right)u=2\ii k\frac{\partial
u}{\partial z}\; .
\end{equation}
This equation is mathematically of a fundamentally different type
than the Helmholtz equation: whereas the former is an elliptic
partial differential equation, the latter is parabolic. The
difference is that solutions of an elliptic equation are determined
by solving a \textit{boundary-value} problem, while solutions of
\eqref{eq:parabolic} are specified by \textit{initial conditions}
along a line $z = z_{0}$, for example. If we try to impose boundary
conditions on \eqref{eq:parabolic} by specifying the values of $u$
on a closed surface in $x,\,y,\,z$ (e.g.\ a microresonator), there
will in general be no solution for any $k$, whereas the original
eigenvalue problem  \eqref{eq:helmholtz} always exhibits solutions
satisfying the boundary conditions for a discrete set of
$k$-values. This well-known consequence of the theory of
characteristcs \cite{SOM47} by which second-order partial
differential equations are classified explains why paraxial
approximations in general do not allow to find all the modes of a
resonator.

Let us put it in other words: Equation \eqref{eq:parabolic} is
formally analogous to the time-dependent Schr{\"o}dinger equation
in which we substitute $z\leftrightarrow t$ and $k\leftrightarrow
m/\hbar$. We know that this equation uniquely determines the time
evolution of any initial wave packet prepared at time $t=t_{0}$.
This corresponds to specifying the solution of \eqref{eq:parabolic}
at some $z=z_{0}$. Clearly, if in addition to the initial condition
we also want to impose an arbitrary ``final distribution'' of $u$
at some other time $t=t_{1}$, then, in general, a contradiction
will arise to the unique time evolution. Only for special choices
of initial distributions can the final condition be met. Boundary
conditions on a closed surface in the ``spacetime'' spanned by
$x,\,y,\,z$ are therefore not generally consistent with
\eqref{eq:parabolic}. Now, based on this parabolic equation how can
we find the modes of a closed resonator?

\subsection{Gaussian beams in free space}

There are various schemes by which this can be accomplished for
some subset of modes, and the vast majority of conventional
resonator calculations in optics are based on such methods
\cite{SIE86}. In order to discuss these approaches and their
limitations for microcavities, we first recall how the propagation
along $z$ (or the ``time evolution'') can be expressed in terms of
the retarded Green function $G$:
\begin{equation}
\label{eq:greenuse}
 u(x',y';z')=\int G(x',y',z';\,x,y,z_{0})\,u(x,y;z_{0})\,dx\,dy\; .
\end{equation}
The Green function of the free particle time dependent
Schr{\"o}dinger equation, also called its propagator \cite{GUT90},
is well known from quantum mechanics \cite{SIF68} as well as from
the theory of thermal conduction \cite{SOM47} (where it appears
with an imaginary time scale):
\begin{equation}
\label{eq:gutzwillerdimgreen}
 G(\vec{r}\,',t';\,\vec{r},t)=
 \left[\frac{k}{2\pi\,\ii\,(t'-t)} \right]^{d/2}\,
\ee^{{\rm i}k(\vec{r}\,'-\vec{r})^2/2(t'-t)}\quad(k\equiv m/\hbar).
\end{equation}
Here, $d=1,2,3$ is the number of spatial dimensions. This result
reminds us of another important feature of the parabolic equation:
The above function describes how a wavepacket localized at a single
point $\vec{r}$ at time $t$ spreads with time. Such pulse spreading
is completely absent in the time dependent Green function of
Maxwell's wave equation, $(\nabla^2-\partial^2/\partial(c t)^2)\,
u=0$, where the time dependence is simply given by the effect of
retardation, $G\propto\delta(t'-t-\sqrt{x^2+y^2+z^2}/c)$ (in three
dimensions).

The discussion in quantum mechanics textbooks usually applies to
three space dimensions ($d=3$). However, we require the analogous
objects for the case of two spatial dimensions ($d=2$) -- in the
paraxial approximation, $z$ takes the role of a time parameter and
hence leaves only $x$ and $y$ as spatial coordinates. The reduced
dimensionality introduces different prefactors in Green's function
but leaves the basic implications of the paraxial approximation
(including pulse spreading) unchanged. With $d=2$,
\eqref{eq:greenuse} describes the transverse profile of a paraxial
beam, as it propagates along the $z$ axis. The equation is
identical to the \textit{Fresnel wave propagation} formula
\cite{TUR97} for the field $\psi$, if we use the definition of $u$
in \eqref{eq:envelopeansatz}.

The Green function in \ref{eq:greenuse}, alias the Fresnel
propagation kernel, defines an integral equation for the transverse
mode functions $u(x,y)$, which is a frequent starting point for
determining resonator modes in the paraxial approximation. Before
we turn to the problem of resonator modes, it is useful to consider
first the eigenfunctions of \eqref{eq:greenuse} in free space which
are called \textit{Gaussian beams}. The derivation of the
fundamental Gaussian beam profile follows the analogy between
\eqref{eq:parabolic} and the time dependent Shr{\"o}dinger
equation. An important solution is the \textit{minumum-uncertainty
wavepacket},
\begin{equation}
\label{eq:minumumuncert}
 u(x,z)=\frac{1}{(2\pi)^{1/4}}\,
 \frac{1}{\sqrt{\sigma+\frac{{\rm i}\,z}{2k\sigma}}}\,
 \exp\left[-\frac{x^2}{4\sigma^2+2\ii z/k}\right],
\end{equation}
where $\sigma$ is the width of the wavepacket in $x$ direction at
``time'' $z=0$. An analogous form can be found for the other
transverse direction, $y$, and the full wave function requires
taking the product of both profiles. This separability is a
consequence of the fact that we have imposed no boundary conditions
on the free-space propagation. We can therefore restrict ourselves
without loss of generality to wave functions that depend only on
one transverse coordinate ($x$) for now.

Note that \eqref{eq:minumumuncert} is a special minimum uncertainty
wavepacket with zero average $x$-momentum. By the uncertainty
relation, $\sigma$ is related to to the spread in wavenumber
$k_{x}$ through
\begin{equation}
\label{eq:uncert}
 \sigma\,\Delta k_{x}=1/2\; .
\end{equation}
The wavepacket therefore has a nonzero width $\Delta k_{x}$ in
Fourier space, which will lead to divergence as a function of
``time'' $z$. By decomposing prefactor and exponent into their real
and imaginary parts, it is straightforward to show that
\eqref{eq:minumumuncert} describes a standard Gaussian beam
\cite{YAR75}, if we reinstate the definition $\psi=u\,\exp(\ii kz)$
and interpret $2\,\sigma$ as the minumum spot size at $z=0$. The
divergence, or angular beam spread, is given by
\begin{equation}
\theta=\arctan\left(\frac{1}{k\,\sigma}\right)\; .
\end{equation}
It must be emphasized that the Gaussian beam is \textbf{not} an
exact solution of the free-space Helmholtz equation
\eqref{eq:helmholtz}, but becomes exact only in the paraxial limit.
The beam spreading is, however, not an artefact of this  parabolic
equation method. It correctly describes the conjugate relation
between spot size and transverse wavenumber which is a consequence
of diffraction.

\subsection{Gauss-Hermite beams}

The spot size $2\,\sigma$ is a parameter at our disposal, by which
we can  determine the relative uncertainties in position and
momentum according to \eqref{eq:uncert}. This freedom is important
when one attempts to construct approximate resonator modes by
properly adjusting Gaussian beams, as we will see below. Further
generalizations of the fundamental Gaussian beam are possible in
free space, by allowing the transverse amplitude to vary. An
important special case are the \textit{Gauss-Hermite} beams which
are constructed as follows. If we set $z=0$ in
\eqref{eq:minumumuncert}, then the resulting instantaneous Gaussian
\begin{equation}
u_{0}(x)=\frac{1}{(2 \pi)^{1/4}}\,\frac{1}{\sqrt{\sigma}}\,e^{-x^2/
4\sigma^2}
\end{equation}
can be formally interpreted as the ground state wavefunction of a
one dimensional harmonic oscillator,
\begin{equation}
-\frac{\hbar^2}{2m}\;\frac{d^2u_{n}}{dx^2}+\frac{1}{2}\,m\omega^2\,u_{n}=
E_{n}\,u_{n},
\end{equation}
in which formally identify the oscillator length with the beam spot
size,
\begin{equation}
\ell=\sqrt{\frac{\hbar}{m\omega}}\equiv\sqrt{2}\,\sigma\; .
\end{equation}
Higher order eigenfunctions can be generated by acting on $u_{0}$
with the formal creation operator\footnote{Note that the operators
we define now are motivated solely the mathematical structure of
the problem. They do not represent quantum mechanically relevant
entities like position or momentum, nor are they physically related
to the operators of section \protect\ref{sect-mode}.}
\begin{equation}
\label{eq:creationop}
 a^{\dagger}\equiv
\frac{1}{\sqrt{2}}({\hat Q}-\ii\,{\hat P}),
\end{equation}
where we defined the operators
\begin{eqnarray}
 {\hat Q}&\equiv& \frac{x}{\ell}=\frac{x}{\sqrt{2}\sigma}\; ,\\
 {\hat P}&\equiv& \frac{\ell}{\ii}\,\frac{d}{dx}=
 \frac{\sqrt{2}\sigma}{\ii}\,\frac{d}{dx}\; .
\end{eqnarray}
The solutions are products of Hermite polynomials and the Gaussian
$u_{0}$. From this construction at $z=0$, however, it is not
directly obvious how the beam shape changes as it propagates along
$z$. To see this we will define a generalized creation operator
$c^{\dagger}$ which properly describes the spreading (divergence)
of the beam, and which, when applied to the $z$-dependent $u(x,z)$
of \eqref{eq:minumumuncert}, generates functions which are
solutions of the parabolic equation \eqref{eq:parabolic}.

\medskip
For propagation in the $x-z$ plane to which we restricted our
attention, the transverse beam coordinate is $x$, whereas $z$ is
the propagation direction playing the role of a time variable. Let
us define the following dimensionless quantities
\begin{eqnarray}
\label{eq:complextraj}
 p & \equiv & \frac{1}{2k\sigma}\; , \\
 q(kz) & \equiv & p\,kz-2\,\ii p\,(k\sigma)^2\; , \nonumber
\end{eqnarray}
where $k$ and $\sigma$ are the parameters from
\eqref{eq:minumumuncert} for which we thus obtain
\begin{equation}
\label{eq:minumumuncert3}
 u(kx,kz)=\frac{1}{(2 \pi)^{1/4}}\,\sqrt{\frac{-\ii k}{q(kz)}}\,
 \exp\left[\ii p\frac{(kx)^2}{2\,q(kz)}\right].
\end{equation}
We can now define a generalized form of the creation operator
\eqref{eq:creationop} as
\begin{eqnarray}
\label{eq:creationop2}
 c^{\dagger} & \equiv & p^*k\,x + \ii\,q^*\frac{\partial}{\partial (kx)}\; ,\\
 c & \equiv & p\,k\,x - \ii\,q\frac{\partial}{\partial (kx)}\; ,
 \label{eq:creationop3}
\end{eqnarray}
which in the limit $z=0$ reduces to \eqref{eq:creationop} because
then $q=-\ii k\sigma$. One can verify that \eqref{eq:creationop2}
and \eqref{eq:creationop3} lead to the correct commutation relation
$[c,c^{\dagger}]=1$, as a direct consequence of the fact that the
quantities defined in \eqref{eq:complextraj} satisfy the relation
\begin{equation}
q^*p-q\,p^*=\ii
\end{equation}
for all $z$. We included complex conjugation of $p$ even though it
is real, in order to highlight the similarity to the quantum
commutation relation between conjugate operators -- and indeed, $p$
can be interpreted formally as the conjugate momentum of the
\textit{complex trajectory} $q(kz)$ because it satisfies the
equation of motion $p=dq/d(kz)$ with dimensionless time $kz$.

For $c^{\dagger}$ generating Gauss-Hermite beams, this operator
must exhibit two additional qualities which we check in the
following.
\begin{itemize}
\item First, we verify by direct application of
  \eqref{eq:creationop3} that $c$ annihilates the fundamental beam
  $u(x,z)$ given in \eqref{eq:minumumuncert}.
\item Second, we verify that $(c^{\dagger})^m\,u(x,z)$ with
  $m=1,2\ldots$ is again a solution of the free-space parabolic
  equation \eqref{eq:parabolic}.
\end{itemize}
This latter property requires that $c^{\dagger}$ commutes with the
differential operator
\begin{equation}
\label{eq:parabolicop}
 {\sf L}\equiv \frac{\partial^2}{\partial (kx)^2}+
 2\ii\frac{\partial}{\partial (kz)}
\end{equation}
which in fact corresponds to \eqref{eq:parabolic} written in
operator form. Thus, in operator notation the parabolic equation
simply reads as
\begin{equation}
{\sf L}\,u=0,
\end{equation}
where we still assume that the $y$ dependence can be separated off.
Using the definitions in Eqs.\ (\ref{eq:creationop2}) and
(\ref{eq:complextraj}), we obtain
\begin{equation}
c^{\dagger}\,{\sf L}-{\sf L}\,c^{\dagger}=2\,\left(
\frac{\dd\,q^*}{\dd(kz)}\,\frac{\partial}{\partial (kx)}-p^*\,
\frac{\partial}{\partial (kx)}\right)\; ,
\end{equation}
where the righthand side vanishes because of the equation of motion
$\dd\,q^*/\dd(kz)=p^*$. This result means that if $u$ is a solution
with ${\sf L}\,u=0$, then $c^{\dagger}u$ is also a solution,
because
\begin{equation}
{\sf L}(c^{\dagger}u)=c^{\dagger}({\sf L}\,u)=0.
\end{equation}
Thus, we have shown that the set of Gauss-Hermite beams are
solutions of the free space wave equation in paraxial
approximation.

\subsection{Resonator modes in the parabolic equation approximation}

The construction of resonator modes using the results obtained so
far requires us first to identify what is meant by the $z$ axis
along which the beams propagate, if the system is bounded by some
surface. For simplicity, we first assume that this boundary is
perfectly reflecting, what corresponds to Dirichlet boundary
conditions. We can approach this situation from two logically
distinct directions.

Traditionally it is assumed that the mirrors are shaped such that
they precisely match the wavefronts of a Gaussian beam for some
value of spot size $\sigma$ and wavenumber $k$. On the other
logical side one can start with a desired field configuration for
which the shape of suitable mirrors is sought. The goal is to find
modes with a predefined form which is optimally adapted to a given
practical problem. In contrast to the traditional approach this
procedure has the form of an \textit{inverse boundary-value
problem}. It has been addressed in Ref.\ \cite{ANG96}, and plays a
role whenever beams have to be matched to a given situation.

The molecular sieve crystals on which this review focuses, as well
as all other self-assembled realizations of dielectric
microresonators such as aerosol or polymer droplets, pose instead a
well defined boundary value problem with little or no freedom of
adjusting the cavity shape. This forces us to take a different
route if we wish to extend the results of the previous subsections.

\begin{figure}[!ht]
\begin{center}
 \resizebox{1.0\textwidth}{!}{%
 \includegraphics{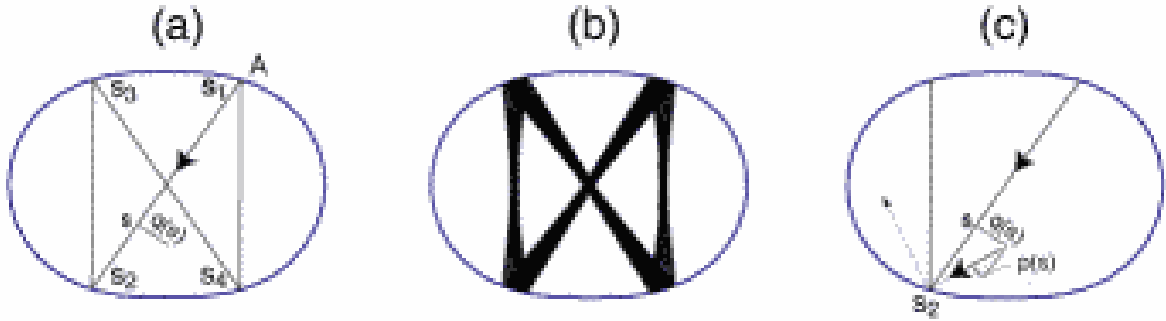}
 }
\end{center}
\caption{ \label{fig:bowtie} (a) A stable periodic orbit in an oval
cavity. The path can be parametrized by the arc length $s$, i.e.\
the longitudinal distance along the ray from the (arbitrary)
starting point A. reflections occur at lengths $s_k$. The stability
of this bow tie shaped pattern is apparent in (b) where a ray is
launched with initial position and direction slightly off the
periodic path: the trajectory is nevertheless confined to the
vicinity of the bow tie, performing an oscillation transverse to
the trajectory in (a). The transverse deviation in position can be
measured by a coordinate $q$ as shown in (c), and likewise $p$
denotes a deviation in momentum (or direction) of the initial ray.
The outcome of the reflection for such a perturbed ray is shown
dashed. Deviations are exaggerated for clarity. }
\end{figure}

The application of the parabolic equation method in the diffraction
theory of resonators is presented in great detail in Ref.\
\cite{BAB72}, where, however, the salient features are at times
obscured by the mathematical apparatus that is necessary to treat
wave propagation in three dimensions with inhomogenous refractive
index. In the folloeing we do not take such complications into
account, but approach the solution of the Helmholtz equation
\eqref{eq:helmholtz} with only $x$ and $z$ as variables, and
Dirichlet boundary conditions on the cavity wall which in view of
the situation with microresonators, we assume to be a two
dimensional closed surface. The idea of this approach is that since
the Gaussian beam is an approximate short wavelength solution near
some $z$ axis, we can piece together a resonator mode by using
Gaussian beams which follow a \textit{closed ray trajectory} in the
cavity. An example for such a closed orbit is shown in Fig.\
\ref{fig:bowtie}. It is a nontrivial example where a ray based mode
analysis of the present type leads to accurate results
\cite{GMA98}.
Each straight segment of the trajectory between two successive
reflections defines a $z$ axis for which the analysis of the
previous subsections can be carried out. Assuming a solution of the
form \ref{eq:minumumuncert3} in each segment, we determine $p$ and
$q$ in a generalization of \eqref{eq:complextraj} such that two
requirements are satisfied: The boundary conditions must be met to
the accuracy of the paraxial approximation, and the full solution
must be single valued (i.e.\ reproduce itself) after a complete
round trip around the ray orbit (constructive interference
condition). We retain the form given in \eqref{eq:complextraj},
with one modification. Without modification, the minimum width of
the Gaussian beam is located always at $z=0$. We can shift this
location by allowing $p$ to be complex, and still maintaining the
relation $q^*p-p\,q^*=\ii$ underlying the construction of the
Gauss-Hermite modes. Thus we define
\begin{equation}
\label{eq:complextraj2}
\begin{split}
 p & \equiv A_{1}\; ,\\
 q(kz) & \equiv A_{1}\,kz+A_{0}\; ,
\end{split}
\end{equation}
where $A_{0,1}$ are complex but restricted to ${\rm
Re}(A_{0}\,A_{1})=0$.

To follow the wave around the closed trajectory, we introduce a new
longitudinal coordinate $s$ which measures the distance along the
whole ray loop from some arbitrary starting point, say point A in
Fig.\ \ref{fig:bowtie}. Then the individual reflections occur after
the lengths $s_{k}$, $k=1,\ldots N$, where $N$ is the number of
bounces. Furthermore, let $L$ be the length of the close path. In
order for the wave function to be single-valued, the parameter $q$
governing the spreading in the transverse wave function must be
periodic as a function of $s$ with period $L$, for example
\begin{equation}
\label{eq:periodiccond}
 q(s+L)=q(s)
\end{equation}
for all $s$ in the interval of straight propagation we consider.

\subsection{Monodromy matrix}

From the boundary condition of vanishing wavefunction at the point
of reflection, we get additional relationships for $p$ and $q$.
Firstly, the phase of the Gaussian beams before and after
reflection must exhibit a jump by $\pi$. Moreover, considering  a
reflection at the point $s=s_{k}$, one finds that the vector
$(q,p)$ before and after reflection (at $s=s_{k}-\Delta s'$ and
$s=s_{k}+\Delta s$, respectively) must obey the equation
\begin{equation}
\label{eq:matrixmonodromy}
\begin{pmatrix}
    q(s_{k}+\Delta s) \\
    p(s_{k}+\Delta s)
\end{pmatrix}
 = M^{(k)}(\Delta s,\Delta s')\,
\begin{pmatrix}
    q(s_{k}-\Delta s') \\
    p(s_{k}-\Delta s')
\end{pmatrix} \; ,
\end{equation}
where, $M^{(k)}(\Delta s,\Delta s')$ is a $2\times 2$ matrix which
depends on the distances $\Delta s,\Delta s'$ from the reflection
point. It is called the \textit{monodromy matrix} of the given
reflection $k$. Its explicit form can be stated in terms of the
local radius of curvature $R=R^(k)$ of the reflecting surface and
the angle of incidence $\chi=\chi^(k)$ with respect to the local
normal \cite{STO99} as,
\begin{equation}
M^{(k)}(\Delta s,\Delta s') =
 \begin{pmatrix}
  \frac{2\Delta s}{2\cos\chi}-1 &
  \frac{(\Delta s)\,(\Delta s')} {\alpha\cos\chi} \\
  \frac{2}{R\,\cos\chi} &
  \frac{2 \,\Delta s'}{R\,\cos\chi}-1
 \end{pmatrix} \; ,
\end{equation}
where
\begin{equation}
\alpha=\left(\frac{2}{R}-\frac{\cos\chi}{\Delta s'}-
       \frac{\cos\chi}{\Delta s}\right)^{-1} \; .
\end{equation}
The linear matrix equation \eqref{eq:matrixmonodromy} holds only
within the paraxial approximation, which for the reflections means
that the transverse spreading at the reflection points shall be
small, as measured by the real part of the exponent of
\eqref{eq:minumumuncert3}, which is
\begin{equation}
\label{eq:spreadq}
 \frac{1}{2}\,(k x)^2{\rm Im}\frac{p}{q} =
 \frac{1}{4}\,(k x)^2\abs{q}^{-2},
\end{equation}
where we used $q^*p-q\,p^*=\ii$. A small extent in $x$ relies
therefore on $|q|/k$ being small. Away from the beam waist
$\abs{q}$ grows with $z$, and the growth is smaller when $k$ is
large. Hence, the required limit is again that of large $k$. For
details of the calculations, the reader is referred to Ref.\
\cite{BAB72}.

\subsection{Round trip stability of periodic orbits}

As is expected from a short-wavelength aproximation, the monodromy
matrix itself has a ray interpretation. Equation
\eqref{eq:matrixmonodromy} is precisely the transformation that
determines how a small deviation from the periodic ray path changes
the outcome of the reflection. As noted earlier, $q$ and $p$ behave
like conjugate variables, and here this fact emerges again. If we
interpret formally $q$ as a cartesian coordinate transverse to the
ray direction along which $s$ varies, and $p$ as the conjugate
transverse momentum, then the true periodic orbit would be
described by $q=p=0$ and $s=0\ldots L$. Nonvanishing $p$ and $q$ at
some point $s=s_{k}-\Delta s'$ denote transverse deviations from
the periodic orbit. This leads to straight line trajectories
hitting the surface slightly off the locations where the exactly
periodic orbit has its reflections. The outcome of the reflection
of this displaced ray is then a ray deviating from the periodic
path by $q(s_{k}+\Delta s)$ in position and $p(s_{k}+\Delta s)$ in
momentum. To show that these classical quantities are related by
\eqref{eq:matrixmonodromy} in the linear way of a Taylor expansion
for small deviations requires straightforward but tedious
trigonometry \cite{BAB72,STO99}.

The connection between the two requirements of
\eqref{eq:periodiccond} and \eqref{eq:matrixmonodromy} becomes
apparent, if we extend the monodromy matrix from a single
reflection to include all $N$ reflections of a full round trip, and
ask for the cumulated deviation of the ray, as a function of the
initial displacement in position $q$ and direction (momentum) $p$.
As a result we can see that in a linear approximation we simply can
replace $M^{(k)}(\Delta s,\Delta s')$ in \eqref{eq:matrixmonodromy}
with the product $M(s)$ of the $N$ individual monodromy matrices.
Assume that we start a ray at longitudinal coordinate $s$ in the
first leg of the ray path, $s_{1}\leq s<s_{2}$, with deviation $q$
and $p$ from the periodic orbit [cf.\ Fig.\ \ref{fig:bowtie}(a)].
We are then interested in the deviation incurred along one round
trip $s+L$.

The distance from $s$ to the first reflection is $\Delta
s'=s_{1}-s$, and the distance between the first and second
reflection accordingly is $\Delta s=s_{2}-s_{1}$. If we denote the
length between reflections at $s_{k}$ and $s_{k+1}$ by $l_{k}$,
then the round-trip deviation is obtained from
\begin{equation}
\label{eq:matrixmonodromyloop}
 \begin{pmatrix}
    q(s+L) \\
    p(s+L)
 \end{pmatrix}
 = M(s)\,
 \begin{pmatrix}
    q(s) \\
    p(s)
 \end{pmatrix}
\; ,
\end{equation}
where for the product $M(s)$ of the $N$ individual monodromy
matrices is given by
\begin{equation}
 \begin{split}
  M(s)
  & = M^{(N)}(s-s_{N},l_{N-1})\times M^{(N-1)}(l_{N-1},l_{N-2}) \\
  & \quad \times \ldots \times M^{(2)}(l_{2},l_{1})\times
  M^{(1)}(l_{1},s_{1}-s) \; .
 \end{split}
\end{equation}
The periodicity condition \eqref{eq:periodiccond} requires that the
vector $\vec{d} = (q(s),\,p(s))$ is an eigenvector of matrix
$M(s)$. Up to now we retained the dependence on the starting
coordinate $s$. However, the investigation of the eigenvalue
problem becomes simpler when we eliminate this dependence. This is
possible because we know how any given trajectory depends on $s$:
Rays in the cavity follow straight lines, so that $q$ is a linear
function of $s$, whereas $p$ is in fact a constant. (This can be
compared to the similar assumption we made in
\eqref{eq:complextraj2} with $kz$ instead of $s$ as the
longitudinal coordinate). For the vector $\vec{d}$ we can therefore
write
\begin{equation} \label{eq:pidef}
 \begin{pmatrix} q \\ p \end{pmatrix}
 =
 \begin{pmatrix} 1 & s \\ 0&1 \end{pmatrix}
 \begin{pmatrix} A_{0} \\ A_{1} \end{pmatrix}
 =
 \Pi(s)\, \begin{pmatrix} A_{0} \\ A_{1} \end{pmatrix} \; ,
\end{equation}
where the matrix $\Pi(s)=\bigl(\begin{smallmatrix}1 & s \\ 0&1
\end{smallmatrix}\bigr)$ appearing here obviously has unit determinant,
independently of $s$.  Thus we can rewrite the periodicity
condition \eqref{eq:matrixmonodromyloop} as
\begin{equation}
 \begin{pmatrix} A_{0} \\ A_{1} \end{pmatrix}
 = \Pi^{-1}(s+L)\,M(s)\Pi(s)\,
 \begin{pmatrix} A_{0} \\ A_{1} \end{pmatrix} \; ,
\end{equation}
which is an eigenvalue equation. We know that the coefficient
vector $\bigl( \begin{smallmatrix} A_{0} \\ A_{1}
\end{smallmatrix} \bigr)$ here is by construction $s$-independent,
and as a consequence, the $s$ dependence of the matrix product $E =
\Pi^{-1}(s+L)\,M(s)\Pi(s)$ must cancel out. The matrix $E$ is the
monodromy matrix of one whole periodic orbit, because it
characterizes the propagation of small deviations along one round
trip irrespective of where exactly we started. Recall, however,
that we set the starting point $s$ on a specific branch of the
periodic orbit. Clearly, the same arguments would apply with $s$
starting in any other branch, but then the resulting matrix
equation will in general be different.

\subsection{Eigenvalues of the monodromy matrix}\label{sec:monodromeig}

Since $E$ is a $2\times 2$ matrix, its eigenvalues and eigenvectors
are easily classified, especially since $E$ can also be shown to
have unit determinant \cite{BAB72}. This follows from ${\rm
det}(\Pi)=1$ together with classical area preservation (Liouville's
theorem) for the ray dynamics; the latter implies that also ${\rm
det}(M(s))=1$. As ${\rm det}\,E=1$ the two eigenvalues
$\lambda_{1,2}$ satisfy
\begin{equation}
\label{eq:eigenprod}
 \lambda_{1}\lambda_{2}=1 \qquad \text{and} \qquad
 \lambda_{1}+\lambda_{2}={\rm Tr}(E)\equiv t
\end{equation}
so that determinant and trace of $E$ fix its eigenvalues via the
quadratic equation
\begin{equation}
\lambda_{1,2}=\frac{1}{2}\,t\pm\frac{1}{2}\sqrt{t^2-4}.
\end{equation}
Three cases can be distinguished depending on the value of $t$:
\begin{description}
\item[$t>2$:] The eigenvalues are real and have
different magnitude.
\item[$t=2$:] Both eigenvalues are degenerate,
 $\lambda_{1}=\lambda_{2}=1$.
\item[$t<2$:] The eigenvalues are complex and both have unit magnitude,
\begin{equation}
\label{eq:eigenphase}
 \lambda_{1,2}=\frac{1}{2}\,t\pm\frac{\ii}{2}\sqrt{4-t^2}
 = e^{\pm {\rm i}\phi}\; .
\end{equation}
In this case the eigenvectors $\vec{h}_{1,2}$ are also complex
conjugates of each other.
\end{description}
The periodic ray orbit under consideration is called
\textit{linearly stable} if and only if case $t<2$ holds. If this
is satisfied in one branch of the path, one can show that it also
holds in all others.

\medskip
The difference between these cases can be appreciated if one writes
a general vector $\vec{d} = (q,\,p)$ as a linear combination of the
eigenvectors,
\begin{equation}
\vec{d}=a\,\vec{h}_{1}+b\,\vec{h}_{2}
\end{equation}
and asks for its evolution under repeated application of the round
trip mapping $E$. For a $\nu$-fold round trip we obtain
\begin{equation}
 E^{\nu}\vec{d} = a\,E^{\nu}\vec{h}_{1}+b\,E^{\nu}\vec{h}_{2}
=a\,\lambda_{1}^{\nu}\vec{h}_{1}+b\,\lambda_{2}^{\nu}\vec{h}_{2}=
a\,e^{i\nu\phi}\vec{h}_{1}+b\,e^{-i\nu\phi}\vec{h}_{2}.
\end{equation}
If $\vec{d}$ is a real vector -- as it should be if it describes
real ray trajectories -- then $E^{\nu}\vec{d}$ is also real because
$E$ is a real matrix, describing only the trigonometry of the
successive reflections and the straight line ray motion in between
(cf.\ the definitions of $\Pi$ and $M(s)$). Therefore, we know that
in the last equation there can be no imaginary part, and thus
nothing is changed by taking the real part. If
$\vec{h}_{1}=(A_{0},\,A_{1})$, then we have
\begin{equation}
\label{eq:elliptorb}
 E^{\nu}\vec{d}=
\begin{pmatrix}
 {\rm Re}[(a+b^*)A_{0}]\,\cos\nu\phi -
 {\rm Im}[(a+b^*)A_{0}]\,\sin\nu\phi \\
 {\rm Re}[(a+b^*)A_{1}]\,\cos\nu\phi-
 {\rm Im}[(a+b^*)A_{1}]\,\sin\nu\phi
\end{pmatrix}.
\end{equation}
Remember that $\phi$ is a fixed eigenphase, characteristic of the
given branch of the periodic orbit. If we assume for the moment
that $\nu$ is not an integer but instead varies continuously, then
the above equation describes an \textit{ellipse} in the plane
spanned by $q$ and $p$. For integer $\nu$, the trajectory simply
visits discrete points on this elliptical curve every time it
completes a loop. This means that the small initial deviation
$\vec{d}$ from the periodic ray orbit leads to a trajectory which
stays close to this original periodic path for all times. The
perturbed ray only performs a small oscillation in transverse
position and momentum around this original path.

In case $t>2$, the curve described by $E^{\nu}\vec{d}$ is also a
conic section, but in this case a hyperbola instead of an ellipse.
To see this, we go back to \eqref{eq:elliptorb} but with real
eigenvalues of the form $\lambda_{1}=1/\lambda_{2}$, cf.\
\eqref{eq:eigenprod}. Since for real eigenvalues the eigenvectors
of the real matrix $E$ can also be chosen real, we obtain with
$\xi\equiv \lambda^{\nu}$
\begin{equation}
\label{eq:hyperorb}
E^{\nu}\vec{d}=\xi\,a\,\vec{h}_{1}+\frac{1}{\xi}\,b\,\vec{h}_{2} \;
,
\end{equation}
which describes a hyperbola if $\xi$ is formally allowed to vary
continuously. This construction tells us that the perturbed ray
deviates more and more from the periodic path with increasing
$\nu$. Hence a periodic orbit whose monodromy matrix has trace
$t>2$ is unstable, and for $t<2$ it is stable.

The question of stability is crucial for the applicability of the
paraxial approximation in constructing resonator modes. Only if a
ray orbit is stable can the paraxial approximation be justified
over the whole round trip. In this case one can say that the mirror
configuration as seen by the closed ray trajectory and determined
by the radii of curvature $R$ at the successive reflection points
acts in a focussing way. Recall that the spreading of the Gaussian
beam in each ray segment is given by \eqref{eq:spreadq}. This will
be unchanged upon a round-trip if $q$ acquires only a phase, not a
change in absolute value. If we determine $q$ according to
\eqref{eq:pidef} with $\vec{h}=(A_{0},A_{1})$ being an eigenvector
of the monodromy matrix $E$, then $q(s)$ indeed changes only by the
eigenvalue $e^{{\rm i}\phi}$ when after a round trip we return to
the starting point $s$. This is a mere phase change provided that
the orbit is stable. For unstable orbits, the eigenvalues are real
and different from unity, making it impossible to recover a beam
with the original spread after one period of the orbit.

\subsection{Resonator eigenfrequencies in the parabolic approximation}

The use of the parabolic (Gaussian beam) approximation requires
that we identify the stable periodic ray orbits of the resonator at
hand. This can in general be a difficult task in itself, because
their number and location can depend rather sensitively on the
shape of the boundary. Furthermore, it must be kept in mind that
there may also be modes which are not in any way associated with
stable ray orbits at all.

Assuming that we have found a set of stable orbits, we can then use
the Gaussian beam solutions to construct the corresponding
eigenfrequencies (within the parabolic approximation). This is done
by recalling that the longitudinal coordinate $s$ along the orbit
plays the role of a time-like variable due to the approximation
made in \eqref{eq:parabolic}. In order for the modal wave function
to be single-valued upon a round-trip, the dependence on $s$ has to
be periodic. The imposition of periodic boundary conditions
introduces a longitudinal ``quantum number'' $m$ counting the nodes
along the ray path. The equations determining the wave solutions
are themselves periodic in $s$, and hence the problem is analogous
to that of a periodically driven time dependent system to which
\textit{Floquet's theorem} applies \cite{BAB72}, or equivalently to
Bloch's theorem governing the band structure of spatially periodic
solids.  The central observation there is that the wavefunction
acquires only a phase factor when a translation amounting to the
periodicity interval is carried out. This is precisely the content
of \eqref{eq:eigenphase} which however holds only when a stable
orbit is present.

Allowing for transverse excitations according to the harmonic
oscillator ladder as described by the generalized operators
\eqref{eq:creationop2}, an additional transverse mode number $n$
labels the number of nodes in the two-dimensional Gauss-Hermite
beam perpendicular to the propagation axis. The resulting quantized
wavenumber is found to be \cite{BAB72}
\begin{equation}
k\,L=2\,\pi\,\left(m+\frac{N}{4}\right)+
\left(n+\frac{1}{2}\right)\,\phi.
\end{equation}
where $L$ is the round-trip path length of the stable orbit. $N$ is
the number of reflections the ray undergoes, and this number enters
here to take into account the phase shifts associated with
reflection. The phase $\phi$ multiplying the transverse excitation
state is the eigenphase of the monodromy matrix as defined in
\eqref{eq:eigenphase}. A simple application of this formula would
consist in determining the longitudinal mode spacing. The
intuitively expected result is that an integer number of
wavelengths must fit around the ray path.

\subsection{Polygonal resonators}\label{sect:PolygonRes}

\subsubsection{Marginally stable orbits}

The stability considerations of section \ref{sec:monodromeig}
provide a conceptual background helping us to understand why
pathological effects can occur when the resonator is bounded by a
polygon with straight sides. The example most relevant here are the
hexagonal facets of a molecular sieve crystal we will present in
section \ref{Sect-realiz}. In case of the hexagon, the boundary
allows neither stable nor unstable orbits. We have precisely the
\textit{marginal} case $t=2$, where the round-trip monodromy matrix
exhibits ${\rm Tr}(E)=2$, so that both of its eigenvalues are
unity. The sides of the cavity are then neither focussing nor
defocussing, and the rays deviating slightly from a closed ray path
do not describe an elliptical transverse envelope as they would for
a stable ray. One obtains this result by evaluating the monodromy
matrix for a plane interface. The resonator modes for a straight
sided cavity are thus not well approximated by Gaussian beams, as
can also be seen from the simple example of a rectangular cavity,
for which the solutions of the scalar wave equation are (for
propagation in the $xy$ plane)
\begin{equation}
\label{eq:rectanglereso}
 \psi(x,y)\propto \sin k_{x}(x-x_{0})\,\sin k_{y}(y-y_{0})
\end{equation}
with suitably chosen discrete wavenumbers $k_{x,y}$ and offsets
$x_{0}$, $y_{0}$.

This simple example bears no resemblance to Gaussian modes but can
be solved exactly because the problem is separable in cartesian
coordinates. However, not all polygonal resonator geometries permit
a separation of variables. This applies in particular to hexagonal
resonators -- the geometry exhibited by the facets of the molecular
sieves microcrystals with which microlasers were realized; cf.\
\ref{Sect-realiz}. For a polygon with precisely $120^{\circ}$
angles between adjacent sides, any ray lauched at some angle to the
surface will go through only a finite number of different
orientations\cite{HOB75}, just as in the more familiar rectangular
resonator where there are at most two non-parallel orientations for
any ray path. In the hexagon, a ray encounters the interface with
at most three different angles of incidence. Despite this apparent
simplicity, there exists no orthogonal coordinate system in which
the wave equation for the hexagonal cavity can be solved by
separation of variables. This property is not surprising, since
there is only a finite number of suitable coordinate systems
available, and on the other hand there are infinitely many shapes
one can think of.

\subsubsection{Triangular and hexagonal resonators}\label{SubSect:HexRes}

Even when separability is lacking, the resonator problem may
sometimes be solvable by an appropriate ansatz. An example for this
is a resonator composed of three straight sides forming an
equilateral triangle on which the fields are assumed to vanish
(Dirichlet boundary conditions). The wave solutions in cartesian
coordinates are given by \cite{RIC81}
\begin{equation}
\label{eq:triangleansatz}
 \begin{split}
 \psi_{m,n}(x,y) =& \sin\left(\frac{2 \pi}{3}(2 m-n)\, x\right)\,
 \sin\left(\frac{2 \pi}{\sqrt{3}}\,n\, y\right) - \\
 &  \sin\left(\frac{2 \pi}{3}(2 n-m)\, x\right)\,
 \sin\left(\frac{2 \pi}{\sqrt{3}}\,m\, y\right) +\\
 & \sin\left(\frac{-2 \pi}{3}(m+n)\, x\right)\,
 \sin\left(\frac{2 \pi}{\sqrt{3}}(m-n)\, y\right) \; .
 \end{split}
\end{equation}
This solution is slightly more complicated than the one for the
rectangular resonator of \eqref{eq:rectanglereso}. In
\eqref{eq:triangleansatz} there is a degenerate set of solutions
${\tilde\psi}_{m,n}$ of different symmetry type which are obtained
by replacing $\sin$ with $\cos$ in the $x$-dependent factors. We
get identical wavefunctions if we exchange the quantum numbers $m$
and $n$, replace $n$ by $m-n$, or replace $m,\,n$ by $-m,\,-n$
(simultaneous sign change); the solutions differ only in sign so
that they do not represent different eigenfunctions. There are,
however, true degeneracies as well. They are characterized by
combinations of $m$ and $n$ giving the same wavenumber
\begin{equation}
k_{m,n}=\frac{4\pi}{3}\,\sqrt{m^2 + n^2- m\, n}\; .
\end{equation}

\begin{figure}[!ht]
\begin{center}
 \resizebox{1.0\textwidth}{!}{%
 \includegraphics{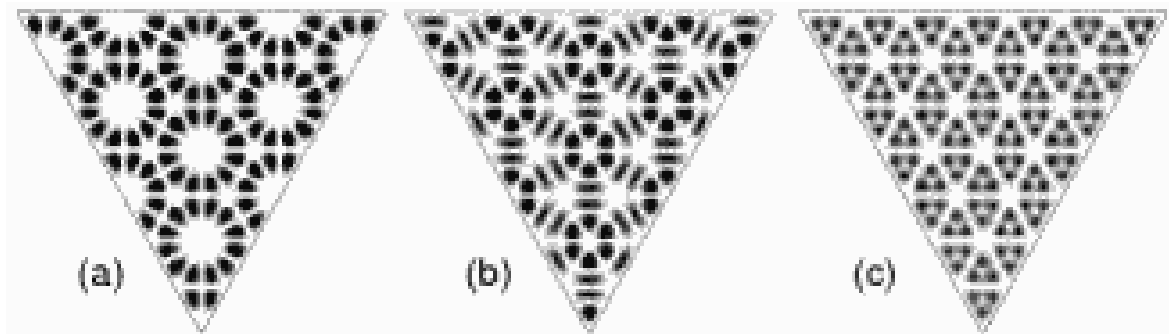}
 }
\end{center}
\caption{ \label{fig:trianglesuperpos} In the equilateral triangle
with Dirichlet boundary conditions, degenerate wave functions (a)
and (b) with $m=14,\,n=21$ are superimposed according to
\protect\eqref{eq:superimpos} to obtain a completely different
spatial intensity pattern (c). Plotted is the absolute square of
the wave function in a grayscale representation with black as the
highest intensity. }
\end{figure}

Since any linear combination of eigenfunctions with the same $k$ is
also an eigenfunction, there is a considerable ambiguity in the
spatial distribution of the wave function whenever there are
degeneracies. For example, for any given $m,\,n$ we can form the
superposition
\begin{equation}
\label{eq:superimpos}
 \Psi_{m,n}=\psi_{m,n}+\ii\,{\tilde\psi}_{m,n}\; .
\end{equation}
Shown in Fig.\ \ref{fig:trianglesuperpos} are the three functions
$\psi_{m,n}$, ${\tilde\psi}_{m,n}$ and $\Psi_{m,n}$ for the
particular case $m=14,\,n=21$.

The wavefunction shown in Fig.\ \ref{fig:trianglesuperpos} (c) has
nodal lines along the edges of a set of smaller triangles inscribed
into the original one. The self similarity exhibited by this
solution bears an intriguing resemblance to the intensity patterns
recently reported in unstable laser resontors with a mirror in the
shape of an equilateral triangle \cite{WOE99}. Although further
investigations of this emerging field of open, unstable resonators
are needed, it seems possible that the transverse mode structure
observed in the above laser experiment is given by a two
dimensional wave equation with solutions similar to the ones shown
in Fig.\ \ref{fig:trianglesuperpos}. The particular linear
combinations appearing in any given experimental realization can be
a result of the measurement setup or perturbations that lift the
degeneracies. In particular, complex-valued superpositions as given
by \eqref{eq:superimpos} become relevant in open resonators such as
those in Ref.\ \cite{WOE99}. Whereas for a closed resonator with
Dirichlet boundary conditions one can always choose the solutions
to be real without losing any modes, this is not generally possible
in open systems, when travelling waves are assumed to describe the
field outside the resonator.

The subdivision into smaller triangles in Fig.\
\ref{fig:trianglesuperpos}(c) suggests that modes of other
polygonal resonators can be constructed in an analogous way if the
boundary can be decomposed into eqal equilateral triangles. This is
the case in particular for the hexagon, which is composed of six
equilateral triangles. A hexagon mode can hence be pieced together
from six copies of any given solution of the triangle problem.
However, solutions obtained in this way do not represent all the
possible modes of the larger hexagon resonator. This is because the
triangle solutions necessarily satisfy Dirichlet boundary
conditions on all the sides -- also on those which form the
diagonals in the hexagon composition. Clearly, we can expect that
there should also be solutions with even parity under reflection at
such a diagonal. These would then have vanishing derivative on the
respective diagonals, i.e., satisfy Neumann boundary conditions.
Unfortunately, there is no exact ansatz such as
\eqref{eq:triangleansatz} for these different symmetry classes of
the hexagon.

The hexagon with Dirichlet boundary conditions on its faces
therefore is an example of how not only separation of variables
breaks down, but also the more general concept of {\em
integrability} fails. The latter essentially means that one can
find as many (globally valid) ``good quantum numbers'' as there are
confined degrees of freedom; we managed to do this in the
equilateral triangle but cannot label the solutions of the hexagon
using only these same numbers. Integrability carries over to the
classical ray dynamics in the form of \textit{constants of the
motion}. Nonintegrability is known to be one of the central
characteristics exhibited by wave equations whose classical (short
wavelength) limit exhibits \textit{chaos} \cite{GUT90}. When the
ray dynamics is chaotic, this does not imply that the rays move in
a stochasitc way. They are still governed by the deterministic laws
of total internal reflection and refraction, and in particular
infinitely many periodic orbits can be found. However, if for the
extreme case of a fully chaotic system we calculate the monodromy
matrices for these periodic orbits, they all turn out to yield
${\rm Tr E}>2$ and hence describe unstable (hyperbolic) orbits.

Hexagonal resonators are therefore an example of a ``pathological''
intermediate class of microresonators for which the ray dynamics is
not chaotic, but not integrable as well. These systems have been
termed {\em pseudo-integrable} \cite{RIC81}. It is interesting to
observe that the modes with which the pathological nature of the
problem manifests itself are precisely those which are not required
by symmetry to vanish on the diagonals of the hexagon. Or with
other words, parity with respect to the diagonal does not in itself
predetermine the value of the wave function at the corners of the
hexagon. Thus, the corners are the root of pseudointegrabilty in
these polygons with Dirichlet boundary conditions, because they
give rise to \textit{corner diffraction}. Corners are singularities
at which the tangent direction changes discontinuously. Diffraction
arises here because the length scale over which the tangent
direction changes (the radius of curvature) is ideally zero, and
hence certainly much smaller than the wavelength \cite{SIE97}. The
angle subtended by the corners determines whether or not a
pseudointegrable cavity is created. In the rectangular cavity, as
well as in the equilateral triangle, the effects of diffraction at
the corners cancel each other out.

An interesting wavelength effect was reported in Ref.\
\cite{VEG95}. Sharp corners of a polygonal billiard are smoothed
over the distance of one wavelength. This means that the system
becomes indistinguishable from a slightly ``streamlined'' billiard
that would be obtained by rounding the corners in such a way as to
obtain an everywhere smooth boundary. This leads to an apparent
contradiction: classically chaotic billiards with a continuously
varying tangent (``smooth'' walls) can be infinitesimally well
approximated by polygonal domains that are themselves never
chaotic, but at most pseudointegrable. From a stability analysis of
the periodic orbits in either the original chaotic system or the
inscribed polygonal approximant, one finds that the former exhibits
unstable periodic orbits, whereas the latter displays only
marginally stable periodic paths. Although one should thus expect
the resulting mode structure to be qualitatively different, the
spectral structure in a fixed frequency interval can become
indistinguishable for the two cases. In section \ref{reson} we will
discuss numerical solutions of the Helmholtz equation, and how this
smoothing on a wavelength scale manifests itself in specific
realisations of molecular sieve microresonators.

Because we are interested in dielectric materials forming
\textit{open resonators}, we shall not go into further detail
concerning the intriguing problem of semiclassically quantizing
pseudointegrable cavities, but instead we discuss briefly the
relevance of corners in the presence of \textit{leaky} boundary
conditions. The breakdown of the short wavelength approximation
near corners will also manifest itself in the emission from
hexagonal dielectric cavities. The concept of pseudointegrability
loses its significance in the case of an open resonator, because
dielectric interfaces can destroy integrability as easily as sharp
corners do. An example is the rectangular cavity made up of a
homogenous dielectric surrounded by air. The dielectric constant
can be written as $\epsilon(x,\,y)=1+ \epsilon_{m}\Theta(\vert
a-x\vert)\,\Theta(\vert b-y\vert)$ where $\epsilon_{m}$ is the
permittivity of the medium. Due to the term $1$, the Helmholtz
equation cannot be reduced to two one-dimensional problems for the
$x$ and $y$ coordinate separately, as was the case in
\eqref{eq:rectanglereso}. In the limiting case of large
permittivity the offending $1$ can be dropped, which restores
separability, because this corresponds precisely to the case of a
closed resonator. As was observed in Ref.~\cite{BIS98}, the
boundary conditions of a polygonal resonator (determined in this
case by the spacial distribution of the refractive index) can be of
greater significance than diffraction for essential (in that study
statistical) properties of the spectrum. It is thus not clear at
present if dielectric resonators are suitable to search for
signatures of pseudointegrability as it was defined for closed
resonators.

\section{Actual realizations of microlasers based on molecular sieve-dye compounds}
\label{Sect-realiz}

After the discussion in previous sections we can now venture to
give a definition of what a \textit{microlaser} is. A microlaser is
a structure in which only a few (in the limit one) mode of the
light field interacts with a collection of atoms or molecules, so
that spontaneous emission processes are enhanced or inhibited, or
in which lasing occurs without a visible threshold. It is in
principle possible to observe these \textit{cavity} effects in
large ($10^3\dots 10^4\,\lambda$) resonators with small mode volume
(confocal resonators). But because of the small frequency spacing
(free spectral range) of the longitudinal resonator modes, cavity
effects in large resonators can only occur when the linewidth of
the atomic transition is smaller than the free spectral range of
the cavity -- a condition which can only be met by very dilute
atomic systems such as an atom beam, but not by atoms or molecules
in a condensed state. As the condensed state linewidth of atoms or
molecules is in the order of 1~nm to 100~nm, cavity effects will
only be observable with resonators exhibiting a correspondigly
large mode frequency spacing, that means resonator sizes of a one
half to few wavelengths. In the past, wavelength size resonators
were realized with semiconductors as microdisks \cite{MCC92,LEV93}
or VCSELs (Vertical Cavity Surface Emitting Lasers)
\cite{KOY88}--\cite{CHA93}, and with organic dyes in planar
resonators as Langmuir-Blodgett \cite{DRE74} or liquid films
\cite{MAR88}, or embedded in polymer spheres
\cite{TZE84}--\cite{KUW92}. This work was discussed recently in
several reviews \cite{BER94}-\cite{RAR96} so that here we will
concentrate on microlasers based on organic dye molecules embedded
in molecular sieve microcrystals \cite{VIE98}. Before we turn to
the proper optical and laser properties, we will shortly discuss
the synthesis of the compounds.

\begin{table}[ht]
\caption{Lattice constants and free pore diameter $\phi$ of
molecular sieves with linear channels \protect\cite{MEI96}.}
\label{tab_pores}
\begin{center}
\begin{tabular}{lcccc}
 \hline\noalign{\smallskip}  & $a$/nm & $b$/nm & $c$/nm & $\phi$/nm  \\
 \noalign{\smallskip}\hline\noalign{\smallskip}
  & \multicolumn{4}{c}{hexagonal}\\
 mazzite & 1.84 & & 0.76 &0.74 \\
 AlPO$_4$-5 & 1.34 && 0.84 & 0.73 \\
 zeolite L & 1.84 && 0.75 & 0.71 \\
 gmelinite & 1.38 && 1 & 0.7 \\
 offretite & 1.33 && 0.76 & 0.68 \\
 CoAPO-50 & 1.28 && 0.9 & 0.61 \\
 cancrinite & 1.28 && 0.51 & 0.59 \\
  & \multicolumn{4}{c}{orthorhombic}\\
 AlPO$_4$-11 & 1.35 & 1.85 & 0.84 & 0.63$\times$0.39\\
 mordenite & 1.81 & 2.05 & 0.75 & 0.70$\times$0.65\\
   \noalign{\smallskip}\hline
\end{tabular}
\end{center}
\end{table}

\subsection{Molecular sieve crystals as host material for microlasers}
\label{host}

\begin{figure}[!ht]
 \begin{center}
 \resizebox{0.6\textwidth}{!}{%
 \includegraphics{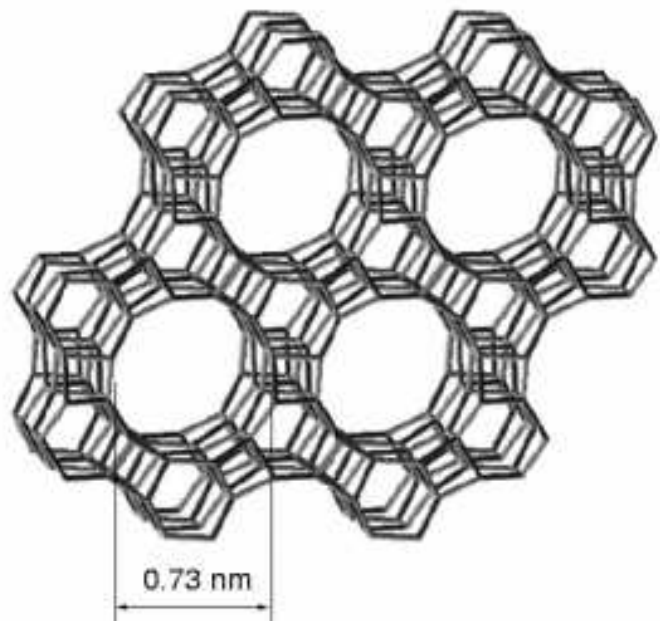}
 }
 \end{center}
 \caption{\label{Fig-AlPO-STRUC}
 Structure of the hexagonal AlPO$_4$-5 molecular sieve. The
 corners of the polygons are occupied alternately with aluminum and
 phosphorus, while the polygon sides represent oxygen. The hexagonal
 $c$-axis is oriented along the channel pores.}
\end{figure}

As is reviewed in the previous chapter \textit{Nanoporous compound
materials for optical applications -- Material design and
properties}, molecular sieve materials are characterized by a
crystallographically defined framework of regularly arranged pores.
In Table \ref{tab_pores} we list some sieves with wide channel
pores, in which optically effective organic molecules can be
inserted. Among the listed materials especially the
aluminophosphate AlPO$_4$-5 (molecular mass 1463.4~g/mol) can be
synthesized with good optical transparency and low internal
scattering losses. In addition, the AlPO$_4$-5 channel pores
exhibit a diameter of 0.73~nm, fit to accommodate a variety of
laser active organic dye molecules. The structure is shown in
Fig.~\ref{Fig-AlPO-STRUC}. AlPO$_4$-5--crystals are crystallized
from aqueous or alcoholic solutions under hydrothermal conditions,
with the addition of an organic structurizing agent, called
\textit{template}. The template is a necessary device in the
synthesis of molecular sieves to direct the chemical reactions
towards the desired crystal structure. For the synthesis of the
AlPO$_4$-5 laser crystals tri-n-propylamine was used as template.
The preferred pH range for the synthesis is mildly acidic to mildly
basic, while the source of phosphor is mostly orthophosphoric acid,
and the sources of aluminum are pseudoboeh\-mite or alkoxides
\cite{WIL91}. It was shown that single crystals with nearly perfect
morphology can be grown using hydrofluoric acid \cite{GIR95}, and
with a specially prepared and aged aluminum hydroxide gel crystal
sizes around  1~mm in $c$-axis direction were obtained
\cite{SUN96}. In addition, microwave heating proved to increase the
crystallization rates by more than one order of magnitude
\cite{GIR95,DU97}.

Pure AlPO$_4$-5 crystals are optically transparent from below
400~nm to above 800~nm and exhibit a refractive index of
$n(500\,{\rm nm})=1.466$. After removing the template (usually by
heating) they show practically no birefringent properties. X-ray
diffraction revealed patterns which are consistent with space group
$\rm P \frac{6}{\rm m}cc$ as well as P6cc. The latter group,
however, corresponds to a polar structure: In fact, the 4 in the
formula  AlPO$_4$-5 is the result of the strict alternation of Al
and P in the tetrahedral nodes of the framework, which prevents the
corner-sharing oxygen tetrahedra to occur with odd numbers, and
which leads to an alternating stacking of Al and P in the direction
of the channels ($c$-axis); cf.\ Fig.~\ref{Fig-AlPO-STRUC}. This
alternance is assumed to cause the crystallographic polar nature
\cite{BEN83} of the framework \cite{MAR94}. The macroscopic polar
nature of AlPO$_4$-5 single crystals was proven recently in
scanning pyroelectric microscopy investigations \cite{KLA99}, and
it was observed that AlPO$_4$-5 crystals are usually twinned. The
murky stripes inside the pyridine~2-loaded crystals shown in
Fig.~\ref{fig-dichro-pyridine} and their slightly bowed side faces
could well be a result of this kind of twinning. At this moment it
is not clear, however, to what extent twinning should affect the
properties relevant for luminescence and lasing discussed here.

\medskip
Up to now laser action was obtained in two different compounds,
namely in AlPO$_4$-5 loaded with 1-ethyl-4-(4-({\sl
p}-dimethylamino\-phen\-yl)-1,3-butadienyl)-pyridinium perchlorate
(pyr\-i\-dine~2 \cite{BRA94}), and with a modified rhodamine~B dye.
In the following we will discuss the two materials in more detail.

\begin{figure}[!ht]
\begin{center}
\resizebox{0.6\textwidth}{!}{%
 \includegraphics{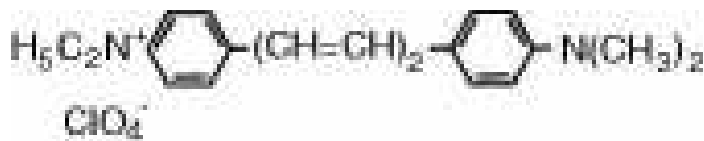}
 }
 \end{center}
 \caption{\label{pyrid2}
 Structure formula of the dye 1-ethyl-4-(4-({\sl
p}-dimethylamino\-phen\-yl)-1,3-butadienyl)-pyridinium perchlorate
(pyr\-i\-dine~2 \protect\cite{BRA94}); molecular mass 378.9~g/mol.}
\end{figure}

\begin{figure}[tbh]
\begin{center}
\resizebox{0.436\textwidth}{!}{%
 \includegraphics{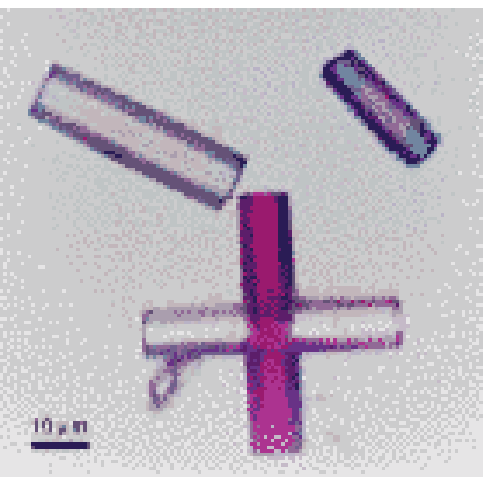}
 }
 \hfill
\resizebox{0.55\textwidth}{!}{%
 \includegraphics{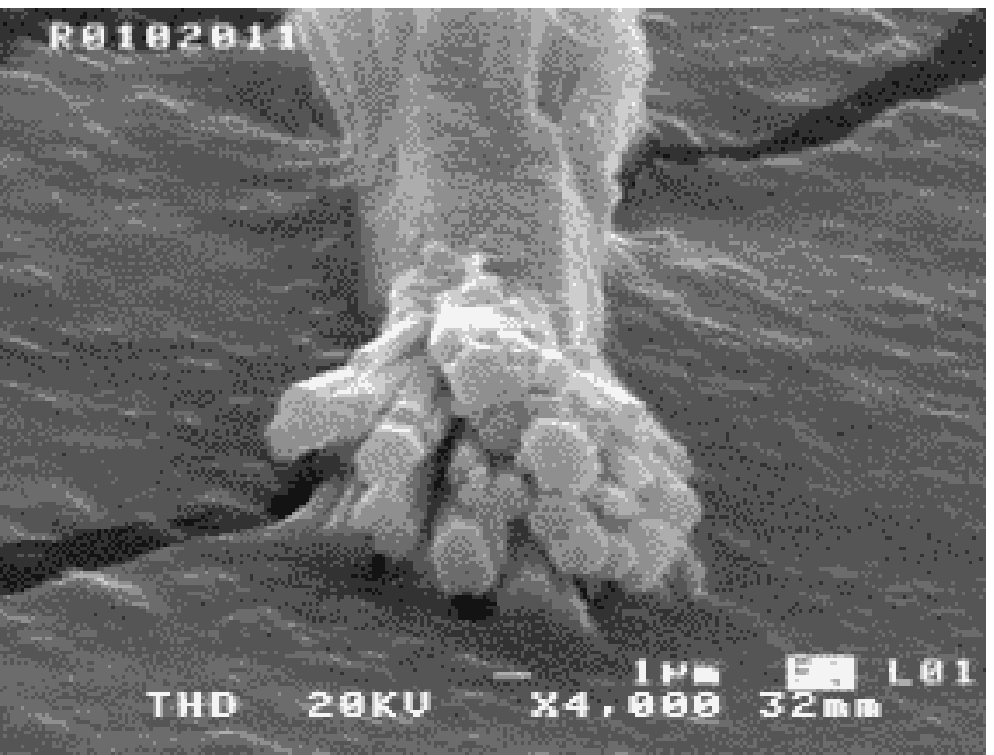}
 }
 \end{center}
\caption{\label{Fig-Morphology} Morphology of typical pyridine
2/AlPO$_4$-5 crystals; \textbf{left:} without lasing properties,
\textbf{right:} with lasing properties.}
\end{figure}

\subsubsection{AlPO4-5/pyridine2 compound} \label{pyrid}

\begin{figure}[!bt]
 \begin{center}
 \resizebox{1.0\textwidth}{!}{%
  \includegraphics{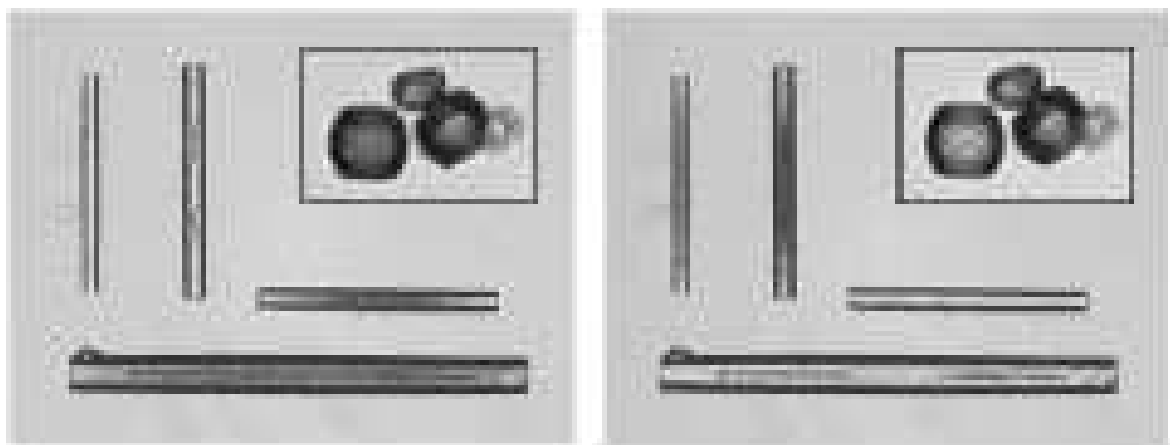}
  }
 \end{center}
 \caption{\label{fig-dichro-pyridine}
 Transmission micrographs of the dichroism in dye-loaded
AlPO$_4$-5 crystals; the rod-shaped crystals contain ca.~0.1~wt-\%
or 1 pyridine~2 molecule per 260 unit cells, while the
barrel-shaped ones (shown in the inset) enclose rhodamine~BE50
(ca.~0.5~wt-\% or 1 molecule per 75 unit cells). Only the
polarization component parallel to the optical transition moment of
the molecules is absorbed. In the rod-shaped crystal the pyridine~2
dyes are completely aligned, whereas with the rhodamine~BE50 dye in
the barrel-shaped crystals we observe only a weak dependence of the
color upon the incident polarization. {\bf Left:} incident light
horizontally polarized; {\bf Right:} incident light vertically
polarized.}
\end{figure}

These AlPO$_4$-5--dye compounds were synthesized \cite{IHL99}
following a procedure in which the dye 1-ethyl-4-(4-({\sl
p}-dimethylamino\-phen\-yl)-1,3-butadienyl)-pyridinium perchlorate
(trade name pyr\-i\-dine~2 \cite{BRA94}), cf.\ Fig.~\ref{pyrid2},
is added to the template or to the aluminum hydroxide suspension
\cite{HOP93,DEM95,IHL98}. After 1~h of hydrothermal synthesis dark
red crystals with a length of up to 100~$\mu$m are obtained. As the
linear dye molecules fit snugly into the 0.73~nm wide channel pores
of the nanoporous AlPO$_4$-5 host, they become aligned along the
crystal $c$-axis. As a result the compound exhibits strong
dichroism, and the emitted fluorescence light is polarized parallel
to the $c$-axis. This is documented in
Fig.~\ref{fig-dichro-pyridine}. It is also observed that with the
inclusion of pyridine~2 the entire compound acquires pyroelectric
properties and an optical second order susceptibility \cite{VIE98}.
%
%
Even though the size of the dye molecules is well compatible with
the channel diameter, the dye content visibly affects the crystal
morphology. Regular hexagonal crystals with a rodlike form, as
e.g.\ the ones shown in Fig.~\ref{Fig-Morphology}(left), or
Fig.~\ref{fig-dichro-pyridine}, are obtained when the dye content
is low, around 0.1~wt\% or 1 molecule per 260 unit cells. At higher
concentrations the dye accumulates in the middle of the crystal,
and the disturbances grow: at a dye content of
$\gtrapprox$~0.2~wt\%, crystals with a characteristic fascicular
shape resulted; cf.\ Fig~\ref{Fig-Morphology}(right). Given the
small size of the crystals the dye content was determined by
chemically dissolving them. With this method, however, it is not
possible to accurately determine the spatial distribution of the
dye. Therefore the content was also evaluated qualitatively by
comparing the depth of the color. It is only with these fascicled
pyridine~2-loaded crystals where laser emission was observed.
Apparently, in the undisturbed rod-shaped crystals the low
concentration of dye does not spoil the growth, but it is not
sufficient to provide the necessary optical gain for compensating
all losses either.

\subsubsection{AlPO4-5/rhodamine BE50 compound} \label{rhod}

\begin{figure}[!ht]
\begin{center}
\resizebox{0.4\textwidth}{!}{%
 \includegraphics{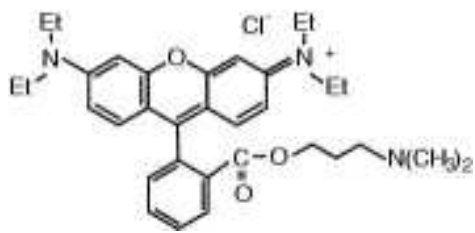}
}
 \caption{Structure formula of the new dye rhodamine BE50 (ethanaminium,
$N$-[6-(diethylamino)-9-[2-($N,N$-di\-meth\-yl-3-amino-1-%
propoxycarbonyl)phenyl]-3H-xanthen-3-y\-li\-dene]-$N$-ethyl-chloride)
molecular mass 564~g/mol \protect\cite{BOC98}. }
 \label{rhbe50}
\end{center}
\end{figure}
The second molecular sieve dye compound which exhibited lasing
properties was obtained by including a modified rhodamine~B dye in
an AlPO$_4$-5 crystal. The size of rhodamine molecules exceeds the
dimensions of the channel pores in AlPO$_4$-5 crystals. As a result
the rhodamine~B is normally not accepted inside crystal pores but
adheres at the exterior. The group of Schulz-Ekloff and W\"ohrle
started an attempt to accomplish rhodamine inclusion in AlPO$_4$-5
crystals, even so the pores are too small. Their idea is to modify
the dye molecule in a way that makes the dye molecule electrically
resemble a template molecule. Their new derivative rhodamine~BE50
(ethanaminium,
$N$-[6-(diethylamino)-9-[2-($N,N$-di\-meth\-yl-3-amino-1-%
propoxycarbonyl)phenyl]-3H-xanthen-3-y\-li\-dene]-$N$-ethyl-chloride;
cf.\ Fig.~\ref{rhbe50}) was synthesized by esterification of
rhodamine~B (Rh~B) with 3-dimethyla\-mi\-no-1-propanol
\cite{BOC98}. It was shown that the concentration of rhodamine BE50
(Rh~BE50) achievable by crystallization inclusion in AlPO$_4$-5
exceeds the possible Rh~B concentration by a factor of 3--4. This
was attributed to the different molecule structures, i.e., to the
zwitterionic nature of Rh~B on one hand, and to the additional
positive charge of a protonated aliphatic amino group of Rh~BE50 on
the other hand \cite{BOC98}. The latter molecules with the
localized positive charge are more compatible with the AlPO$_4$-5
framework than the delocalized charge of Rh~B. As a consequence
they observe that at a given dye concentration  Rh~BE50 inclusion
leads to a better crystal morphology than Rh~B inclusion
\cite{BRA99}.

\begin{figure}[!ht]
 \resizebox{0.48\textwidth}{!}{%
 \includegraphics{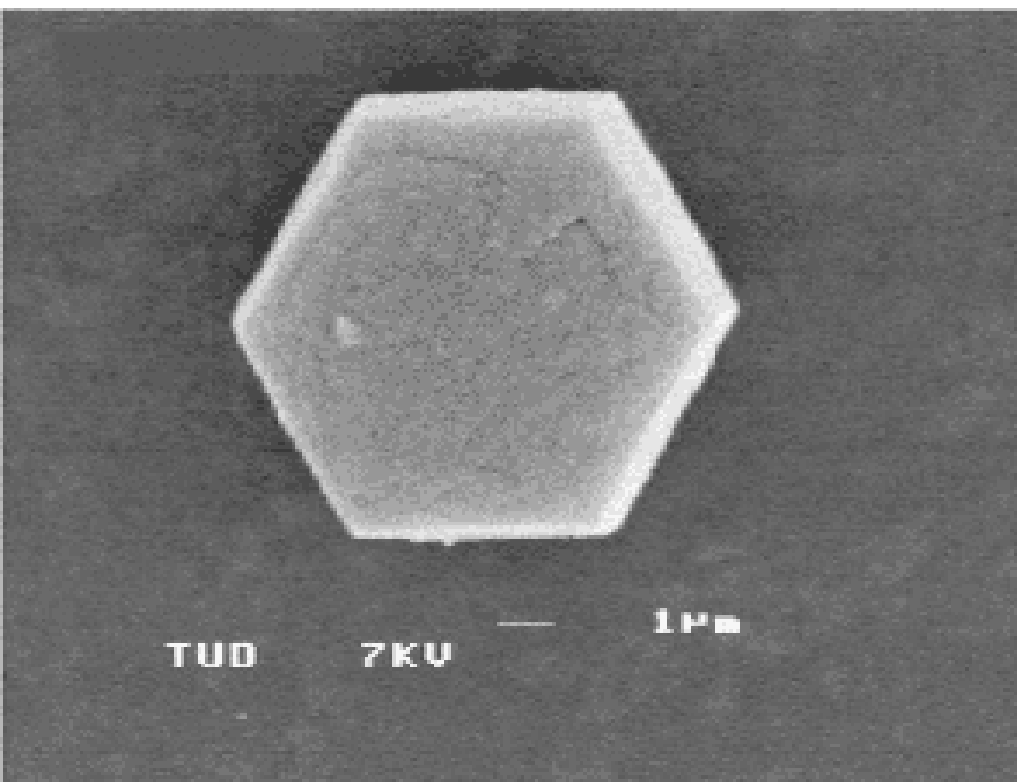}
 }
 \hfill
 \resizebox{0.48\textwidth}{!}{%
 \includegraphics{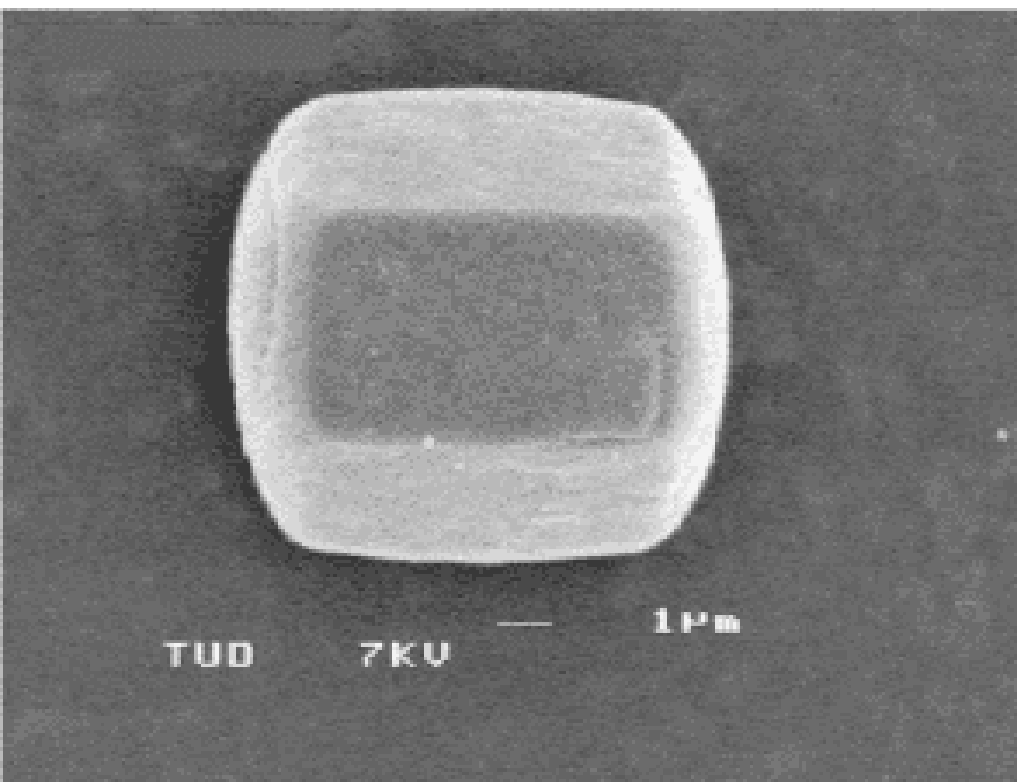}
 }
 \caption{Morphology of typical rhodamine BE50/AlPO$_4$-5 crystals
  with lasing properties; dye content ca.~0.5~wt-\% or 1 dye molecule
per 75 unit cells.}
 \label{BE50morph}
\end{figure}
The gel for the synthesis of the AlPO$_4$ crystals is prepared
according to recipes \cite{BEN83,WIL82} which are modified for the
purpose of crystallization inclusion of dyes \cite{WOH92}. To a
suspension of 61.6~mmol $\rm Al_2O_3$ as aluminum source
\cite{AL2O3} and 75~g deionized water, 61.6~mmol $\rm P_2O_5$
\cite{P2O5} in 11.3~g deionized water is added under mechanical
stirring. After 5~min a uniform gel formed and then 92.4~mmol
tripropylamine \cite{P3N} is added slowly. Subsequently, the
appropriate amount (0.1--10~mmol) of Rh~BE50 dye powder is mixed
with the gel. The synthesis of the Rh~BE50/AlPO$_4$-5 crystals is
performed by microwave heating \cite{BRA98}, which proved to be
superior in respect of avoiding damage of sensitive dyes like
coumarines \cite{BRA97} as well as of reducing the time of
synthesis \cite{BOC98}. Unlike the pyridine~2 molecules, which with
a diameter of  0.6~nm fit into the 0.73~nm wide pores of the
AlPO$_4$-5 host, the Rh~BE50 molecules with dimension of
0.91~$\times$~1.36~nm$^2$ are accommodated in defect sites, or
\textit{mesopores}, of the host crystal. Remarkably, up to
concentrations of 1 molecule per 75 unit cells this remains without
any visible negative consequences for the crystal morphology, as is
documented in Fig.~\ref{BE50morph}.

In Fig.~\ref{fig-dichro-pyridine} the dichroic properties of the
compound are illustrated. In comparison with the
pyridine-2/AlPO$_4$-5 compound, the dichroism of the Rh~BE50
com\-pound is reduced and the fluorescence emission is only
partially polarized. This is an indication that the electrical
anisotropy of the host structure does not fully carry through the
mesopores, resulting in a weak dye alignment in the statistical
average.

\subsection{Microresonator structure} \label{reson}

As is visible in Fig.~\ref{fig-dichro-pyridine} -- and as the
polarization of the emitted fluorescence of the compounds indicates
-- the absorption, as well as the emission dipole moment of the
included dyes are oriented preferentially along the crystal
$c$-axis (in the pyridine 2/AlPO$_4$-5 compound the orientation is
complete). As dipole emission along the dipole axis is not
possible, the emission parallel to a plane perpendicular to the
prevailing dipole orientation (i.e.\ the hexagonal axis) is
enhanced. Here a bundle of emission directions meets the condition
for total internal reflection (TIR) at the hexagonal side faces
inside the crystal. In a whispering-gallery-mode-like way the
corresponding emission can circulate sufficiently often to
accumulate the gain required to overcome the lasing threshold; cf.\
Fig.~\ref{TIR} in which this intuitive model is illustrated for a
particular ray bundle of high symmetry. However in section
\ref{microresonators} we have shown that for resonators with a
sizes in the order of a few wavelengths the naive ray picture does
not correctly represent the field modes. E.g.\ the ray picture
insinuates a mode concentration in the center of the faces but
field-free corners. This, however, is not consistent with the
experimental evidence, which clearly shows that the emission occurs
at the corners; cf.\ Fig.~\ref{nearfield}. As we already pointed
out in section \ref{sect:PolygonRes}, the main feature that
distinguishes the hexagonal resonator from other common
whispering-gallery type cavities such as microdroplets \cite{MEK95}
or semiconductor disk lasers \cite{GMA98,NOE97b}, is that the
latter do not exhibit sharp corners and flat sides. There we
described that portions of the boundary in convex resonators act as
focussing or defocussing elements, whereas the straight sides of a
hexagon are neither one nor the other. As a result we have shown
that ray paths in the hexagon display a degree of complexity that
cannot be classified as chaotic and was called  pseudointegrable.

The closed ray path underlying Fig.~\ref{whisp} is only one member
of an infinite family of periodic orbits of the hexagon billiard
that all have the same length, $L=3\times WoF$. Long-lived cavity
modes exist only if the corresponding rays satisfy the condition of
TIR at the interface, $\sin\chi>1/n$, where $\chi$ is the angle of
incidence with respect to the surface normal. In a naive ray
approach one would furthermore obtain the spectrum of modes by
requiring an integer number of half wavelengths to fit into $L$,
leading to constructive interference on a round-trip. As we shall
see shortly, this estimate is justified, even though a proper
treatment of the ray-wave connection has to take into account that
any given mode is in fact made up of a whole \textit{family} of
different ray paths. The ultimate breakdown of the ray model,
however, occurs when a ray orbit hits the corners where the
classical laws of refraction and reflection are no longer defined.
To characterize the size of the samples, we can specify either the
radius $R$ of the hexagon at the corner points or -- more
conveniently -- the width over flats (WoF) characterized by $R= WoF
/ \sqrt{3}$.

\subsubsection{Wave picture: spectral properties}\label{WavPic}

\begin{figure}[bt]
\begin{center}
\resizebox{0.48\textwidth}{!}{%
  \includegraphics{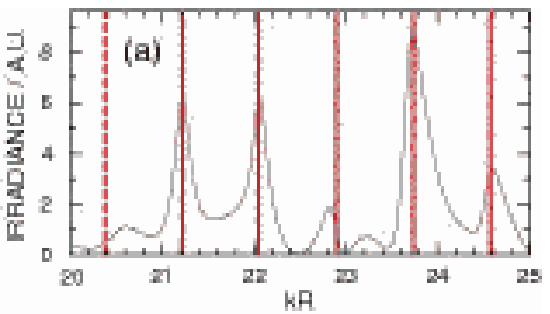}
 }
  \vspace*{5mm}
\resizebox{0.48\textwidth}{!}{%
  \includegraphics{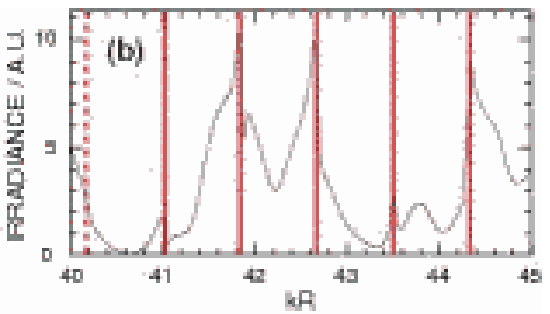}
  }
\end{center}
\vspace*{-1cm}
\caption{\label{fig:k20spec} Calculated scattering intensity
spectra of a hexagonal cylinder for plane-wave incidence at
$15^\circ$ to a side face and detection at $60^{\circ}$ from
incidence. (a) corresponds to a spectral interval $\lambda\approx
653\ldots 816$~nm for width over flats (WoF) $4.5\,\mu$m; (b)
covers the interval $\lambda\approx 605\ldots 680$ nm for WoF
$7.5\,\mu$m. Vertical lines are guides to the eye, indicating
narrow resonances. The spacing between resonances is
$\Delta(kR)\approx0.84$ in (a) and $\Delta(kR)\approx0.83$ in (b),
in good agreement with the characteristic mode spacing
$\Delta(kR)_c\approx0.83$ of a closed hexagonal orbit. Expected
resonances not clearly seen in the above spectra are marked by
dashed lines; they appear at other detection angles.}
\end{figure}

Because the hexagonal faces are neither focussing nor defocussing,
there is no obvious way of determining the weight that should be
given to individual members of a ray family in order to predict the
spatial structure of the resulting mode. Full solutions of
Maxwell's equations have therefore been carried out for the TM
polarized modes of a dielectric hexagonal prism, using methods
previously applied in \cite{GMA98,NOE97b} and discussed in section
\ref{microresonators}. In anticipation of the experimental spectra
discussed in section~\ref{laserprop}, we focus on three different
sample sizes characterized by a WoF of $4.5\,\mu$m, $7.5\,\mu$m,
and $22\,\mu$m. The aim is to understand the observed laser line
spacings and the emission directionality.

To document that orbits of different families determine the
characteristic mode spacing of the cavity, Fig.~\ref{fig:k20spec}
shows scattering spectra for different sample sizes in the vicinity
of the experimental wavelengths. Intensity is plotted versus
dimensionless wavenumber $kR$, where $k=2\pi/\lambda$. This is the
natural scale for comparison with semiclassical predictions because
modes differing by one node along a closed path should then be
equally spaced, with a characteristic separation
$\Delta(kR)_c=2\pi\,R/(n\,L)=2\pi/(3\sqrt{3}n)=0.825$ independent
of the sample size. The expected wavelength spacing of the modes
(free spectral range $FSR$) is $\Delta\lambda=\lambda^2 \times
\Delta(kR)_c / (2\pi R) \approx 23$~nm in (a) and
$\Delta\lambda\approx 11$~nm in (b). For $WoF = 22\,\mu$m, we
obtain $\Delta\lambda\approx 4.9$~nm. Figure \ref{fig:k20spec}
indeed shows a series of resonant features with approximately the
predicted wavevector spacing.

Each of the peaks marked in Fig.~\ref{fig:k20spec}(b) is in fact a
multiplet which is not resolved, because the splittings of the
individual modes comprising the multiplet are smaller than their
passive linewidths. There is evidence for this because several of
the peaks are very asymmetric, and in particular exhibit a steep
slope on one side. For an isolated resonance, the most general
lineshape that could arise is the Fano function (of which the
Lorentzian is a special case), which however does not yield
satisfactory fits here.

\begin{figure}[bt]
\begin{center}
\resizebox{0.48\textwidth}{!}{%
  \includegraphics{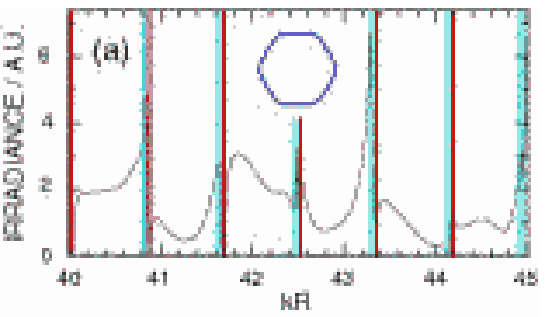}
 }
 \vspace*{5mm}
\resizebox{0.48\textwidth}{!}{%
  \includegraphics{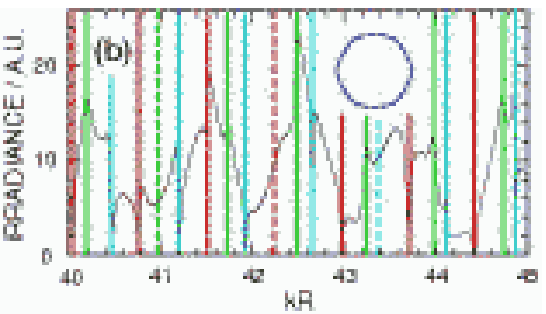}
 }
\end{center}
\vspace*{-1cm}
\caption{\label{fig:k40smoothspec} Calculated scattering intensity
spectra for slightly rounded hexagonal cavities (shapes depicted as
insets). The incoming plane wave is at an angle of $15^{\circ}$ to
a facet in (a) and $30^{\circ}$ in (b); detection occurs at
$60^{\circ}$ from incidence. Spacings between modes of the same
color agree well with $\Delta(kR)_c\approx 0.83$, cf.\
Fig.~\ref{fig:k20spec}. All resonances in (a) appear as doublets.
At $kR\approx 42.5$ the doublet structure is seen most clearly. In
(b), stronger deviation from hexagonal shape leads to further
lifting of degeneracies. Dashed lines mark expected resonances not
seen at this observation angle.}
\end{figure}

To further expose the multiplet structure, we modeled deviations
from the ideal hexagonal shape which could lead to narrower
individual linewidths and increase the multiplet splitting. Shape
perturbations were chosen that preserve the $D_{6h}$ point group
symmetry and hence remove only ``accidental'' quasi-degeneracies.
The actual perturbation that is present in the samples of
Figs.~\ref{pyr2spectr} and \ref{BE50spectr} eluded experimental
characterization, so that a model calculation can reproduce only
generic features which are insensitive to the precise type of
perturbation. One such feature is the {\em average} mode spacing
after degeneracies have been lifted sufficiently.

Figure \ref{fig:k40smoothspec}(a) shows the spectrum of a rounded
hexagon where the radius of curvature at the corners is $\rho
\approx 0.9\,\lambda$ (assuming $\lambda\approx 610$ nm for
definiteness). No qualitative difference to
Fig.~\ref{fig:k20spec}(b) is seen, except that the resonant
features have become somewhat narrower, thus enabling us to
identify two distinct series of modes with caracteristic spacing
$\Delta(kR)_c$. This indicates that departures from sharp corners
are not resolved in the wave equation when their scale is smaller
than $\lambda$. A qualitatively different spectrum is observed in
Fig.~\ref{fig:k40smoothspec}(b) where $\rho\approx 3.7\,\lambda$.
Here, the perturbation reveals three well-separated,
interpenetrating combs of modes, again with period $\Delta(kR)_c$.
There are $21$ distinct resonances in the wavelength interval of
Fig.~\ref{fig:k40smoothspec}(b), which translates to an average
mode spacing of $\Delta\lambda\approx 3.6$ nm for a $WoF =
7.5\;\mu$m resonator, well in agreement with the experiment; cf.\
Fig.~\ref{BE50spectr}.

In order to verify that no further modes will be revealed by other
choices of deformation, an independent estimate of the average
density of modes can be made based on semiclassical considerations
\cite{NOEun}:
\begin{eqnarray}
\left\langle\frac{dN}{d(kR)}\right\rangle&=&\frac{n^2\,k\,R}{4}\nonumber\\
&\times&\left[1-\frac{2}{\pi}\,\left(\arcsin{
\frac{1}{n}}+\frac{1}{n}\, \sqrt{1-\frac{1}{n^2}}\right)\right]
\label{wgweylformtseqn}
\end{eqnarray}
Here, $dN$ is the number of modes in the interval $d(kR)$. The
result is $\langle\frac{dN}{d(kR)}\rangle\approx 4.6$, and hence we
expect $\approx 22$ modes in the interval of
Fig.~\ref{fig:k40smoothspec}(b), again in good agreement with the
actual count.

\subsubsection{Wave picture: intensity profile}

There is one class of quasi-degeneracies that is not removed by any
of the perturbations in Fig.~\ref{fig:k40smoothspec}. Their
physical origin is time reversal symmetry for the ray motion inside
the cavity. Any of the periodic orbits can be traversed either
clockwise or counterclockwise, and the same holds for more general
ray paths. The different propagation directions can be linearly
combined in various ways to obtain nearly-degenerate standing-wave
patterns that differ only in their parity with respect to some of
the crystal's reflection axes. A minute splitting does exist
because the nonintegrability of the ray motion implies that the
propagation direction itself is not a ``good quantum number'',
i.e.\ reversals of the sense of rotation are unlikely but not
impossible in the wave equation. This is analogous to quantum
tunneling and hence leads only to exponentially small splittings
that can be neglected on the scale of the individual resonance
linewidths \cite{DAV81}. These multiplets have been counted as one
resonance in (\ref{wgweylformtseqn}).

\begin{figure}[!hbt]
\begin{center}
\resizebox{0.48\textwidth}{!}{%
 \includegraphics{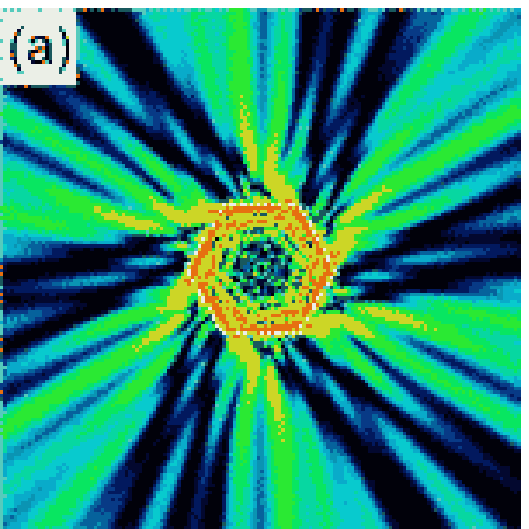}
 } \vspace*{5mm}
\resizebox{0.48\textwidth}{!}{%
  \includegraphics{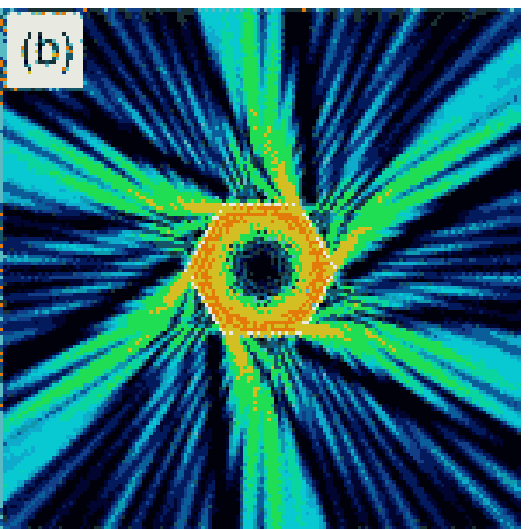}
 }
\end{center}
\vspace*{-1cm}
\caption{\label{fig:waksharp} False-color representation of the
cross-sectional intensity in the ideal hexagon for a mode with (a)
$kR=22.89$ [cf.\ Fig.~\ref{fig:k20spec} (a)] and (b) $kR= 42.78$
[cf.\ Fig.~\ref{fig:k20spec} (b)]. The resonance width is
$\delta(kR)=0.10$ in (a) and $\delta(kR)=0.04$ in (b).}
\end{figure}

Following this reasoning, in Fig.\ \ref{fig:waksharp} the
\textit{traveling-wave} patterns belonging to one of the resonances
in Fig.~\ref{fig:k20spec}(a) and (b), respectively, is plotted.
High intensity ridges inside the resonator form a
whispering-gallery-like pattern that decays from the interface into
the cavity center. The number of ridges in the radial direction
(perpendicular to a side face) provides an approximate analogue of
a transverse mode order, however upon closer examination one sees
that the number of ridges and nodal lines is not uniquely defined,
in particular along a diameter joining opposite corners. The modes
can therefore not be properly labeled by ``good quantum numbers''
characterizing the number of radial and azimuthal nodes -- this is
a direct consequence of the nonintegrability of the problem. The
most significant difference to the whispering-gallery modes of a
circular cavity is clearly the anisotropic emission. High intensity
is seen to emanate predominantly from the corners and is directed
almost parallel to an adjacent crystal facet. The overall emission
pattern is very similar in both modes despite the large difference
in size (or $kR$) between the two hexagons.

\subsection{Laser properties} \label{laserprop}

The dye loaded molecular sieve microcrystals presented in section
\ref{pyrid} and \ref{rhod} were excited with 10~ns pulses from the
532~nm second harmonic of a Nd:YAG-laser, and the emitted
luminescence was collected with a 42$^\circ$ aperture lens relaying
the microlaser emission to a spectrometer and an imaging system
consisting of a cooled low noise CCD-camera.

\subsubsection{Pyridine2 AlPO4-5 compound}
\begin{figure}[!ht]
\begin{center}
\resizebox{0.7\textwidth}{!}{%
\includegraphics{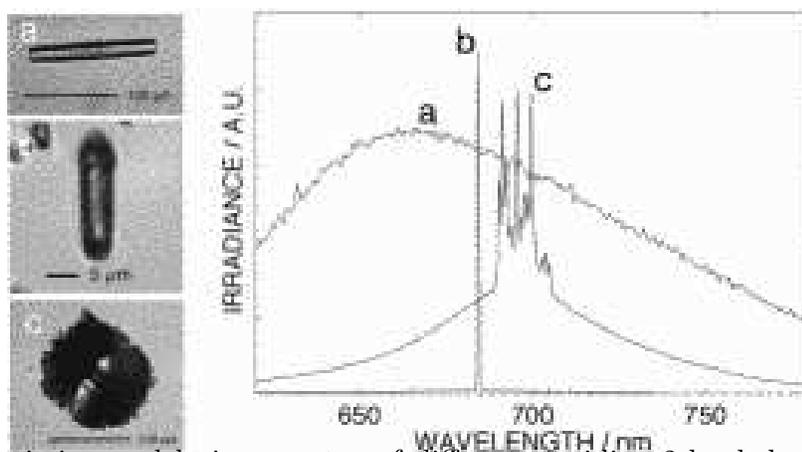}
 }
 \end{center}
\vspace*{-1cm}
\caption{Emission and lasing spectra of different pyridine~2-loaded
compounds with from {\bf a} to {\bf c} increasing amounts of
included dye. The dye concentration was estimated empirically based
on the sample color depth. The width over flats of sample {\bf b}
is 4~$\mu$m. The free spectral range (FSR) of this resonator is so
large (ca.~25~nm) that only one emission mode acquires the
available gain, resulting in single line emission. On the other
hand, sample {\bf c} is ca.\ 5 times larger. In this case the FSR
is around 5~nm so that laser emission can occur on a multitude of
modes simultaneously.
 }
 \label{pyr2spectr}
\end{figure}

In Fig.~\ref{pyr2spectr} the emission and lasing spectra of three
pyridine~2-loaded compounds with different dye content are
compared. Fig.~\ref{pyr2spectr}a represents the class of regularly
shaped samples with a low dye content around 0.1~wt\% [cf.\ also
Fig.~\ref{Fig-Morphology}(left)]. In this class the fluorescence
emission maximum was observed between 645~nm to 665~nm, where the
shift to longer wavelength correlates with the increase of dye
content that was assessed by the saturation of the red color. In
none of these rod-shaped crystals laser emission was observed.

However, narrow laser emission peaks were observed in fascicled
samples with a detected linewidth of ca.\ 0.3~nm which was limited
by the spectrometer resolution. Here the emission maxima were
observed at wavelenghts up to 695~nm, and again, increasing dye
concentration correlated with increasing redshift (cf.\
Figs.~\ref{pyr2spectr}b and c). As already mentioned, together with
the increasing dye content the crystal morphology becomes more and
more disturbed. The observation of a disturbed morphology and the
red-shift of the emission spectrum are consistent with the
hypothesis of a host-guest interaction which increases with dye
content. In fact both, the pyridine~2 molecules, as well as the
AlPO$_4$-5 framework, carry a static dipole moment
\cite{BEN83,MAR94}. As a consequence, the increased buildup of
electrostatic energy in the crystal lattice has to be compensated
by an increasing amount of stacking faults. On the other side, the
mechanism of the redshift is not unequivocally identified, yet, but
is probably related to the one discussed in \cite{FOE49,FOE51}.

\begin{figure}[!ht]
 \begin{center}
 \resizebox{0.6\textwidth}{!}{%
 \includegraphics{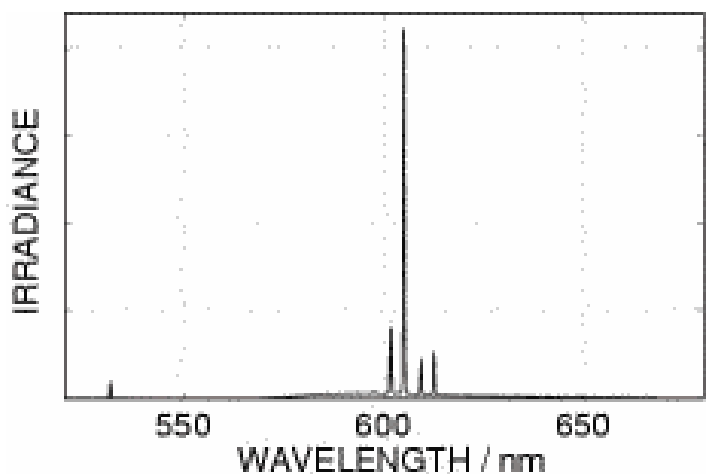}
 }
 \end{center}
\caption{Lasing spectrum of a rhodamine~BE50-loaded AlPO$_4$-5
microcrystal with a concentration around 75 unit cells per dye
molecule and size of 7.5~$\mu$m width over flats.}
 \label{BE50spectr}
\end{figure}

\subsubsection{Rhodamine BE50 AlPO4-5 compound}

Also with these samples the same correlation between the emission
wavelength and dye concentration was observed; cf.\ also
\cite{BOC98}. In contrast to the pyridine 2-loaded samples, the
fluorescence emission was not completely polarized. The observed
polarization contrast $c_p=
\frac{I_\parallel-I_\perp}{I_\parallel+I_\perp}$ was around 10\%,
indicating that in the average the Rh~BE50 molecules are only
weakly aligned with respect to the host crystal. Laser emission was
observed in samples with a dye concentration around 75 unit cells
per dye molecule, corresponding to 0.5~wt\%; cf.\
Fig.~\ref{BE50spectr}.

\subsubsection{Laser properties of molecular sieve dye compounds}

\begin{figure}[!ht]
\begin{center}
 \resizebox{0.3\textwidth}{!}{%
 \includegraphics{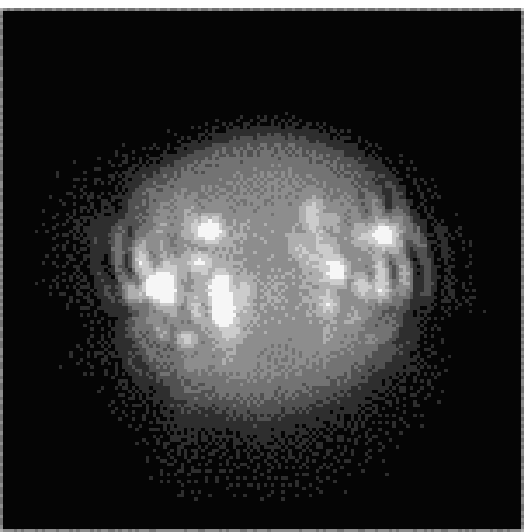}
 }
 \resizebox{0.3\textwidth}{!}{%
 \includegraphics{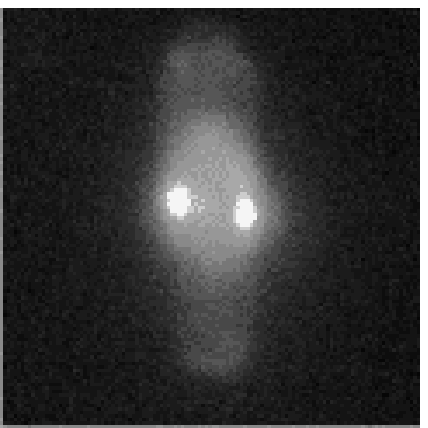}
 }
\end{center}
\vspace*{-.5cm}
 \caption{Patterns of the laser emission show that the emission
originates from regions along side edges. {\bf Left:}
rhodamine~BE50/AlPO$_4$-5 compound; width over flats 7.5~$\mu$m. An
electron micrograph of the sample is shown in
Fig.~\protect\ref{BE50morph} with horizontal $c$-axis. Here the
$c$-axis orientation is nearly vertical. The corresponding emission
spectrum is shown in Fig.~\protect\ref{BE50spectr}. {\bf Right:}
pyridine~2/AlPO$_4$-5 compound; width over flats 4.5~$\mu$m. The
corresponding sample and emission spectrum is represented in
Fig.~\protect\ref{pyr2spectr}~{\bf b}.
  }
 \label{nearfield}
\end{figure}

Independent of the type of loading, in most microcrystals with $WoF
\gtrapprox$~8~$\mu$m lasing was observed to occur on several sharp
lines with instrument resolution limited width (a typical example
is shown in Fig.~\ref{BE50spectr}). In general the laser emission
lines are not equally spaced. In fact, the free spectral range
($FSR$) of 11~nm corresponding to the resonator size of
$WoF=8\;\mu$m is far above the line spacing of 3.2, 4.3 and 3.4~nm
shown in Fig.~\ref{BE50spectr}. This is in agreement with the
theoretical model of paragraph~\ref{WavPic}, in which the average
lasing mode spacing (after lifting the quasi-degeneracies in the
ideal hexagon) was estimated to be $\Delta\lambda =$ 3.6~nm. Also
in agreement with the theoretical discussion are the regions where
the laser light leaves the hexagonal resonators.
Figure~\ref{nearfield}(left) shows the laser emission as bright
spots. Clearly the emission is concentrated along the crystal
edges. The complex emission distribution is compatible with the
simultaneously recorded spectrum (cf.\ Fig.~\ref{BE50spectr}) which
reveals emission on four modes.

While the average line spacing of 3.6~nm observed in the sample of
Fig.~\ref{BE50spectr} is not compatible with the free spectral
range of $\Delta\lambda = 12$~nm of the corresponding resonator,
the 4.2~nm spacing of the 3 dominant peaks in the sample shown in
Fig.~\ref{pyr2spectr}c is in accord with the $FSR$ resulting from
the 22-$\mu$m-WoF hexagonal resonator. On the other hand, the
theoretical model for the $22\,\mu$m-WoF hexagonal resonator (cf.\
Eq.~(\ref{wgweylformtseqn})) yields an average mode spacing of
$\Delta\lambda\approx$ 0.5~nm which is close to the spectrometer
resolution. This high spectral density explains the large
background in the lasing spectrum of Fig.~\ref{pyr2spectr}: It is
likely that not all the individual lasing modes in this sample are
resolved, and hence part of the shoulder on which the three peaks
of curve c sit is probably shaped by a series of closely spaced
lasing modes.

On the other hand, samples with smaller resonator ($WoF
\gtrapprox$~4~$\mu$m), as e.g.\ the one shown in
Fig.~\ref{pyr2spectr}b, emitted one single laser line. Thus the
emission is unadulterated by hole burning induced multimode beating
and interference, and this results in the simple emission pattern
shown in Fig~\ref{nearfield}(right), where two ca.~1~$\mu$m-spots
($\approx$ microscope resolution limit) mark the region of laser
emission, which, again, is located at the crystal edges.
Remarkably, compared to larger samples, the ratio of the line peak
to the underlying fluorescence shoulder of these small lasers is an
order of magnitude higher (cf.\ Figs.~\ref{pyr2spectr} and
\ref{threshold}).

\subsubsection{Laser threshold}
\begin{figure}[!ht]
\begin{center}
\resizebox{0.5\textwidth}{!}{%
  \includegraphics{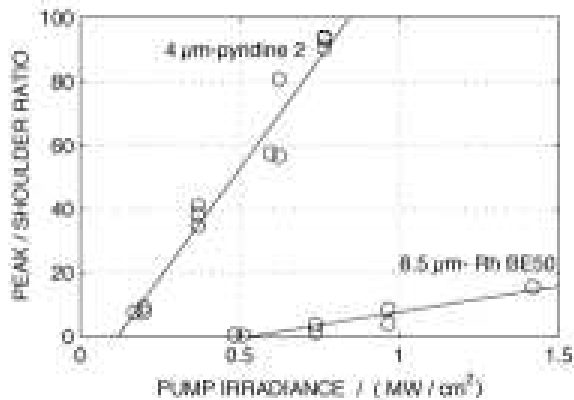}
 }
 \end{center}
\caption{Lasing threshold and differential efficiency of typical
AlPO$_4$-5/dye compounds. Shown is the peak of the laser emission
spectrum normalized by the fluorescence shoulder as a function of
the pump power density for the sample shown in
Figs.~\protect\ref{pyr2spectr}b and \protect\ref{BE50spectr}.
 }
 \label{threshold}
\end{figure}

Figure~\ref{threshold} illustrates the differential efficiency
behavior of a typical microlaser with $WoF < 10\,\mu$m and one with
$WoF > 10\,\mu$m. Lasing threshold for the latter size samples was
around 0.5~MW/cm$^2$, regardless of the type of dye loading. On the
other side, crystals of smaller size ($WoF=4\;\mu$m) from the same
synthesis batch revealed a considerably smaller threshold
(0.12~MW/cm$^2$) and a factor of $>7$ larger differential gain. We
assume that this is a consequence of the quantum size effects which
we described in section \ref{sect-stimem}.

It is informative to compare the threshold of molecular sieve
microlasers with vertical cavity surface emitting lasers (VCSELs).
For that we convert the pump irradiance $I$ (incident optical power
per unit area) into a flux of photons $\phi$ (number of pump
photons per second incident on the mode cross section $A$)
\begin{equation}
\phi = \frac{I\; A}{h\,\nu}=\frac{I\; A\; \lambda_p}{h\, c} \; .
 \label{NUM-OF-PHOT-AT-THRESH}
\end{equation}
If $\eta$ denotes the efficiency with which an incident pump photon
actually contributes to an excitation, then $\phi\times\eta$
corresponds to the number of quantum processes per second required
to reach a given level of inversion. Multiplication of this number
with the charge of an electron $q_e=1.6\times10^{-19}$As gives the
electric current which corresponds to the pump current flowing in a
comparable electric device, such as a VCSEL. According to
\eqref{NUM-OF-PHOT-AT-THRESH} the threshold power density of
0.12~MW/cm$^2$ incident on the surface of the molecular sieve laser
shown in Fig.~\ref{pyr2spectr}b of $1 \times 4\;\mu$m (cf.\
Fig.~\ref{nearfield}b) corresponds to a flux of $1.4 \times
10^{16}$~s$^{-1}$ 532-nm-photons. If we take into account that the
pump field was not polarized, but the dye molecules are aligned,
then only 50\% of the pump photons could contribute to the
inversion. In addition the absorption length for 532~nm radiation
is longer than the crystal size so that only a fraction of the
incident pump is absorbed by the dye molecules. If we assume that
50\% of the pump are absorbed, then the effects of polarization and
absorbtion length together result in a quantum efficiency of
$\eta=0.25$. With this quantum efficiency we obtain a comparable
electric threshold current of 560~$\mu$A, which compares well with
the threshold current of VCSELs of comparable size. Thus, in terms
of elementary (quantum) pump processes needed to reach lasing
threshold the molecular sieve lasers are as good as actual VCSELs.

\subsection{Photostability} \label{phstab}

Photostability is usually a critical issue with organic dyes.
Corresponding investigations were carried out with pyridine~2, as
well as rhodamine~BE50 loaded samples, revealing some unexpected
results.

\subsubsection{Photostability of pyridine~2 compounds}
\begin{figure}[!ht]
 \begin{center}
\resizebox{.6\textwidth}{!}{%
 \includegraphics{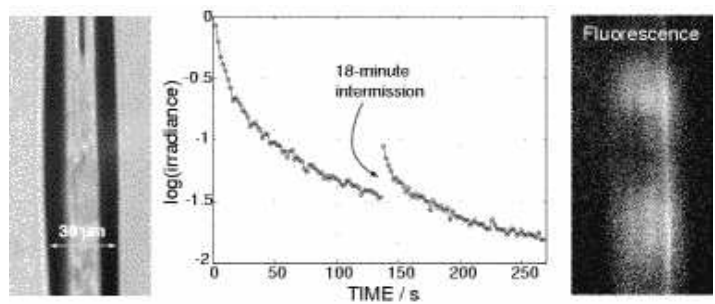}
 }
 \end{center}
\caption{Fluorescence activity of a pyridine~2-loaded AlPO$_4$-5
sample under bleaching laser irradiation. {\bf Left:} Micrograph of
the sample crystal. {\bf Center:} Fluorescence activity as a
function of time: After a first bleach period of 140~seconds the
bleach beam is interrupted for 18~minutes. During this intermission
the fluorescence recovers to start the second bleach period with
$3\times$ stronger emission. {\bf Right:} The bleaching laser is
incident from the left and concentrated in the center of the
crystal. Shown is the fluorescence distribution at the end of the
first bleach period of 140~s. Clearly visible is the bleached hole
in the center, where the bleach beam was concentrated.}
 \label{pyr2bleach}
\end{figure}

The photostability of pyridine~2 compounds was studied with samples
exhibiting an undisturbed morphology, similar to the one shown in
Fig.~\ref{fig-dichro-pyridine}. The samples were irradiated with
10~Hz trains of 10~ns pulses of the 532~nm second harmonic of a
Nd:YAG-laser and a power density of 5~MW/cm$^2$.
Figure~\ref{pyr2bleach} illustrates the wane of fluorescence
activity of a pyridine~2-loaded AlPO$_4$-5 sample under such
bleaching irradiation. After a bleaching period of 140~seconds the
exposure was interrupted for 18~minutes. Then the bleaching
procedure was resumed. The figure shows that the fluorescence
recovers during the intermission.

The physical origin of this unexpected fluorescence recovery is not
clear yet. If we assume that bleaching consists in breaking bonds
of the dye molecules, then we have to consider bond energies in the
eV-range. Even if we assume that the dye debris stay encaged in
their pores, spontaneous or thermally activated self healing of
broken eV-bonds seems not very probable. Considering the
stereometrically restricted possibilities inside the molecular
sieve framework together with diffusion distances of several
$\mu$m, and the observed recovery time in the range of minutes, we
believe that diffusion of new, intact dye molecules into the
irradiated crystal volume is more plausible than self healing.

As bleaching reduces the concentration of dye, a blue\-shift of the
fluorescence is expected with increasing photobleaching
\cite{BEN83,MAR94}. However, the 656-nm-fluorescence emission
maximum of this sample is already at the shortest observed
wavelength (cf.\ Fig.~\ref{pyr2spectr}), corresponding to a low dye
concentration, and consequently, to weak dipole interactions. Thus,
the blueshift under these circumstances must be rather small. This
explains that a blueshift was not observed with these
pyridine~2-loaded samples.

\subsubsection{Photostability of rhodamine~BE50 compounds}
\begin{figure}[!ht]
 \begin{center}
\resizebox{0.75\textwidth}{!}{%
 \includegraphics{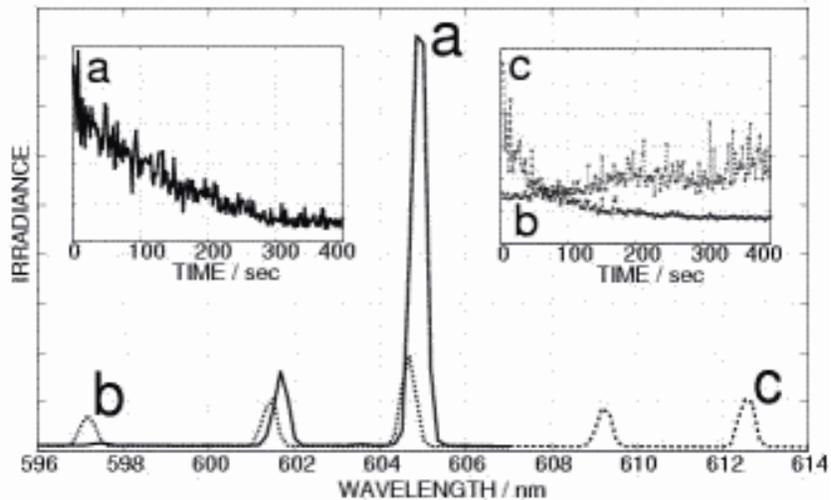}
 }
 \end{center}
\caption{Effect of photobleaching on the laser emission intensity:
short wavelength line b grows with progressive bleaching, while
longer wavelength lines a and c decrease.}
 \label{be50bleach}
\end{figure}

The rhodamine~BE50-loaded samples under investigation contained dye
at a concentration of around one Rh~BE50 molecule per 75 unit
cells, and therefore bleaching caused a detectable 4~nm shift of
the fluorescence towards the blue. The 532~nm bleach irradiance was
0.5~MW/cm$^2$ with this samples. In contrast to the pyridine~2
samples these crystals exhibited laser emission. Observing the
laser emission spectrum while bleaching the samples, a further
consequence of the blueshift was revealed: Blueshift of the
fluorescence reduces the overlap of the fluorescence band with the
absorption spectrum, and as a result, laser modes at lower
wavelengths will suffer less losses with increasing bleaching. This
is documented in Fig.~\ref{be50bleach}, where the intensity of mode
b with the shortest oscillating wavelength increases, while
longer-wavelength-modes a and c decrease during the bleach
procedure. At the same time a 0.2~nm blueshift of the oscillation
wavelength was detected. We attribute this to a weak decrease of
the refractive index of the resonator material due to the smaller
polarizability of the dye debris. In contrast to the pyridine 2
samples, however, recovery of the fluorescence was not detected. As
we mentioned in section \ref{rhod} the Rh~50BE molecules are
considerably larger than pyridine 2 molecules, and are encaged in
mesopores. As a result their mobility in the molecular sieve
framework is severly hampered. So, diffusion of intact molecules
into the bleached volume occurs -- if ever -- on timescales
considerably larger than minutes. This is probably the reason why
fluorescence recovery was not observed in rhodamine~BE50 compounds.

\section{Conclusion}

This chapter was centered on microscopic lasers realized with
nanoporous materials, especially molecular sieve dye compounds. We
discussed the effects of a microresonator on spontaneous and
stimulated emission properties, and we showed that wavelength scale
laser resonators may exhibit lasing without threshold. Such effects
are not just thoughtful observations, but were investigated in real
devices as well, and we reviewed observations of threshold
reduction in molecular sieve lasers.

The microresonators of the molecular sieve lasers which were
presented up to now exhibit a hexagonal resonator geometry.
Hexagonal microresonators in which the light field is confined by
total internal reflection at the dielectric boundary define a new
class of pseudointegrable optical structure of which we reviewed
the properties in some detail.

As mentioned, molecular sieve microlasers can be fabricated in
large amounts with the same large scale processes which are used to
produce the molecular sieve crystallites in the petrochemical
industry. Although applications of such \textit{laser powders} have
not been reported yet, many practical uses can be envisaged, such
as for example in efficient fluorescent, or even lasing paint
pigments or lasing pixels.

\subsection{Acknowledgements}

We thank the following individuals for their help in preparing this
text as well as for their contributions on the way to realize
lasing in molecular sieves: P. Behrens, I. Braun, G. Ihlein, O.
Krau\ss, F. Marlow, G. Schulz-Ekloff, F. Sch\"uth, U. Vietze, \"O.
Weiss, and D. W\"ohrle. Some of the work reviewed here was
supported by the ``Deutsche Forschungsgemeinschaft'' and the
``Max-Planck-Gesellschaft''.


\begin{thebibliography}{999}


\bibitem{KRA98} J. Kr\"ager, K. Hahn, V. Kukla, and C. R\"odenbeck,
Phys.\ Bl., \textbf{54}, 811 (1998)

\bibitem{GFE98} N. Gfeller, S. Megelski, and G. Calzaferri, J. Phys.\ Chem.\ B
\textbf{102,} (1998) 2433;  J. Phys.\ Chem.\ B \textbf{103,} 1250
(1999)

\bibitem{VIE98} U. Vietze, O. Krau\ss, F. Laeri, G. Ihlein, F.
Sch\"uth, B. Limburg, and M. Abraham, Phys.\ Rev.\ Lett.,
\textbf{81,} 4628 (1998)





\bibitem{FEY64} R. P. Feynman, R. B. Leighton, and M. Sands, \textit{The Feynman
Lectures on Physics Vol.\ II}, Addison-Wesley, Reading
(Massachusetts), 1964, p.\ II-27-5

\bibitem{MER70} E. Merzbacher, {\em Quantum Mechanics}, 2nd.\ edn.,
Wiley, New York, 1970, S. 356

\bibitem{HEI54} W. Heitler, {\em The Quantum Theory of Radiation}, 3rd.\
edn., Oxford University Press, London, 1954

\bibitem{LOU73} W. H. Louisell, {\em Quantum Statistical Properties of
Radiation}, Wiley, New York, 1973

\bibitem{GLA63} R. J. Glauber, Phys.\ Rev., {\bf 131}, 2766 (1963)

\bibitem{BAS97} T. Basch\'e, W. E. Moerner, M. Orrit, and U. P. Wild, eds.,
\textit{Single-Molecule Optical Detection, Imaging, and
Spectroscopy}, VCH-Verlagsgesellschaft, Weinheim (Germany), 1996

\bibitem{MOE99} W. E. Moerner and M. Orrit, Science {\bf 283}, 1670 (1999).





\bibitem{ALL75} L. Allen and J. H. Eberley, \textit{Optical resonance and
two-level atoms and ions}, Wiley, New York, 1975

\bibitem{RAB37} I. I. Rabi, Phys.\ Rev., {\bf 51}, 652 (1937)

\bibitem{BLO46} F. Bloch, Phys.\ Rev., {\bf 70}, 460 (1946)

\bibitem{SIF68} L.~I.~Schiff, \textit{Quantum mechnics}, McGraw-Hill,
New York, 1968

\bibitem{HEC87} E. Hecht, \textit{Optics; 2nd ed.}, Addison-Wesley,
Reading, 1987, p.~322.

\bibitem{FEY57} R. P. Feynman, F. L. Vernon, and R. W. Hellwarth,
J. Appl.\ Phys., {\bf 28}, 49 (1957)

\bibitem{WAL77} H. Walther, Phys.\ Bl., {\bf 33}, 653 (1977)

\bibitem{DAG78} M. Dagenais and L. Mandel, Phys.\ Rev.~A, {\bf 18},
2217 (1978)

\bibitem{SHO90} B. W. Shore, P. Meystre, and S. Stenholm, J. Opt.\
Soc.\ B., {\bf 8}, 903 (1991)

\bibitem{WEI30} V. Weisskopf and E. P. Wigner, Z. Phys., {\bf 63},
54 (1930)

\bibitem{MEY91} P. Meystre ans M. Sargent III, \textit{Elements of Quantum
Optics, 2nd ed.}, Springer, Berlin, 1991

\bibitem{GIS92} N. Gisin and I. Percival, Phys.\ Rev.\ A, {\bf 46},
4382 (1992)

\bibitem{MOE93} K. Moelmer, Y. Castin, and J. Dalibard, J.
Opt.\ Soc.\ Am.\ B, {\bf 10}, 524 (1993)




\bibitem{PUR46} E. M. Purcell, Phys.\ Rev., {\bf 69}, 681 (1946)

\bibitem{KLE81} D. Kleppner, Phys.\ Rev.\ Lett., {\bf 47}, 233
(1981)

\bibitem{RAI89} M. G. Raizen, R. J. Thompson, R. J. Brecha, H. J.
Kimble, and H. J. Carmichael, Phys.\ Rev.\ Lett., {\bf 63}, 240
(1989)

\bibitem{BER92} F. Bernardot, P. Nussenzweig, M. Brune, J. M.
Raimond, and S. Haroche, Europhys.\ Lett., {\bf 17}, 33 (1992)

\bibitem{SIE86} A. E. Siegman, \textit{Lasers}, University Science
Books, Mill Valley(CA), 1986

\bibitem{PAR87} J. Parker and C. R. Stroud, Phys.\ Rev.\ A, {\bf 35},
4226 (1987)

\bibitem{COO87} R. G. Cook and P. W. Milonni, Phys.\ Rev.\ A, {\bf 35},
5071 (1987)

\bibitem{CAR89} H. J. Carmichael, R. J. Brecha, M. G. Raizen, H. J.
Kimble, and P. R. Rice, Phys.\ Rev.\ A, {\bf 40}, 5516 (1989)

\bibitem{JAY63} E. T. Jaynes and F. W. Cummings, Proc.\ IEEE, {\bf
51}, 89 (1963)



\bibitem{MES85} D. Meschede, H. Walter, and G. M\"uller, Phys.\
Rev.\ Lett., {\bf 54}, 551 (1985)

\bibitem{YOK92} H. Yokoyama, K. Nishi, T. Anan, Y. Nambu, S. D. Brorson,
E. P. Ippen, and M.~Suzuki, Opt.\ Quantum Electron., {\bf 24}, S245
(1992)

\bibitem{MAR88} F. De Martini and G. R. Jacobovitz, Phys.\ Rev.\
Lett., {\bf 60}, 1711 (1988)




\bibitem{BLO90} K. J. Blow, R. Loudon, S. J. D. Phoenix, and T. J. Shepherd,
Phys.\ Rev.\ A, \textbf{42}, 4102 (1990)

\bibitem{JAC98} J. D. Jackson, \textit{Classical electrodynamics, 3rd ed.,}
Wiley, New York, 1998

\bibitem{GLA91} R. J. Glauber and M. Lewenstein, Phys.\ Rev.\ A, \textbf{43},
467 (1991)

\bibitem{SEL98} S. Scheel, L. Kn\"oll, and D. G. Welsch, Phys.\ Rev.\ A,
\textbf{58}, 700 (1998)

\bibitem{HOU94} R. Houdr{\'e}, C. Weisbuch, R. P. Stanley, U. Oesterle,
P. Pellandini, and M. Ilegems,  Phys.\ Rev.\ Lett., \textbf{73},
2043 (1994)

\bibitem{KLI99} V. V. Klimov, M. Ducloy, and V. S. Letokhov, Phys.\ Rev.\ A,
\textbf{59}, 2996 (1999)

\bibitem{SEN95} T. B. A. Senior and J. L. Volakis, \textit{Approximate boundary
conditions in electromagnetics}, IEE Electromagnetic waves series
41, London, 1995

\bibitem{NUS92} L. G. Guimares, H. M. Nussenzveig, and W. J. Wiscombe,
Opt.\ Commun., \textbf{89}, 363 (1992)

\bibitem{HOH93} M. C. Cross and P. C. Hohenberg, Rev.\ Mod.\ Phys.,
\textbf{65}, 851 (1993)

\bibitem{WEI92} C. O. Weiss, M. Vaupel, K. Staliunas, G. Slekys,
V. B. Taranenko, Appl.\ Phys.\ B, \textbf{68}, 151 (1999); C. O.
Weiss, Phys.\ Rep., \textbf{219}, 311 (1992)

\bibitem{HUL81} H. C. van de Hulst, \textit{Light scattering by small particles},
Dover Publications, New York, 1981

\bibitem{KER69} M. Kerker, \textit{The scattering of light and other
Electromagnetic Radiation}, Academic Press, New York, 1969

\bibitem{BAR90} P. W. Barber and S. C. Hill, \textit{Light scattering by particles:
Computational methods}, World Scientific, Singapore (1990)

\bibitem{SNY91} A. W. Snyder and J. D. Love, \textit{Optical
waveguide theory}, Chapman and Hall, London (1991)

\bibitem{POO98} A. W. Poon, R. K. Chang, and J. A. Lock, Opt.\ Lett., \textbf{23},
1105 (1998)

\bibitem{AHN99} J. C. Ahn, K. S. Kwak, B. H. Park, H. Y. Kang, J. Y. Kim,
and O'Dae Kwon, Phys.\ Rev.\ Lett., \textbf{82}, 536 (1999)



\bibitem{COU68} R. Courant and D. Hilbert, \textit{Methoden der Mathematischen
Physik I}, Springer, Berlin, 1968

\bibitem{JOH93} B. R. Johnson, J. Opt.\ Soc.\ Am.\ A, \textbf{10}, 343 (1993)

\bibitem{NOE96} J. U. N\"ockel and A. D. Stone, in: \textit{Optical
processes in microcavities}, edited by R.~K.~Chang and
A.~J.~Campillo, World Scientific, Singapore, 1996

\bibitem{NOE97} J. U. N\"ockel, PhD Thesis, Yale University (1997)

\bibitem{FAN61} U. Fano, Phys.\ Rev., \textbf{124}, 1866 (1961)

\bibitem{CHI96} E. S. C. Ching, P. T. Leung, and K. Young, in
\textit{Optical processes in microcavities}, R. K. Chang and A. J.
Campillo, eds., World Scientific, Singapore, 1996

\bibitem{NOE94} J. U. N\"ockel and A. D. Stone, Phys.\ Rev.\
B, \textbf{50}, 17415 (1994)


\bibitem{BAB72} V. M. Babi\v{c} and V. S. Buldyrev, \textit{Short-wavelength
diffraction theory}, Springer, Berlin, 1972

\bibitem{GUT90} M. C. Gutzwiller, \textit{Chaos in classical and quantum
mechanics}, Springer, New York, 1990

\bibitem{SOM47} A. Sommerfeld, \textit{Vorlesungen {\"u}ber Theoretische
Physik VI: Partielle Differentialgleichungen der Physik},
Akademische Verlagsgesellschaft, Leipzig, 1947

\bibitem{FOX61} A. G. Fox and T. Li, Bell Syst.\ Tech.\ J.,
\textbf{40}, 453 (1961)

\bibitem{TUR97} J. Turunen and F. Wyrowski, eds., \textit{Diffractive optics
for industrial and commercial applications}, Wiley-VCH, Berlin,
1997

\bibitem{YAR75} A. Yariv, \textit{Quantum electronics}, Wiley, New York,
1975

\bibitem{ANG96} G. Angelow, F. Laeri, and T. Tschudi, Opt.\ Lett., \textbf{21},
1324 (1996)

\bibitem{STO99} H. J. St\"ockmann, \textit{Quantum chaos -- an
introduction}, Cambridge University Press, 1999

\bibitem{GMA98} C. Gmachl, F. Capasso, E. E. Narimanov, J. U. N\"ockel,
A. D. Stone, J. Faist, D. L. Sivco, and A. Y. Cho, Science,
\textbf{280}, 1556 (1998)

\bibitem{RIC81} P. J. Richens and M. V. Berry, Physica, \textbf{2D}, 495 (1981)

\bibitem{HOB75} A. Hobson, J. Math.\ Phys., \textbf{16}, 2210 (1975)

\bibitem{WOE99} G. P. Karman, G. S. McDonald, G. H. C. New, and J. P. Woerdman,
Nature, \textbf{402}, 138 (1999)

\bibitem{SIE97} J. B. Keller, J. Opt.\ Soc.\ Am., \textbf{52}, 116
(1962); M. Sieber, N. Pavloff, and C. Schmit, Phys.\ Rev.\ E,
\textbf{55}, 2279 (1997)

\bibitem{VEG95} J. L. Vega, T. Uzer, and J. Ford, Phys.\ Rev.\ E, \textbf{52},
1490 (1995)

\bibitem{BIS98} D. Biswas, Phys.\ Rev.\ E \textbf{57}, R3699 (1998)




\bibitem{MCC92} S. L. McCall, A. F. J. Levi, R. E. Slusher, S. J.
Pearton, and R. A. Logan, Appl.\ Phys.\ Lett., {\bf 60}, 289 (1992)

\bibitem{LEV93} A. F. J. Levi, R. E. Slusher, S. L. McCall, S. J.
Pearton, and W. S. Hobson, Appl.\ Phys.\ Lett., {\bf 62}, 2021
(1993)

\bibitem{KOY88} F. Koyama, S. Kinoshita, and K.Iga, Trans.\ Inst.\
Electron.\ Inf.\ Commun.\ Eng.\ E, \textbf{E71}, 1089 (1988)

\bibitem{JEW89} J. L. Jewell, S. L. McCall, Y. H. Lee, A. Schere,
A. C. Gossard, and J. H. English, Appl.\ Phys.\ Lett., {\bf 55},
1400 (1989)

\bibitem{SER89} A. Schere, J. L. Jewell, Y. H. Lee, J. P. Harbison,
and L. T. Florez,  Appl.\ Phys.\ Lett., {\bf 55}, 2724 (1989)

\bibitem{GEE90} R. S. Geels and L. A. Coldren,  Appl.\ Phys.\
Lett., {\bf 57}, 1605 (1990)

\bibitem{CHA93} C. J. Chang-Haanain, Y. A. Wu, G. S. Li, G.
Hasnain, K. D. Choquete, C. Ceneau, and L. T. Florez,  Appl.\
Phys.\ Lett., {\bf 63}, 1307 (1993)

\bibitem{DRE74} K. H. Drexhage, in \textit{Progress in Optics,
Vol.\ XII}, edited by E. Wolf, North Holland, Amsterdam, 1974,
p.~165

\bibitem{TZE84} H.-M.\ Tzeng, K. F. Wall, M. B. Long, and R. K.
Chang, Opt.\ Lett., {\bf 9}, 499 (1984)

\bibitem{CAM91} A. J. Campillo, J. D. Eversole, and H.-B. Lin,
Phys.\ Rev.\ Lett., {\bf 67}, 437 (1991)

\bibitem{KUW92} M. Kuwata-Gonokami, K. Takeda, H. Yasuda, and K. Ema,
Jpn.\ J. Appl.\ Phys., {\bf 31}, L99 (1992)

\bibitem{BER94} P. R. Berman, ed., \textit{Cavity quantum
electrodynamics}, Academic Press, Boston, 1994

\bibitem{YOK95} H. Yokoyama and K. Ujihara, eds., \textit{Spontaneous
Emission and Laser Oscillation in Microcavities}, CRC Press, Boca
Raton, 1995

\bibitem{CHA96} R. K. Chang and A. J. Campillo, eds., \textit{Optical
processes in microcavities}, World Scientific, Singapore, 1996

\bibitem{DUC96} M. Ducloy and D. Bloch, eds., \textit{Quantum optics
of confined systems}, (Proc.\ of the NATO ASI Ser.~E, Vol.~314)
Kluwer, Dordrecht, 1996

\bibitem{RAR96} J. Rarity and C. Weisbuch, eds., \textit{Microcavities
and photonic bandgaps: Physics and applications}, (Proc.\ of the
NATO ASI Ser.~E, Vol.~324) Kluwer, Dordrecht, 1996


\bibitem{MEI96} W. M. Meier, D. H. Ohlson, and C. Baerlocher,
\textit{Atlas of Zeolite Structure Types}, Elsevier, London, 1996.
cf.\ also the URL:
http://www.iza-sc.ethz.ch/IZA-SC/Atlas/AtlasHome.html

\bibitem{WIL91} S. T. Wilson, in H. van Bekkum, E. M. Flanigen, and J.
C. Jansen (Eds.), \textit{Introduction to Zeolite Science and
Practice}, Elsevier, Amsterdam, 1991,  Stud.\ Surf.\ Sci.\ Catal.,
vol.~58, p.~137

\bibitem{GIR95} I. Girnus, K. Jancke, R. Vetter, J. Richter-Mendau,
and J. Caro, Zeolite, \textbf{15,} 33 (1995); I. Girnus, M. Poll,
J. Richter-Mendau, M. Schneider, M. Noack, D. Venzke, and J. Caro,
Adv.\ Mater., \textbf{7,} 711 (1995)

\bibitem{SUN96} S. A. Schunk, D. G. Demuth, B. Schulz-Dobrik, K. K.
Unger, and F. Sch\"uth, Microporous Mater., \textbf{6,} 273 (1996);
\"O. Weiss, G. Ihlein, and F. Sch\"uth, Micrporous and Mesoporous
Mater., in press.

\bibitem{DU97} H. Du, M. Fang, W. Xu, X. Meng, and W. Pang, J. Mater.\
Chem., \textbf{7,} 551 (1997)

\bibitem{BEN83}J. M. Bennett, J. R. Cohen, E. M. Flanigen, J. J.
Pluth, and J. V. Smith, in \textit{Intrazeolite Chemistry}, edited
by G. D. Stucky and F. G. Dwyer, ACS Symp.\ Series, vol.~218, Am.\
Chem.\ Soc., Washington DC, 1983, p.~109

\bibitem{MAR94} J. A. Martens and P. A. Jacobs, in \textit{Advanced
Zeolite Science and Applications}, edited by J. C. Jansen, M.
St\"ocker, H. G. Karge, and J. Weitkamp, Elsevier, Amsterdam, 1994,
Stud.\ Surf.\ Sci.\ Catal., vol.~85, p.~653

\bibitem{KLA99} G. J. Klap, S. M. van Klooster. M. W\"ubbenhorst. J. C.
Jansen, H. van Bekkum, and J. van Turnhout, in \textit{Proc.\
12$^{\rm th}$ Intern.\ Zeolite Conf.}, edited by M. M. J. Treacy,
B. K. Marcus, M. E. Bisher, and J. B. Higgins, Materials Research
Society, Warrendale(PA), 1999, vol.~3, p.~2117

\bibitem{BRA94} U. Brackmann, \textit{Lambdachrome Laser Dyes},
Lambda Physik, G\"ottingen, 1994, p.~198

\bibitem{IHL99} G. Ihlein, PhD Thesis, University of Frankfurt,
Frankfurt a. M., 1998

\bibitem{HOP93} R. Hoppe, G. Schulz-Ekloff, D. W\"ohrle, E. S.
Shpiro, P. P. Tkachenko, Zeolites, \textbf{13,} 222 (1993)

\bibitem{DEM95} D. Demuth, G. D. Stucky, K. K. Unger, F. Sch\"uth,
Microporous Mater., \textbf{3,} 473 (1995)

\bibitem{IHL98} G. Ihlein, F. Sch\"uth, O. Krau\ss, U. Vietze, F.
Laeri, Adv.\ Mater., \textbf{10,} 1117 (1998)

\bibitem{BOC98} M. Bockstette, D. W\"ohrle, I. Braun, G.
Schulz-Ekloff, Microporous and Mesoporous Mater., \textbf{23,} 83
(1998)

\bibitem{BRA99} I. Braun, G. Schulz-Ekloff, G. Schnurpfeil. D.
W\"orle, K. Hoffmann, in preparation.

\bibitem{WIL82} S. T. Wilson, B. M. Lok, C. A. Messina, T. R.
Cannan, E. M. Flanigen, J. Am.\ Chem.\ Soc., \textbf{104,} 1146
(1982)

\bibitem{WOH92} S. Wohlrab, R. Hoppe, G. Schulz-Ekloff, D.
W\"ohrle, Zeolites, \textbf{12,} 862 (1992); D. W\"ohrle, A.K.
Sobbi, O. Franke, G. Schulz-Ekloff, Zeolites, \textbf{15,} 540
(1995)

\bibitem{AL2O3} 8.44~g; Pural~SB, Condea Chemie

\bibitem{P2O5} 14.20~g phosphoric acid; 85~wt\%, p.a.\ Merck

\bibitem{P3N} 13.25~g $\rm Prop_3N$, Merck

\bibitem{BRA98} I. Braun, G. Schulz-Ekloff, D. W\"ohrle, W.
Lautenschl\"ager, Microporous and Mesoporous Mater., \textbf{23,}
79 (1998)

\bibitem{BRA97} I. Braun, M. Bockstette, G. Schulz-Ekloff, D.
W\"ohrle, Zeolites \textbf{19,} 128 (1997)

\bibitem{MEK95} A. Mekis, J. U. N\"ockel, G. Chen, A. D. Stone,
and R. K. Chang, Phys.\ Rev.\ Lett., \textbf{75,} 2682 (1995)

\bibitem{NOE97b} J. U. N\"ockel and A. D. Stone, Nature,
\textbf{385,} 45 (1997)

\bibitem{NOEun} J. U. N\"ockel and A. D. Stone, unpublished

\bibitem{DAV81} M. J. Davis, J. E. Heller, J. Chem.\ Phys., \textbf{75,}
246 (1981)

\bibitem{FOE49} T. F\"orster, Z. Naturforsch.\ A, \textbf{4,} 321
(1949)

\bibitem{FOE51} T. F\"orster, \textit{Fluoreszenz Organischer
Verbindungen} (Vandenhoeck \& Ruprecht, G\"ottingen, 1951) p.~139ff

\end{thebibliography}
\end{document}